\documentclass[letterpaper, 12pt]{article}

\usepackage{amsmath}
\usepackage{amssymb}
\usepackage{amsfonts}
\usepackage{amsthm}
\usepackage{dsfont}
\usepackage{graphicx}
\usepackage{color}
\usepackage{ushort}
\usepackage{xurl}
\usepackage{lscape}
\usepackage{subcaption}
\usepackage{pdflscape}
\usepackage{bbm}
\usepackage{epstopdf}
\usepackage[onehalfspacing]{setspace}
\usepackage[longnamesfirst]{natbib}

\shortcites{gmbc2007}
\usepackage{enumerate,enumitem}
\usepackage{libertine}
\usepackage[T2A, T1]{fontenc}
\usepackage[utf8]{inputenc}
\usepackage[english]{babel}
\usepackage{soul}
\usepackage{tcolorbox}
\usepackage{xcolor,colortbl}
\usepackage{threeparttable,multirow}
\usepackage{xfrac}
\usepackage{threeparttable}
\usepackage{rotating}
\usepackage{tabularx,ragged2e,booktabs}
\usepackage{subfiles}
\usepackage[margin=1in]{geometry}
\usepackage{bm}
\usepackage[title]{appendix}
\usepackage{setspace}

\definecolor{red}{rgb}{0.9,0.1,0.4}

\DeclareMathOperator{\argmax}{argmax}

\DeclareMathOperator{\Prob}{Pr}

\newtheorem{proposition}{Proposition}
\newtheorem{theorem}{Theorem}

\newtheorem{lemma}{Lemma}
\newtheorem{assumption}{Assumption}

\theoremstyle{remark}

\theoremstyle{plain}

\definecolor{Blue}{RGB}{0,32,216}
\usepackage[colorlinks=true,allcolors=Blue]{hyperref}
\hypersetup{pdftitle={Contracting for Experimentation with Private Findings},pdfauthor={Hongcheng Li},bookmarksnumbered=true,bookmarksdepth=2}
\makeatletter
\def\proofref#1{\hyperref[app:proof \proofref@name#1\@nil]{\ref*{#1}}}\def\proofref@name#1:#2\@nil{#2}
\makeatother
\usepackage{xcolor}
\usepackage{rotating}

\title{Contracting for Experimentation with Private Findings\thanks{I am indebted to Marina Halac, Navin Kartik, and Elliot Lipnowski for their invaluable advice. I thank Sandeep Baliga, Theo Durandard, Georgy Egorov, Tan Gan, Ying Gao, Olivier Gossner, Yuhta Ishii, Ilia Krasikov, Xiao Lin, Tianhao Liu, Chiara Margaria, Sofia Moroni, Ellen Muir, Ferdinand Pieroth, Sara Shahanaghi, Nicholas Wu, and Haoyuan Zeng for their helpful discussions and comments. I also thank audiences at Yale.}}
\author{Hongcheng Li\thanks{
        Hongcheng Li:  Department of Economics, Yale University (e-mail: \href{mailto:hongcheng.li@yale.edu}{hongcheng.li@yale.edu}; address: 27 Hillhouse Ave, B04, New Haven, CT 06511, USA).}}

\date{October 31, 2026}

\begin{document}
\onehalfspacing
\maketitle

\vspace{-12mm}

\begin{center}
\href{https://drive.google.com/file/d/1kCb7pSahbTwF9bkAlopU_Az2qtgRztbN/view?usp=sharing}{Click here for the latest version}
\end{center}

\vspace{2mm}

\begin{abstract}
    \noindent A principal hires an agent to work on a project of unknown quality. Success is contractible, but before it occurs, the agent may privately obtain a favorable finding. For example, a scientist developing a drug may observe a promising microscope reading. An informed agent may conceal the finding; an uninformed agent may fake one. The agent's private information could guide both the project's duration and his pay. Yet I show that an optimal contract pays only success bonuses tied to time, not reported findings, consistent with time-based payment schedules often used in research funding. Dynamically arriving private information drives this simplicity: the agent can conceal good news to retain higher bonuses, so deterring concealment limits pay differences across report histories. Bonuses are high and rising in a research stage that induces learning, then fall to a low, constant level in a development stage where only an informed agent continues working. Rising bonuses reflect a novel force: rents from private learning motivate effort but shrink as fewer research periods remain.
\end{abstract}



\newpage

\tableofcontents

\newpage

\section{Introduction}\label{sec:intro}

Experimentation often generates information about a project's prospects that only those doing the work observe. Consider a pharmaceutical firm that hires a scientist to develop a drug of uncertain viability. The firm seeks a \emph{success}, such as demonstrating the drug's effectiveness in a trial. Before success, however, the scientist may obtain \emph{findings} about the drug's viability, such as promising test results or microscope readings that only he can interpret. Only the scientist knows whether and when he obtains such findings, and his incentives to report them depend on how his reports affect compensation. He may conceal a finding if he expects to earn more by remaining quiet, or fake one if he expects to earn more by reporting good news. Similar issues arise when government agencies fund investigators or when firms form research alliances with universities.

In these settings, findings could guide both how long the principal runs the project and how much she pays the agent. A favorable finding raises the perceived probability of success, making continued work more attractive to the principal and allowing her to motivate the agent's effort with a smaller bonus upon success. In principle, the principal can offer a rich contract with success bonuses, wages, severance payments, and payments upon a report. Each of these transfers can depend on whether and when the agent reports a finding, tailoring his incentives to the information he holds.

In practice, however, research funding often takes a simpler form: reports guide whether a project continues into later phases, but payment amounts depend on the time since project approval rather than on the content of the reports. The National Institutes of Health (NIH), for example, requires annual progress reports from the investigators it supports. Yet the amount allocated for each project year is generally specified when the awards are first issued, rather than adjusted to the findings described in the reports.\footnote{Progress reports ``usually are required annually as part of the non-competing continuation award process,'' and NIH staff ``may require additional information to evaluate the project for continued funding'' \citep*[Section 8.4.1.1]{nigp2026}. The initial Notice of Award lists an amount for each future year as ``[r]ecommended future year total cost support, subject to the availability of funds and satisfactory progress of the project'' \citep*[p.~3]{nisn2024}. Continuation awards made in FY 2026 ``will generally be issued at the commitment level indicated on the Notice of Award'' \citep*{nifp2026}.} NIH instead evaluates the reports to decide whether funding should continue.\footnote{A related example arises in university licensing, where faculty inventors disclose their inventions but receive licensing income according to revenue-sharing rules established by university policy. Yale, for example, requires prompt written disclosure and specifies a schedule for the inventors' share of net revenue \citep*[Sections 3(a) and 4(d)]{yupp2023}. See also \citet*{jeth2001} on university licensing and inventor revenue shares.}

This contrast hints at a tension between the value of tailoring incentives to an agent's private information and the cost of eliciting this information. I therefore ask: How do private findings shape optimal incentives for experimentation? Should the agent's compensation depend on his reports?

To study these questions, I build a model in which a principal hires an agent to work on a project. The project's quality is either high or low; this quality state remains fixed over time and is initially unknown to both the principal and the agent. Both parties are risk neutral and share a common discount factor. In each period, the agent privately chooses whether to work at a cost. Work may produce a contractible success, such as an effective drug, which yields revenue to the principal and ends the relationship. Success is possible in either state but is more likely in the high state. Work may also produce a private, unverifiable finding. Unlike success, a finding is purely informational: it yields no payoff by itself and does not end the relationship. I assume that the finding arrives only in the high state, so a finding reveals that the project's quality is high.\footnote{The finding is conclusive good news about the high state. When findings are instead bad news, Section~\ref{sec:bad news} identifies conditions under which a simple contract is optimal: the principal stops the project after bad news, and the agent has no incentive to misreport, so privacy plays no role. With good-news findings, both informed and uninformed agents can be induced to work, and eliciting the agent's private information becomes more difficult. Section~\ref{sec:inconclusive} discusses inconclusive findings.} I also assume that working is efficient in the high state but inefficient in the low state.

The agent is \emph{informed} once he obtains a finding and \emph{uninformed} before then. Working without either a success or a finding makes an uninformed agent more pessimistic, because this outcome is less likely in the high state. Shirking, by contrast, leaves his belief unchanged. The principal commits to a contract. I show that she can, without loss, ask the agent to report when he obtains a finding and make both the recommended duration of work and all transfers depend on whether and when he reports. Two frictions shape the contract: hidden effort and a limited-liability constraint requiring all transfers to the agent to be nonnegative. Absent either friction, the principal could implement the first-best work schedule and extract the full surplus.

To isolate the effect of privacy, I first study a benchmark setting in which findings are contractible. After a finding is publicly observed, the principal can motivate work with a low success bonus: the agent knows that the state is high, so the expected bonus payment need only cover his effort cost. Before a finding, however, motivating work is more costly. An agent who works and receives no news becomes more pessimistic, whereas one who shirks remains optimistic and thus values later bonuses more. To satisfy \emph{effort obedience}, early bonuses must therefore more than cover, in expectation, the effort cost of an obedient agent, namely one who has worked as intended. Theorem~\ref{thm:cf} shows that the principal optimally sets bonuses so that effort obedience binds in every period before a finding, and pays the low bonus after a finding. The optimal contract induces work for finitely many periods if no finding arrives, but until success if a finding is obtained. Because the bonus offered before a finding also exceeds the bonus offered after one, the contract conditions both the duration of work and the agent's compensation on the finding.

When findings are private, this contract fails to induce truthful reporting. An agent who obtains a finding early would conceal it to keep working for the higher success bonuses offered without a report. Faking a finding, by contrast, would give an uninformed agent access only to the low bonus, at which he would shirk and earn no rent. Privacy thus adds the \emph{no-concealing} constraint: an informed agent must receive at least as much from reporting a finding as from concealing it and continuing to work for the bonuses offered to the uninformed agent. The very reduction in compensation made possible by a favorable finding therefore gives the agent an incentive to hide it. The no-concealing constraint captures the tension between tailoring incentives to private information and eliciting that information.

My main result, Theorem~\ref{thm:main result}, shows that this tension eliminates the gains from making the agent's compensation depend on his reports. Although the principal can use all the transfers described above and condition them on whether and when the agent reports a finding, an optimal contract pays only success bonuses tied to time alone. The principal optimally gives an agent who reports a finding exactly what he could earn by concealing it, so no-concealing holds with equality. She can deliver this payoff without asking for reports at all, simply by posting a single bonus sequence in advance. I call this the \emph{posted-bonus-sequence (PBS) contract}.\footnote{The name draws an analogy with a posted price, which screens buyers' information only through who buys. The PBS contract likewise posts a bonus in each period and screens the agent's information only through who keeps working, without menus or report-dependent pay.}

\begin{figure}[t]
    \centering
    \includegraphics[width=0.45\textwidth,trim=284 0 0 0,clip]{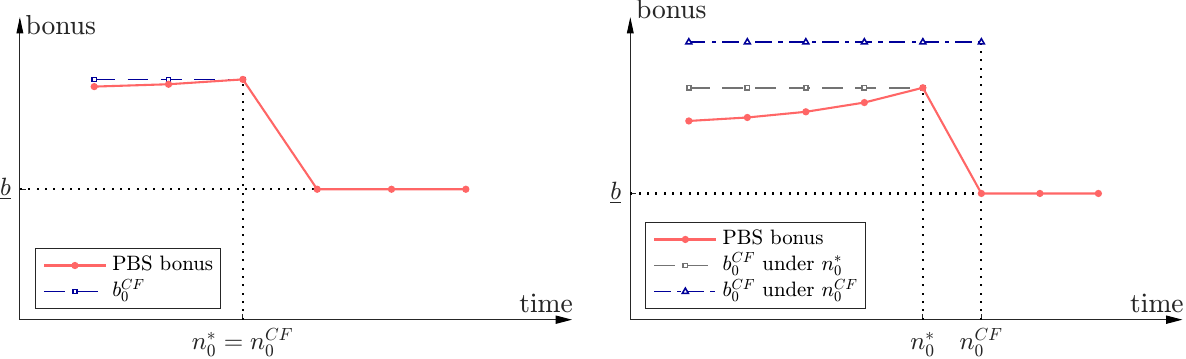}
    \caption{The PBS bonuses (red) and the contractible-finding (CF) bonuses $b_0^{CF}$, without discounting. Under the PBS contract, the research stage lasts $n_0^\ast$ periods, after which the bonus falls to the low bonus $\underline b$. The gray line is $b_0^{CF}$ for a research stage of the same duration, and the blue line is $b_0^{CF}$ at its own optimal research stage of $n_0^{CF}$ periods. The primitives are in Figure~\ref{fig:main result}.}
    \label{fig:intro}
\end{figure}

As Figure~\ref{fig:intro} illustrates, the bonus sequence has two stages, with the transition date $n_0^\ast$ chosen by the principal. In the \emph{research stage}, up to $n_0^\ast$, bonuses exceed the low level $\underline b$ needed to motivate an informed agent, and they induce even an uninformed agent to work and learn. These bonuses rise strictly over time, even without discounting. In the \emph{development stage}, the bonus falls to $\underline b$ and remains constant, so only an informed agent continues working until success.\footnote{Such sequential stages are common in R\&D. The SBIR/STTR program by default funds a development phase only for projects that clear a feasibility phase \citep[see][]{nisbir}, the accounting standard IAS 38 splits an internal project into a research phase and a development phase \citep*{ias38}, and \citet*{rost2007} document that staged financing is pervasive in biotech alliances.} For any given research-stage duration of more than one period, the PBS contract is the unique optimal contract that does not use reports. A small reporting cost would therefore select it over otherwise optimal contracts that use reports.\footnote{\citet*{grta2016} employ the same rule to select a contract that uses at most two reports, whereas the PBS contract uses none.}

With contractible findings, information guides both how long work continues and the bonus offered at a given date. With private findings, information still determines whether work continues, but conditioning the bonus on reported findings provides no benefit. This feature is consistent with the NIH practice described above: reports guide whether funding continues, while the amount allocated for each project year is specified in advance rather than adjusted to reported findings.

The optimality of the PBS contract is driven by two families of constraints: effort obedience, which deters shirking, and no-concealing, which deters delaying the report of a finding. I show that these constraints determine the principal's highest attainable payoff, while other deviations impose no additional cost.

Why does the principal choose to make no-concealing bind? Consider reducing the continuation payoff promised after a report in a given period and raising that period's success bonus, keeping the obedient agent's expected payoff unchanged. This shift makes both earlier shirking and earlier concealment less attractive. An agent who has shirked is more optimistic, so he expects more findings per success: a finding arrives only in the high state, whereas success can occur in either state. He therefore loses more from the lower payoff after a report than he gains from the higher success bonus. An agent who has concealed a finding obtained in an earlier period also values the payoff after a report more highly relative to the success bonus, since he already holds a finding and collects that payoff as long as no success arrives before he reports.\footnote{An obedient agent who enters that given period uninformed, by contrast, collects this payoff only if his work produces a finding but no success.} Making these earlier deviations less attractive allows the principal to reduce compensation elsewhere. But no-concealing limits how far she can go: a lower payoff after a report makes concealment more attractive to an agent who obtains a finding in that given period. The principal therefore lowers the payoff until no-concealing binds. By contrast, when findings are contractible, no reporting constraint limits this adjustment, so the principal can leave an informed agent zero rent.

A contract that does not use reports suffices to deliver the payoffs implied by binding no-concealing. Once the bonuses offered without a report are fixed, making each informed agent indifferent between concealing his finding and reporting it at once pins down the payoffs he must receive after a report, and hence the agent's expected payoff at the start of the relationship. The principal can deliver these payoffs by offering, in each period of the research stage, the same success bonus whether or not a finding has been reported. No-concealing then holds with equality automatically. The finding's conclusiveness is important: once the agent knows that the state is high, his belief no longer depends on when he learned it. At any given date, all informed agents therefore value the remaining bonuses in the same way. One posted sequence can thus deliver the required payoffs for every possible finding date. Because the bonus never depends on reports, a false report does not change the agent's payoff either. Concealment, not faking, drives the result: the same PBS contract remains optimal when findings are privately observed by the agent but verifiable, in which case the agent cannot fake a finding but can still conceal one. Verifiability alone thus brings the principal no benefit.

Binding effort obedience then uniquely determines the bonuses offered for success. These bonuses rise over the research stage, but not simply because a working agent grows more pessimistic. A novel \emph{learning incentive} is also present. Work offers not only a chance of collecting a success bonus but also a chance of obtaining a finding. An early finding lets the agent earn rents by working through the remaining research-stage periods, whose bonuses more than cover his effort costs in expectation. The prospect of these rents helps motivate the uninformed agent's effort. A later finding leaves fewer research-stage periods in which to earn these rents, so this learning incentive weakens over time. Success bonuses must therefore rise to keep effort obedience binding. When findings are contractible, the agent earns no rent after a finding, so the learning incentive is absent. At a common research-stage duration and without discounting, the contractible-finding bonuses (gray line in Figure~\ref{fig:intro}) are constant, whereas the PBS bonuses (red line) start lower and rise strictly until the two coincide in the final research-stage period.

Next, I examine how findings and their privacy affect welfare. First, making findings more likely in the high state strictly increases first-best surplus whenever the first best involves any work. This comparison includes moving from no finding to a finding. Under optimal contracting, however, more information can hurt both parties, whether findings are contractible or private. When findings are more likely, receiving no news after working is stronger evidence that the state is low. Shirking avoids this unfavorable learning and can become more attractive, raising the rents needed to sustain work. The principal may respond by shortening the research stage. Although this adjustment limits her incentive costs, her payoff can still fall, and the shorter research stage can also reduce the agent's payoff and total surplus.

Second, privacy never benefits the principal, because it restricts the contracts she can offer. It strictly harms her when the optimal research stage with contractible findings lasts more than two periods. Privacy can also hurt the agent. Holding the research-stage duration fixed, privacy can give him additional rents after a finding, but these rents make research costlier to sustain. The principal may then shorten the research stage, and the agent's loss from this cut can outweigh his gain. In Figure~\ref{fig:intro}, privacy shortens research and leaves both parties worse off.

Finally, I study four extensions. First, suppose findings may arrive whether or not the agent works. The same PBS contract remains optimal and induces work after a finding even if the finding was obtained without effort. The learning incentive, however, reverses: work may produce success and end the relationship before the agent obtains a finding that would otherwise arrive for free. The prospect of rents after a finding therefore makes effort harder, rather than easier, to motivate. For a given research-stage duration, the PBS bonuses weakly exceed those that would be optimal with contractible costless findings. Second, suppose success is privately observed by the agent. Without discounting, rising bonuses would then encourage an agent who has succeeded to delay disclosure. Under certain conditions, I show that a nonincreasing posted bonus sequence is optimal.

Third, suppose a finding is inconclusive good news: it is more likely in the high state but can arrive in either state. In this case, the agent continues learning after a finding, so shirking still preserves his optimism, and inducing longer work requires leaving him more rent. The principal must therefore also choose how long work continues after each report. A posted bonus sequence can still be optimal, but a general characterization is harder: the agent can combine shirking and concealment in many ways, and which deviations are most attractive depends on how long work continues after each report. Nevertheless, I show that a posted bonus sequence is close to optimal when findings are nearly conclusive. Fourth, suppose findings are bad news and reveal the low state. I identify conditions under which a posted bonus sequence remains optimal.

\paragraph{Literature review.} This paper relates to studies on dynamic agency with privately observed progress. Relative to these studies, the main departure of this paper is that the agent learns through hidden effort while working toward a contractible success, and that private findings are purely informational. \cite*{maye2022}, \cite*{rodi2022}, \cite*{okna2024}, and \cite*{ashi2025} consider work generating a single hidden outcome whose revelation yields direct payoffs. \cite*{ezha2021}, \cite*{mads2022}, and \cite*{cusi2025} study news that is costless to generate, so these models have no learning through hidden effort. In \cite*{mads2022} and \cite*{cusi2025}, the agent prefers the status quo to what follows a report, and the principal uses a deadline to screen information. \cite*{lili2026} study a forecaster acquiring information over time, where truthful reports are induced with pay conditioned on the reports and the ex-post state.

More closely related papers are \cite*{grta2016} and \cite*{ftwz2024}, in which the agent privately observes progress. In both, this information has no social value---the first best always continues the project until completion---yet optimal contracts use reports. In my setting, findings have social value because they guide continuation decisions, yet reports are unnecessary. Learning also changes effort incentives. Without learning, the effort constraints in those papers reflect only what the agent can gain by shirking over the remaining time. In my setting, learning creates two dynamic forces: shirking preserves the agent's optimism, and the prospect of rents from a finding helps motivate effort. Finally, both papers, unlike mine, give the agent a private benefit from shirking that makes termination of the project a punishment. \cite*{grta2016} use randomized termination to screen the agent. In \cite*{ftwz2024}, as here, the constraint against concealment binds. Their agent, however, delays reporting to collect the shirking benefit, which is exogenous, and their principal adjusts the punishment through randomized termination upon a report. My agent conceals a finding to keep working for bonuses that the principal designs. In contrast to those two papers, the principal here gains nothing from reports or randomization.

Like this paper, \cite*{gema2012} study an agent who learns through hidden effort, but their principal hires him to generate evidence about the state and pays him only if his reports match the state revealed ex post. The concern there is claiming evidence without testing rather than concealing it, and their pooling result concerns types known before contracting rather than a finding arriving over time.

This paper also builds on an extensive literature on contracting for experimentation. A central theme is the adverse selection generated by private learning when effort is hidden. My setting departs from this work in that experimentation can produce a private finding in addition to success. In \cite*{behe1998,behe2005}, \cite*{boho2011}, \cite*{mans2011}, \cite*{hosa2013}, \cite*{klei2016}, and \cite*{bhma2019}, output, such as success, is the only source of learning. \cite*{eygr2025} study a principal who privately observes success and jointly designs effort incentives and the information disclosed to the agent. \cite*{moro2022} allows two sequential, publicly observed successes.

Within this literature, \cite*{hakl2016} and \cite*{khlr2026} study an agent who has private information before contracting. In my setting, by contrast, a finding can arrive in any period, and only the agent observes when it arrives. This creates an additional reporting deviation: an informed agent can conceal his finding and report it later to imitate a type informed in the future. Faking a finding is closer to their static misreports since a faker imitates a type informed today. Such concealment would be profitable under the optimal contract with contractible findings, and how best to deter it when findings are private is the focus of my analysis.

This paper also relates to studies in dynamic mechanism design that identify when some types are pooled. The literature has shown that some pooling emerges when the principal cannot commit, as with the ratchet effect \citep*[e.g.,][]{lati1988}. With commitment, some pooling can also be optimal when the principal knows when the agent's private information arrives \citep*{rize1983,krst2015,guo2016}. Finally, in \cite*{bosk2016}, the timing of buyers' arrivals is uncertain, but the optimal mechanism is the same whether or not the seller observes their arrivals. My paper shows that pooling can be optimal even when none of these conditions holds: the principal commits, only the agent knows whether and when information arrives, and the privacy of findings is costly to the principal.

\section{Model}\label{sec:model}

\paragraph{Environment.} A principal (she) hires an agent (he) to work on a project whose quality state $\omega\in\{L,H\}$ remains fixed over time. The state $\omega$ is never directly observed by either party, and both share the prior $\mu_0$ over $\omega$, with $\mu_0(H)\in(0,1)$. Time is discrete, and in each period $t=1,2,\dots$ before the relationship ends, three things happen in order. First, the agent privately chooses effort $e\in\{0,1\}$ at cost $ce$, where $c>0$. The setting therefore has moral hazard. Next, if he works, a success, a finding, or both may arrive with probabilities detailed below, while shirking generates nothing. Lastly, the agent sends a message to the principal. Both parties are risk neutral and share the discount factor $\delta\in(0,1]$. The agent's payoff equals the transfers he receives from the principal minus any effort costs, and his outside option is zero. The principal's payoff equals her revenue from project success net of the transfers she pays.

\paragraph{Success.} A \emph{success} is the payoff-relevant output of the project, such as a drug shown to work in pharmaceutical R\&D. It is contractible, that is, public and verifiable. When a success arrives, the principal receives revenue $\pi>0$ and the relationship ends. In a period in which the agent works, a success arrives with probability $p_\omega:=\Prob(\text{success}\mid\omega)$ in state $\omega$, where
\[
1>p_H>p_L>0 .
\]
This assumption defines the states: work is more productive in the more promising state $H$ than in the less promising state $L$.\footnote{The main characterizations, Theorems~\ref{thm:cf} and~\ref{thm:main result}, hold under $p_L=0$ (Online Appendix~\ref{app:conclusive}). The baseline setting takes $p_L>0$ to ease exposition because $p_L=0$ expands the set of optimal contracts, which I discuss separately in Section~\ref{sec:discussions}.}

\paragraph{Finding.} While working, the agent may also obtain a \emph{finding}, such as a promising test result that only he can interpret. A finding is pure information, in the sense that it yields no payoff, does not change the success process, and does not end the relationship. It is private to the agent and unverifiable.\footnote{The driving friction is the privacy of the finding, not its unverifiability: Proposition~\ref{prop:verifiable} shows that the main result does not depend on whether the finding is verifiable.} A finding arrives at most once, and conditional on $\omega$, it arrives independently of success. In a period in which the agent works, a finding arrives with probability $q_\omega:=\Prob(\text{finding}\mid\omega)$ in state $\omega$, where
\[
1>q_H>0=q_L .
\]
The finding is therefore conclusive news that the state is high. This assumption is important for keeping the analysis tractable; I discuss the inconclusive case in Section~\ref{sec:inconclusive}.\footnote{One can show that conclusiveness also makes the model equivalent to one in which conditionally independent findings may arrive several times in sequence, because the history of findings matters for the agent's strategy only through his belief.} An agent who has obtained a finding is \emph{informed}, and he is \emph{uninformed} before then. Success ends the relationship even if a finding arrives in the same period.

Call a worked period \emph{silent} if neither a success nor a finding arrives in that period. Since silence is less likely when the state is high, the uninformed agent grows more pessimistic as silent periods accumulate. Formally, each silent worked period multiplies his likelihood ratio of $\omega=H$ to $\omega=L$ by
\[
\rho:=\frac{(1-p_H)(1-q_H)}{1-p_L}\in(0,1),
\]
so the agent's belief falls while he works without news. That is, \emph{no news is bad news}. A shirked period, in contrast, generates nothing and therefore preserves the agent's belief.

\paragraph{Contracts.} In period $0$, the principal commits to a dynamic contract. I first describe the general class of contracts allowed in the model and then restrict attention to direct mechanisms. In a general contract, both parties may communicate as the relationship proceeds. In each period, the contract can make a public announcement before the agent chooses effort and sends a message, with the announcement and the agent's message each drawn from a finite space. The contractible history records calendar time, the announcements and messages sent so far, and whether and when a success has arrived. (This history does not record effort or the arrival of a finding, as these are private to the agent.) The contract specifies transfers adapted to this history, and both announcements and transfers may be random. One constraint I impose is \emph{limited liability} (LL), which requires every transfer from the principal to the agent to be nonnegative.

To justify the focus on direct mechanisms, Online Appendix~\ref{app:general contracts} defines the general contract class formally and shows the following result.

\begin{lemma}[Informal]\label{lem:direct}
    Direct mechanisms achieve the principal's optimal payoff.
\end{lemma}

A contract is a \emph{direct mechanism} if it satisfies the following conditions. The message space in each period is $\{\emptyset,\text{``}F\text{''}\}$, and a truthful agent reports ``$F$'' precisely when the finding arrives and reports $\emptyset$ otherwise. The message history is summarized by the first report period $k$, equal to $\emptyset$ if no report of ``$F$'' has been made. The contract specifies a report-contingent effort \emph{schedule} $(n_0,n_1)$ and recommends work according to it. While no report has been made, the contract is in the \emph{uninformed track}, and work is recommended in periods $t=1,\dots,n_0$. This track is overridden by a report. Following a period-$k$ report, the contract enters the \emph{$k$-informed track}, where work is recommended in periods $t=k+1,\dots,k+n_{1k}$. Therefore, $n_0$ and $n_{1k}$ are the durations of the corresponding tracks; I call $n_0$ the \emph{uninformed duration} and $n_{1k}$ the \emph{$k$-informed duration}. Each duration lies in $\{0,1,\dots\}\cup\{\infty\}$, and $n_1:=(n_{1k})_{k\le n_0}$ collects the informed durations.

The transfers specified by the contract are as follows. On the uninformed track, a success \emph{bonus} $b_0(t)$ is paid upon a period-$t$ success, $t\le n_0$, and a \emph{wage} $w_0(t)$ is paid when the relationship reaches period $t\le n_0+1$. When $n_0<\infty$, the last entry $w_0(n_0+1)$ is interpreted as the severance, which is collected by an agent who completes period $n_0$ on the track with no success and no report. On the $k$-informed track, the success bonus and the wage are $b_{1k}(t)$ and $w_{1k}(t)$ for $t=k+1,\dots,k+n_{1k}$, together with the track severance $w_{1k}(k+n_{1k}+1)$ when $n_{1k}<\infty$. In addition, a \emph{report lump sum} $\ell_k$ is paid upon a period-$k$ report.

By Lemma~\ref{lem:direct}, I will thus focus on contracts of the form
\begin{equation}\label{eq:contract}
\phi:=\big(n_0,\;(n_{1k})_{k\le n_0},\;b_0(t),\;w_0(t),\;\left(b_{1k}(t),\;w_{1k}(t),\;\ell_k\right)_{k\le n_0}\big),
\end{equation}
where every transfer must be nonnegative by (LL). The model maintains $\sup_t b_0(t)<\infty$ and $\sup_t w_0(t)<\infty$ whenever $n_0=\infty$, and the same requirement applies to $b_{1k}$ and $w_{1k}$ when $n_{1k}=\infty$.\footnote{Payoff sums never take the form $\infty-\infty$, because each worked period ends the relationship with probability at least $p_L$, so expected effort costs sum to at most $c/p_L$. A payoff is therefore well defined, possibly equal to $+\infty$, as when the agent shirks forever while collecting wages that sum to $+\infty$.} Throughout the paper, $t$ denotes the current calendar period, $k$ the report period, and $m$ the arrival period of the finding.

Lemma~\ref{lem:direct} relies on more than the revelation principle \citep*{myer1986}. Beyond truthful reporting and obedient effort, the direct mechanism has two further features: the effort schedule never skips a period, and the recommendations, the termination, and the transfers are deterministic. Section~\ref{sec:discussions} explains why these two restrictions entail no loss.

\paragraph{Agent strategy and payoffs.} The agent's private history in period $t$ comprises the report period $k$ and the arrival period $m$ of the finding he observes by $t$, each equal to $\emptyset$ if it has not yet occurred. A \emph{strategy} $(e,r)$ specifies his play at every such history.\footnote{The agent also observes his past efforts, which shape his belief. Leaving them out of the private history loses nothing: the agent chooses a strategy to maximize his ex-ante payoff, and, given a strategy, his past efforts at each history follow from $(t,k,m)$, so conditioning on them cannot raise that payoff.} The \emph{effort rule} $e(t,k,m)\in\{0,1\}$ specifies the agent's effort in each active period. The active periods are $1\le t\le n_0$ if $k=\emptyset$ and $k+1\le t\le k+n_{1k}$ if a report was made at $k$. The \emph{reporting rule} $r(t,m)\in\{\emptyset,\text{``}F\text{''}\}$ specifies the agent's message in each period $t\le n_0$ given no report yet. Any message sent after a report is ignored by the direct mechanism, and the contract ends with the uninformed track if a report does not occur by $n_0$. Hence, the reporting rule matters only for $k=\emptyset$ and $t\le n_0$. The \emph{obedient strategy} $(e^\ast,r^\ast)$ prescribes that the agent work whenever recommended and report a finding as soon as it arrives. That is, $e^\ast\equiv1$, and $r^\ast(t,m)=\text{``}F\text{''}$ if and only if the finding has arrived ($m\neq\emptyset$).

Payoffs aggregate transfers along the realized path. To keep the notation compact, given a realized report period $k\neq\emptyset$ with $k\le n_0$, define the effective wage and the effective bonus
\[
\begin{aligned}
\widetilde w(t,k;\phi)&:=w_0(t)\mathbf 1\{t\le k\}+w_{1k}(t)\mathbf 1\{t>k\}+\frac{\ell_k}{\delta}\mathbf 1\{t=k+1\},\\
\widetilde b(t,k;\phi)&:=b_0(t)\mathbf 1\{t\le k\}+b_{1k}(t)\mathbf 1\{t>k\},
\end{aligned}
\]
and set $\widetilde w(\cdot,\emptyset;\phi):=w_0$ and $\widetilde b(\cdot,\emptyset;\phi):=b_0$ if no report is ever made. Here, the case $\mathbf 1\{t\le k\}$ selects the uninformed-track transfers, while the case $\mathbf 1\{t>k\}$, with a past report at $k$, selects the $k$-informed-track transfers. The entry $\ell_k/\delta$ assigns the report lump sum, which is collected in period $k$, to period $k+1$ at the same present value.

Fix a strategy $(e,r)$ and consider play before success. Each arrival period $m\in\{\emptyset\}\cup\{1,2,\dots\}$ of the finding then pins down a path, which I call the \emph{$m$-path}. Here, $m=\emptyset$ means that the finding never arrives. Write $k_m$ for the realized report period on this path, which is the first period in which the reporting rule $r$ sends ``$F$'', or $\emptyset$ if $r$ never does, and write $e_m(t)$ for the realized effort, extended by $e_m(t)=0$ beyond the scheduled termination. Before the arrival, the $m$-path coincides with the no-arrival path, so its effort there equals $e_\emptyset$. The \emph{reach probability} that the relationship reaches period $t$ on the $m$-path, given the state $\omega$ and the effort rule $e$, is
\[
\Prob\nolimits_\omega(m,t;e):=
\underbrace{\prod_{s=1}^{m-1}\big[(1-p_\omega)(1-q_\omega)\big]^{e_\emptyset(s)}}_{\text{no success or finding while uninformed}}
\cdot\;\underbrace{e_\emptyset(m)\,q_\omega(1-p_\omega)\vphantom{\prod_{s=1}^{m-1}}}_{\text{finding at }m}
\cdot\;\underbrace{\prod_{s=m+1}^{t-1}(1-p_\omega)^{e_m(s)}}_{\text{no success while informed}}
\qquad(m\neq\emptyset),
\]
and $\Prob_\omega(\emptyset,t;e):=\prod_{s=1}^{t-1}[(1-p_\omega)(1-q_\omega)]^{e_\emptyset(s)}$ if the finding never arrives ($m=\emptyset$). For $m\neq\emptyset$, set $\Prob_\omega(m,t;e):=0$ for every $t\le m$. The agent's payoff and total surplus are therefore
\[
U_A(\phi,e,r):=\sum_\omega\mu_0(\omega)\sum_m\sum_t\delta^{t-1}\,\Prob\nolimits_\omega(m,t;e)\Big[\widetilde w(t,k_m;\phi)-c\,e_m(t)+p_\omega e_m(t)\,\widetilde b(t,k_m;\phi)\Big],
\]
\[
S(e,r):=\sum_\omega\mu_0(\omega)\sum_m\sum_t\delta^{t-1}\,\Prob\nolimits_\omega(m,t;e)\,e_m(t)\,(p_\omega\pi-c),
\]
where the $t$-sum runs over the active periods and the severance periods, and the $m$-sum runs over $m=\emptyset$ and $m=1,2,\dots$. Each period is hence counted exactly once: the $m=\emptyset$ term carries period $t$ whenever the finding has not arrived before $t$, and the $m$-term with $m<t$ carries it once the finding arrived in period $m$. In each period, the agent receives the wage $\widetilde w(t,k_m;\phi)$, pays the effort cost $c\,e_m(t)$, and obtains the success bonus $\widetilde b(t,k_m;\phi)$ with probability $p_\omega$ if he works. The agent's work generates expected surplus equal to the success revenue $\pi$, received with probability $p_\omega$, minus his effort cost $c$. The principal's payoff equals total surplus minus the agent's payoff:
\[
U_P(\phi,e,r):=S(e,r)-U_A(\phi,e,r).
\]

\paragraph{Principal's problem.} The principal chooses a contract $\phi$ of the form described in (\ref{eq:contract}). I describe her problem in two steps. The \emph{first step} fixes the schedule $(n_0,n_1)$, which pins down the obedient strategy $(e^\ast,r^\ast)$, and then finds the cheapest transfers that implement this strategy. Since the schedule fixes expected effort costs, this is equivalent to minimizing the agent's payoff:
\begin{equation}\label{eq:first step}
\begin{aligned}
U_A(n_0,n_1):={}&\inf_{\phi\ \text{with}\ (n_0,n_1)}\;U_A(\phi,e^\ast,r^\ast)\\
\text{s.t. }&\quad\text{(LL) }\phi\ge0\\
&\quad\text{(IC) }(e^\ast,r^\ast)\in\argmax_{(e,r)}\,U_A(\phi,e,r).
\end{aligned}
\end{equation}
If, for given $(n_0,n_1)$, this program has no feasible solution, I set its value to $+\infty$. The \emph{second step} of the principal's problem finds the schedule $(n_0,n_1)$ that maximizes her payoff given $U_A(n_0,n_1)$. Since the schedule pins down total surplus, let $S(n_0,n_1)$ denote the total surplus $S(e^\ast,r^\ast)$ induced under the obedient strategy. The principal solves
\[
\max_{(n_0,n_1)}\;\big[\,S(n_0,n_1)-U_A(n_0,n_1)\,\big].
\]

\subsection{Which Constraints Matter?}\label{sec:which constraint matter}

In program (\ref{eq:first step}) above, the incentive constraint (IC) compares the obedient strategy with every other strategy available to the agent, so it comprises a large class of constraints. I next describe the three families of constraints that will be relevant for the analysis.

\emph{Effort obedience.} The uninformed agent must not gain by shirking in a single period while working in every other recommended period. Formally, for $t\le n_0$, denote by $e_{\text{Shirk}}^t$ the effort rule that shirks in period $t$ when no report has been made and works in every other recommended period. Effort obedience requires
\begin{equation}\label{eq:effort obedience}
U_A(\phi,e^\ast,r^\ast)\;\ge\;U_A(\phi,e_{\text{Shirk}}^t,r^\ast),
\qquad t\le n_0 .
\end{equation}
An agent who follows $e_{\text{Shirk}}^t$ is called a \emph{shirker}.

\emph{No-concealing.} An agent informed in period $t\le n_0$ must not gain by delaying his report by one period. Formally, denote by $r_{\text{Conceal}}^{t}$ the reporting rule that sends no report while uninformed and reports a finding in its arrival period, unless the arrival period is $t$, in which case it reports the finding in period $t+1$. When $t=n_0$, the delayed report is never sent, since the reporting window closes at $n_0$, so the agent finishes the uninformed track and collects the severance. No-concealing requires
\begin{equation}\label{eq:no-concealing}
U_A(\phi,e^\ast,r^\ast)\;\ge\;U_A(\phi,e^\ast,r_{\text{Conceal}}^{t}),
\qquad t\le n_0 .
\end{equation}
An agent who delays his report is called a \emph{concealer}.

\emph{No-faking.} An uninformed agent must not gain by reporting a finding he does not have. Formally, for $t\le n_0$, denote by $r_{\text{Fake}}^t$ the reporting rule that reports a finding in its arrival period and, if the agent is still uninformed and has not reported by period $t$, reports ``$F$'' in period $t$. In the latter scenario, the agent is placed on the $t$-informed track while he is in fact uninformed. No-faking requires
\begin{equation}\label{eq:no-faking}
U_A(\phi,e^\ast,r^\ast)\;\ge\;U_A(\phi,e,r_{\text{Fake}}^t),
\qquad t\le n_0,
\end{equation}
for every effort rule $e$ that coincides with $e^\ast$ off the $t$-informed track. The constraint is tightest at the rule that chooses effort on the $t$-informed track to maximize the agent's continuation payoff. An agent who reports a finding he does not have is called a \emph{faker}.

\subsection{First-Best Schedule}\label{sec:first best}

Throughout the paper, I maintain the following assumption.

\begin{assumption}\label{ass:efficiency}
    Working is efficient (inefficient) if the state is known to be high (low): \[p_L\pi<c<p_H\pi.\]
\end{assumption}

The \emph{first-best schedule} $(n_0^{FB},n_1^{FB})$ maximizes total surplus $S(n_0,n_1)$, with ties broken in favor of the shortest uninformed duration.\footnote{Throughout the paper, whenever several uninformed durations are optimal, I select the shortest one. This applies to $n_0^{FB}$ here and to $n_0^{CF}$, $n_0^\ast$, and $n_0^C$ below. One can show that the main characterizations, Theorems~\ref{thm:cf} and~\ref{thm:main result}, remain true when the longest optimal duration is selected instead in every case.} Assumption~\ref{ass:efficiency} pins down the shape of the first-best schedule. On the informed tracks, the finding reveals $\omega=H$, and working is then efficient in every period: $n_{1k}^{FB}=\infty$ for every $k\le n_0^{FB}$. On the uninformed track, no news is bad news. An additional period of uninformed work yields expected success revenue and the prospect of obtaining the finding, net of the effort cost. The value of such a period declines as silent worked periods accumulate and eventually turns negative as it tends to $p_L\pi-c<0$. Consequently, uninformed work stops after finitely many periods: $n_0^{FB}<\infty$. For the same reason, an infinite uninformed duration is also suboptimal in the contracting problems studied in Sections~\ref{sec:cf} and~\ref{sec:pf}.

Whether the principal can extract the entire first-best surplus depends on two frictions present in my model: moral hazard and limited liability. The following result shows that the principal attains the first best absent either friction. Define the agent's participation constraint to require his expected payoff at the start of the relationship to be nonnegative: $U_A(\phi,e^\ast,r^\ast)\ge0$.\footnote{Under (LL), participation is implied by (IC), since shirking in every period guarantees the agent a nonnegative payoff.}

\begin{proposition}\label{prop:frictionless}
    Suppose either that effort is contractible, or that transfers may be negative and the agent's participation constraint is imposed. Then the principal implements the first-best schedule and extracts the entire surplus, leaving the agent zero payoff.
\end{proposition}

\emph{Proof}. See Appendix~\ref{app:proof frictionless}.

\medskip

To see why each relaxation restores the first best, consider the two frictions in turn. Without moral hazard, effort is contractible, so the principal specifies the first-best schedule and reimburses the agent for his effort costs. Without limited liability, the principal can ``sell the project'' to the agent: she charges him an upfront price equal to the first-best surplus and then pays him whatever revenue the project generates during the contract. The agent then faces the planner's problem, and the upfront price leaves him with zero expected payoff.

\section{Contractible Findings}\label{sec:cf}

To isolate the effects of privacy, I first consider the case in which the finding is \emph{contractible}: both parties observe its arrival and transfers can condition on it directly. Moral hazard and limited liability remain. There is no reporting, since a period-$k$ finding directly overrides the uninformed track and starts the $k$-informed track, with $k$ now denoting the arrival period. The agent's deviations concern his effort choices alone.

After the finding, both parties know that $\omega=H$. Define the \emph{high-state bonus}
\[
\underline b:=\frac{c}{p_H}.
\]
Facing $\underline b$, an informed agent is indifferent between working and shirking, so $\underline b$ induces informed work while leaving the agent no rent. Since a worked informed period yields the surplus $p_H\pi-c>0$, the principal wants every informed track to run until success, paying $\underline b$ on it.

Before the finding, the design of the uninformed bonus $b_0$ is less straightforward. The main reason is familiar from other settings where the agent learns from hidden effort: shirking covertly preserves an optimistic belief.\footnote{This source of asymmetric information due to hidden effort is central to other studies of contracting for experimentation. See, for example, \citet*{hakl2016}, whose bonus contract informs the form of the bonuses in (\ref{eq:cf bonus}).} To see how this force shapes the bonus, denote by $P_t$ the success probability perceived by an uninformed agent who has worked $t$ silent periods. Bayes' rule gives
\[
P_t:=\frac{\mu_0(H)\big[(1-p_H)(1-q_H)\big]^{t}p_H+\mu_0(L)(1-p_L)^{t}p_L}{\mu_0(H)\big[(1-p_H)(1-q_H)\big]^{t}+\mu_0(L)(1-p_L)^{t}}.
\]
Since no news is bad news, $P_t$ is strictly decreasing in $t$. On the uninformed track, the obedient agent reaches period $t$ having worked $t-1$ silent periods, so his perceived success probability is $P_{t-1}$.

A shirker reaches period $t$ with a higher perceived probability of success than an obedient agent, so he can earn a rent from later bonuses. To see this, suppose the bonus just covers the effort cost of an obedient agent. That is, the principal offers the \emph{myopic bonus} $b_0(t)=c/P_{t-1}$, thus exactly compensating the obedient agent in every period. An agent who shirks in period $t-1$, however, reaches period $t$ with the more optimistic perception $P_{t-2}>P_{t-1}$, under which the period-$t$ bonus more than covers his effort cost and yields a positive payoff. This shirker, who follows $e_{\text{Shirk}}^{t-1}$, therefore gains, and the bonus must rise above the myopic level to satisfy effort obedience (\ref{eq:effort obedience}). For a finite uninformed duration $n_0$, requiring every effort obedience constraint to hold with equality yields the \emph{contractible-finding bonus}
\begin{equation}\label{eq:cf bonus}
b_0^{CF}(n_0):=\frac{c}{P_{n_0-1}},
\qquad
b_0^{CF}(t):=(1-\delta)\frac{c}{P_{t-1}}+\delta\,b_0^{CF}(t+1)
\quad\text{for }t<n_0 .
\end{equation}
I will sometimes write $b_0^{CF}(t;n_0)$ to make the duration $n_0$ explicit. In the last period $t=n_0$, the contractible-finding bonus equals the myopic bonus, $b_0^{CF}(n_0)=c/P_{n_0-1}$, because an agent who shirks in the last period has no later bonus to collect. In every earlier period, the bonus exceeds the myopic bonus, $b_0^{CF}(t)>c/P_{t-1}$, and it exceeds $\underline b$ in every period. Without discounting, that is, when $\delta=1$, the bonus $b_0^{CF}(\cdot)$ is constant over time.

For a finite $n_0$, the \emph{contractible-finding contract} $\phi^{CF}(n_0)$ sets every informed duration infinite, $n_{1k}=\infty$ for every $k\le n_0$, and pays
\[
b_0=b_0^{CF},\qquad b_{1k}\equiv\underline b,\qquad w_0=0,\quad w_{1k}=0,\quad \ell_k=0 .
\]
Let $n_0^{CF}$ be the smallest uninformed duration attaining the principal's optimal payoff when the finding is contractible.

\begin{theorem}\label{thm:cf}
    When the finding is contractible, the principal optimally offers the contract $\phi^{CF}(n_0^{CF})$, with $n_0^{CF}\le n_0^{FB}$.
\end{theorem}

\emph{Proof}. See Appendix~\ref{app:proof cf}.

\medskip

Under the contract in Theorem~\ref{thm:cf}, both the contract's duration and the agent's pay depend on whether and when a finding arrives. Informed and uninformed agents work for different durations, since the uninformed agent stops after $n_0^{CF}$ periods while an informed agent works until success. They also face different bonuses, since in every period in which both tracks are active, the uninformed track pays a strictly higher bonus than the informed track, $b_0^{CF}(t)>\underline b$.

The uninformed duration $n_0^{CF}$ can fall short of the first best, and never exceeds it. The reason is that the agent's payoff rises with the uninformed duration. A longer uninformed track carries more bonuses above the myopic level, so when the principal chooses $n_0$, she trades off surplus against paying a higher rent to the agent.

Theorem~\ref{thm:cf} identifies one optimal contract, and Online Appendix~\ref{app:cf all} describes properties of all optimal contracts (Proposition~\ref{prop:cf all}). From period $2$ on, every optimal contract pays the bonus $b_0^{CF}$ on the uninformed track. A finding obtained in period $2$ or later leaves the agent no rent: he receives $\underline b$ in every subsequent period and no other transfer. Moreover, no optimal contract pays a wage on the uninformed track. Thus, the only freedom concerns period $1$: the agent's combined expected payoff from a success and a finding is pinned down, but its division between them is not. An optimal contract can therefore raise the payoff after a period-$1$ finding and lower $b_0(1)$ accordingly.

This characterization also shows when the agent's pay must depend on the finding. \emph{Every} optimal contract conditions pay on whether and when a finding arrives if and only if $n_0^{CF}>2$. To see why, consider a contract that offers a single sequence of success bonuses: the uninformed bonuses through period $n_0^{CF}$, followed by $\underline b$, whether or not a finding arrives. An agent who obtains a finding before $n_0^{CF}$ earns a positive payoff from the remaining uninformed bonuses, which exceed $\underline b$. When $n_0^{CF}>2$, a period-$2$ finding would therefore leave him a positive payoff, contradicting the characterization above. When $n_0^{CF}\le2$, however, only a period-$1$ finding can generate such a payoff, and the principal can offset it with a lower initial success bonus.

Why is period $1$ different? From period $2$ on, an agent who shirked earlier can hold a different belief from an obedient agent. Agents with different beliefs value payments upon success and payoffs after a finding differently. How compensation is divided between these two events therefore affects deviation incentives, and optimality pins down this division (Section~\ref{sec:proof sketch} returns to this point). In period $1$, however, no earlier deviation is possible, so every agent who works holds the same prior. Only the combined expected payoff is therefore pinned down, not its division between a success and a finding.

\section{Private Findings}\label{sec:pf}

This section returns to the main setting, in which the finding is private to the agent. Section~\ref{sec:example} describes an example that identifies the binding constraints. Section~\ref{sec:main result} presents the main result, Theorem~\ref{thm:main result}, and discusses the optimal contract. Section~\ref{sec:comparative statics} asks how information and its privacy affect the two parties' payoffs. Section~\ref{sec:proof sketch} sketches the proof of the main result and explains the mechanism behind it.

\subsection{Illustrative Example}\label{sec:example}

Before stating the general theorem, I describe an example that moves in three steps from the contractible-finding contract to the optimal contract when the finding is private. The example identifies the constraints that matter for optimality. The parameters, recorded in Table~\ref{tab:example}, are chosen so that $n_0^{CF}=3$ and $\underline b=2$. Since there is no discounting ($\delta=1$), the contractible-finding bonus is constant over time. In the table, $b_0$ is the bonus before any report and $b_{1k}$ the bonus after a period-$k$ report.

Does the contractible-finding contract remain feasible when the finding is private? It does not, because the agent would conceal his finding. The first panel of Table~\ref{tab:example} displays this contract, whose transfers depend on the agent's report, since $b_{1k}(t)=\underline b<b_0(t)$ for $k<t\le3$. Two of the three families of constraints in Section~\ref{sec:which constraint matter} hold. Every effort obedience constraint holds with equality, by the construction of $b_0^{CF}$, and no-faking holds because a track that pays $\underline b$ is worth zero to any agent, so faking never benefits the agent. No-concealing fails, however. For example, the agent informed in period $1$ is supposed to report at once and enter the $1$-informed track (row $b_{11}$), where he earns zero. But he can instead conceal the finding and work for the uninformed bonus $b_0>\underline b=b_{11}$ to gain a positive payoff. Restoring feasibility therefore requires narrowing the bonus gap $b_0-b_{1k}$ across tracks.

\begin{table}[t]
\centering
\begin{minipage}{0.48\textwidth}
\centering
\begin{tabular}{lccccc}
\toprule
& $t=1$ & $t=2$ & $t=3$ & $t=4$ & $t\ge5$\\
\midrule
$b_0$ & $3.68$ & $3.68$ & $3.68$ & & \\
$b_{11}$ & & $\underline b$ & $\underline b$ & $\underline b$ & $\underline b$\\
$b_{12}$ & & & $\underline b$ & $\underline b$ & $\underline b$\\
$b_{13}$ & & & & $\underline b$ & $\underline b$\\
\bottomrule
\end{tabular}

\medskip
Contractible-finding contract
\end{minipage}
\hfill
\begin{minipage}{0.48\textwidth}
\centering
\begin{tabular}{lccccc}
\toprule
& $t=1$ & $t=2$ & $t=3$ & $t=4$ & $t\ge5$\\
\midrule
$b_0$ & $3.68$ & $3.68$ & $3.68$ & & \\
$b_{11}$ & & {\color{blue}$3.60$} & {\color{blue}$3.50$} & {\color{blue}$2.71$} & $\underline b$\\
$b_{12}$ & & & {\color{blue}$3.60$} & {\color{blue}$2.17$} & $\underline b$\\
$b_{13}$ & & & & $\underline b$ & $\underline b$\\
\bottomrule
\end{tabular}

\medskip
Feasible
\end{minipage}

\bigskip

\begin{minipage}{0.48\textwidth}
\centering
\begin{tabular}{lccccc}
\toprule
& $t=1$ & $t=2$ & $t=3$ & $t=4$ & $t\ge5$\\
\midrule
$b_0$ & $3.68$ & $3.68$ & $3.68$ & & \\
$b_{11}$ & & {\color{blue}$3.68$} & {\color{blue}$3.68$} & $\underline b$ & $\underline b$\\
$b_{12}$ & & & {\color{blue}$3.68$} & $\underline b$ & $\underline b$\\
$b_{13}$ & & & & $\underline b$ & $\underline b$\\
\bottomrule
\end{tabular}

\medskip
Report-free
\end{minipage}
\hfill
\begin{minipage}{0.48\textwidth}
\centering
\begin{tabular}{lccccc}
\toprule
& $t=1$ & $t=2$ & $t=3$ & $t=4$ & $t\ge5$\\
\midrule
$b_0$ & $3.57$ & $3.61$ & $3.68$ & & \\
$b_{11}$ & & $3.61$ & $3.68$ & $\underline b$ & $\underline b$\\
$b_{12}$ & & & $3.68$ & $\underline b$ & $\underline b$\\
$b_{13}$ & & & & $\underline b$ & $\underline b$\\
\bottomrule
\end{tabular}

\medskip
Optimal
\end{minipage}
\caption{Four contracts in the illustrative example. Parameters: $c=1$, $p_H=1/2$, $p_L=1/10$, $q_H=1/10$, $\mu_0(H)=3/4$, $\delta=1$, and $\pi=8$.}
\label{tab:example}
\end{table}

To restore feasibility, I modify the contractible-finding contract by partially narrowing the bonus gap. In particular, I hold $b_0$ fixed and raise the early informed bonuses, shown in blue in the Feasible panel, just enough that no-concealing (\ref{eq:no-concealing}) holds with equality in every period $t\le3$. That is, an agent who obtains a finding in any period $t\le3$ earns the same by reporting it at once as by concealing it for one more period, so he is willing to report.

The modified contract is feasible, but its transfers still depend on the agent's report. Can the same payoff to the agent be achieved without report-dependent bonuses? The answer is yes, since fully closing the bonus gap, as in the Report-free panel, yields the same payoff to the agent. Under both contracts, $b_0$ pins down what the agent gains from a success, and, because no-concealing holds with equality, the value of concealing pins down what he gains from a finding.\footnote{More generally, feasible contracts that pay no wage, satisfy no-concealing with equality and share the same $b_0$ give the agent the same payoff.} The new contract is \emph{report-free}, because it is equivalent to posting the single bonus sequence $(3.68,\allowbreak\,3.68,\allowbreak\,3.68,\allowbreak\,\underline b,\allowbreak\,\underline b,\dots)$, so that reporting changes no transfer. This sequence implements the same effort schedule, since after period $3$ the bonus falls to $\underline b$, which leaves the uninformed agent unwilling to work while an informed agent, indifferent at $\underline b$, keeps working. Since reporting does not affect any transfers, both reporting constraints, no-concealing and no-faking, hold with equality.

The contract constructed so far is still one step away from optimality, because effort obedience is now strictly slack in periods $1$ and $2$. The informed bonuses shown in blue in the Report-free panel exceed $\underline b$, so an agent who obtains a finding in period $1$ or $2$ earns a positive payoff, which he would not earn under the contractible-finding contract. Since working may bring such a finding, the uninformed agent has an extra reason to work, and a success bonus of $3.68$ is more than is needed to induce work in periods $1$ and $2$.

The final step thus removes this slack by lowering the early bonuses until effort obedience (\ref{eq:effort obedience}) binds in every period $t\le3$, which yields the optimal contract in the Optimal panel. This contract is again report-free and equivalent to the posted bonus sequence $(3.57,\,3.61,\,3.68,\,\underline b,\,\underline b,\dots)$. This optimal bonus sequence has two features that my main result in Theorem~\ref{thm:main result} below generalizes. First, the bonus strictly increases over the uninformed track, periods $1$ to $3$. Second, the bonus is strictly lower than the contractible-finding bonus in those periods, except for the last period, $n_0$, where they are equal.

\subsection{Main Result}\label{sec:main result}

The following definition generalizes the optimal contract of the example. For a finite uninformed duration $n_0$, the \emph{posted-bonus-sequence (PBS) contract} $\phi^\ast(n_0)$ sets every informed duration to infinity, $n_{1k}=\infty$ for every $k\le n_0$, and specifies
\[
b_{1k}(t)=
\begin{cases}
b_0(t), & k<t\le n_0,\\
\underline b, & t>n_0,
\end{cases}
\qquad
w_0=0,\quad w_{1k}=0,\quad \ell_k=0 .
\]
In particular, the PBS contract is equivalent to posting the single bonus sequence
\[
\big(b_0(1),\;\dots,\;b_0(n_0),\;\underline b,\;\underline b,\;\dots\big),
\]
without any need for a report. The uninformed-track bonus is $b_0(t)=b_0^\ast(t;n_0)$, where $b_0^\ast(\cdot\,;n_0)$ denotes the unique sequence for which every effort obedience constraint (\ref{eq:effort obedience}) holds with equality:
\[
U_A(\phi^\ast(n_0),e^\ast,r^\ast)=U_A(\phi^\ast(n_0),e_{\text{Shirk}}^t,r^\ast)
\qquad\text{for every }t\le n_0 .
\]
The word ``posted'' in the contract's name draws an analogy with a posted price, which screens buyers' information only through who buys. The PBS contract likewise screens the agent's information only through who keeps working. I compare $b_0^\ast(t;n_0)$ with the contractible-finding bonus $b_0^{CF}(t;n_0)$ of (\ref{eq:cf bonus}) at the same duration. In parallel to $n_0^{CF}$, let $n_0^\ast$ denote the smallest uninformed duration attaining the principal's optimal payoff when the finding is private.

\begin{theorem}\label{thm:main result}
    When the finding is private, the principal optimally offers the PBS contract $\phi^\ast(n_0^\ast)$, with $n_0^\ast\le n_0^{CF}\le n_0^{FB}$, and its bonus satisfies:\begin{enumerate}[nolistsep,label=(\roman*)]
        \item $b_0^\ast(t;n_0^\ast)$ is strictly increasing in $t$ for $t\le n_0^\ast$;
        \item $b_0^\ast(t;n_0^\ast)\le b_0^{CF}(t;n_0^\ast)$ for $t\le n_0^\ast$, strictly iff $t<n_0^\ast$.
    \end{enumerate}
\end{theorem}

\emph{Proof}. See Appendix~\ref{app:proof main result}.

\medskip

Theorem~\ref{thm:main result} shows that the principal loses nothing by ignoring the agent's reports. The model allows the contract to condition every transfer on whether and when a finding is reported, yet a posted bonus sequence is optimal. The PBS contract induces the uninformed agent to work for $n_0^\ast$ periods, no longer than when the finding is contractible, and it keeps an informed agent working until success. Parts (i) and (ii) generalize the two features of the example: over the uninformed duration, the PBS bonus strictly rises, and it lies below the contractible-finding bonus at the same duration, except in the last period, where both equal the myopic bonus $c/P_{n_0^\ast-1}$. The proof sketch in Section~\ref{sec:proof sketch} shows that the principal optimally satisfies the no-concealing constraint with equality. It also shows that, because learning is absent after a finding, a report-free contract suffices to deliver the payoffs that such a binding constraint pins down (see Section~\ref{sec:inconclusive} for a discussion of inconclusive findings).

\begin{figure}[h]
    \centering
    \includegraphics[width=\textwidth]{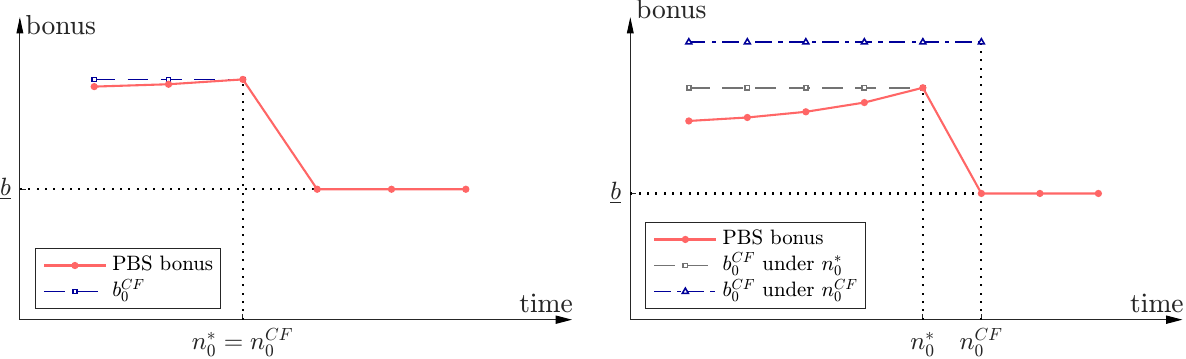}
    \caption{The PBS contract (red) and the contractible-finding contract (blue) in two cases. The left panel depicts the example of Section~\ref{sec:example}. In the right panel, the contractible-finding contract has a longer uninformed track than the PBS contract ($n_0^{CF}>n_0^\ast$). The gray line is the contractible-finding bonus at the duration $n_0^\ast$. The right panel uses $c=1$, $p_H=1/4$, $p_L=1/10$, $q_H=1/5$, $\mu_0(H)=19/20$, $\delta=1$, and $\pi=224/25$.}
    \label{fig:main result}
\end{figure}

Figure~\ref{fig:main result} illustrates the two features in two cases. The left panel is the example of Section~\ref{sec:example}, where $n_0^\ast=n_0^{CF}=3$. The right panel displays another set of primitives, under which the uninformed duration is strictly longer when the finding is contractible: $n_0^\ast=5<n_0^{CF}=6$. In that case, the contractible-finding bonus at its own optimal duration is even higher than at the duration $n_0^\ast$ (gray line), since the sequence is determined by backward induction from a later final period, when the perceived success probability is lower. The two features of the PBS bonus are unchanged.

\paragraph{Contract selection.} Theorem~\ref{thm:main result} identifies one optimal contract, but the general class of contracts in Section~\ref{sec:model} can admit others (Section~\ref{sec:discussions} gives an example). The PBS contract stands out among them because its transfers ignore reports. Formally, a contract is report-free if it is equivalent to offering a single bonus sequence and a single wage sequence. The next result shows that no other report-free contract does as well at any uninformed duration of more than one period.

\begin{proposition}\label{prop:report free}
    For every finite uninformed duration $n_0>1$, the PBS contract $\phi^\ast(n_0)$ is the unique report-free contract that is optimal among the contracts with uninformed duration $n_0$.
\end{proposition}

\emph{Proof}. See Appendix~\ref{app:proof report free}.

\medskip

Whenever $n_0^\ast>1$, a small reporting cost would select the PBS contract over otherwise optimal contracts that use reports. \citet*{grta2016} use this same selection rule to identify an optimal contract with minimal communication.\footnote{In \citet*{grta2016}, this selection rule yields a contract that still uses up to two reports, whereas here the selected PBS contract uses none.} If sending a report costs the agent a small amount, every optimal contract under which reports are sent must bear this cost, whereas posting the PBS bonus sequence asks for no report and thus incurs no reporting cost.

The optimal contracts that use reports give the agent the same payoff as the PBS contract and differ mainly in the form of pay. Online Appendix~\ref{app:pf all} (Proposition~\ref{prop:pf all}) shows that every optimal contract pays the PBS bonus on the uninformed track from period $2$ on. For findings obtained in period $2$ or later, every optimal contract also delivers the same \emph{informed payoff}---the payoff following a truthful report---as the PBS contract. As in Section~\ref{sec:cf}, period $1$ leaves open the division of the agent's expected payoff between a success and a finding. Relative to the PBS contract, a higher payoff after a period-$1$ finding must therefore be offset by a lower bonus $b_0(1)$. In fact, the bonus rises over the uninformed duration in \emph{every} optimal contract: $b_0(1)\le b_0^\ast(1;n_0)<b_0^\ast(2;n_0)=b_0(2)<\dots<b_0(n_0)$. Section~\ref{sec:discussions} returns to optimal contracts that can pay lump sums.

\paragraph{Role of concealing and faking.} The simplicity of the optimal contract comes from the private arrival of the finding. Because only the agent knows when a finding arrives, an informed agent can conceal it and continue on the uninformed track. Eliciting a truthful report therefore requires the payoff after a report to be at least what the informed agent earns without one, so concealment limits the pay gap across report histories. At the same time, reports are valuable only if they change pay so as to better tailor the agent's incentives. Theorem~\ref{thm:main result} then shows that this tension eliminates the value of reporting.

The option to fake a finding, in contrast, plays no role. To see this, suppose the finding is \emph{verifiable}: once disclosed, it comes with hard evidence that cannot be faked, such as a laboratory record or a formal mathematical proof. Formally, the reporting rule may send ``$F$'' only at or after the arrival period, so faking is impossible, while concealment (reporting late or not at all) remains possible.

\begin{proposition}\label{prop:verifiable}
    When the finding is verifiable, the same PBS contract remains optimal.
\end{proposition}

\emph{Proof}. See Appendix~\ref{app:proof verifiable}.

\medskip

Verifiability removes the option to fake but does not improve the principal's payoff. The simplicity of the optimal contract therefore comes from the option to conceal, not from the option to fake, and the main result applies equally to the many applications in which a finding comes with hard evidence.

\paragraph{Endogenous two-stage structure.} The PBS bonus sequence has two stages, separated by the endogenous date $n_0^\ast$. The \emph{research stage} covers periods $t\le n_0^\ast$. There, bonuses above $\underline b$ induce both the uninformed agent and an informed agent to work, and the uninformed agent keeps learning. The \emph{development stage} covers periods $t>n_0^\ast$. In this stage, the low, constant bonus $\underline b$ induces only an informed agent to work, exactly covering his effort cost, since he knows that $\omega=H$ and has nothing left to learn. The development stage has no fixed deadline: each additional period of informed work yields surplus $p_H\pi-c>0$ and, at $\underline b$, leaves the agent no rent. The principal therefore keeps an informed agent working until success.

Organizing research in sequential stages is common in practice. America's Seed Fund, the SBIR/STTR program, by default supports a feasibility phase first and, only for projects that clear this phase, a separate development phase afterward \citep[see][]{nisbir}. The international accounting standard IAS 38 likewise requires firms to split an internal project into a research phase and a development phase \citep*{ias38}, and \citet*{rost2007} document that staged financing is pervasive in biotech strategic alliances.

The two-stage structure also speaks to a notable feature of R\&D practice. A priori, information generated through experimentation could guide both how long the project runs and how much the agent is paid, and the contractible-finding contract of Section~\ref{sec:cf} does condition both the duration and the pay on information. In practice, however, information often guides continuation but not pay. The National Institutes of Health, for example, collects annual progress reports and uses them to decide whether a project continues, yet the award amount for each project year depends on the time since project approval rather than on the content of the reports \citep*{nigp2026,nifp2026,nisn2024}. The PBS contract has the same structure, since the finding determines whether the agent continues working after period $n_0^\ast$, while the bonus varies with calendar time alone. Theorem~\ref{thm:main result} shows that making pay depend on the information has no value when information arrival is \emph{private}.

\paragraph{Rising research-stage bonus.} Why does the bonus rise strictly over the research stage? Effort obedience reflects two forces. The first force, familiar from Section~\ref{sec:cf}, is that shirking preserves an optimistic belief, which makes future bonuses more valuable to a shirker. The second force, the \emph{learning incentive}, is novel. Working may produce a finding, which gives the agent a positive payoff because the PBS contract continues to pay research-stage bonuses above $\underline b$ after a finding. When the finding is contractible, only the familiar force operates, since an informed agent then earns no rent.

This learning incentive weakens over time. Obtaining the finding earlier lets the agent collect the research-stage bonus above $\underline b$ with the success probability $p_H$ for more periods, so the prospect of a finding makes early work more valuable to him. As the research stage approaches its end, fewer periods remain in which the agent can earn rents from a finding. The success bonus must therefore increase to offset this declining learning incentive and keep effort obedience binding.

To see this formally, consider the effort obedience constraints. Recall that $P_t$ is the success probability perceived by an uninformed agent who has worked for $t$ silent periods, and let $Q_t$ denote the probability, as perceived by the same agent, that a period of work produces a finding but no success. Let $F_t$ denote the informed payoff of an agent who obtains a finding in period $t$ under the PBS contract, discounted to period $t$. For every $t<n_0$, the binding effort obedience constraints at the contractible-finding contract and at the PBS contract give, respectively,
\[
\begin{aligned}
b_0^{CF}(t{+}1)-b_0^{CF}(t)&=(1-\delta)\Big[b_0^{CF}(t{+}1)-\frac{c}{P_{t-1}}\Big],\\
b_0^\ast(t{+}1)-b_0^\ast(t)&=(1-\delta)\Big[b_0^\ast(t{+}1)-\frac{c}{P_{t-1}}\Big]+\frac{Q_{t-1}}{P_{t-1}}\,\big(F_t-\delta F_{t+1}\big).
\end{aligned}
\]
The first term, common to both contracts, reflects the familiar force. A shirker postpones his work by one period without changing his belief, so in the next period he evaluates the bonus $b_0(t{+}1)$ relative to the myopic level $c/P_{t-1}$ implied by his unchanged belief. The second term under the PBS contract comes from the novel force. Its factor $F_t-\delta F_{t+1}$ is the value of obtaining the finding one period earlier, which is strictly positive for every $t<n_0$ (Appendix~\ref{app:proof main result}). The ratio $Q_{t-1}/P_{t-1}$, the number of findings the obedient agent expects per success, converts this value into bonus units.

Without discounting, the familiar term vanishes, so the PBS bonus rises strictly not because a working agent grows more pessimistic but because of the learning incentive alone. The contractible-finding bonus is then constant, and since both bonuses equal $c/P_{n_0-1}$ in the last period, summing the second term from period $t$ to period $n_0-1$ gives
\[
b_0^\ast(t;n_0)=b_0^{CF}(t;n_0)-\sum_{s=t}^{n_0-1}\frac{Q_{s-1}}{P_{s-1}}\,\big(F_s-F_{s+1}\big).
\]
The PBS bonus thus falls short of the contractible-finding bonus by the value of obtaining the finding one period earlier, converted into bonus units and summed over the remaining research-stage periods, and this gap shrinks as the research stage runs down. With discounting, the contractible-finding bonus rises as well, and the next result shows that the gap, measured in period-$1$ value, still shrinks over the research stage.

\begin{proposition}\label{prop:gap}
    Fix a finite $n_0$. The discounted difference $\delta^{t-1}\big[b_0^\ast(t;n_0)-b_0^{CF}(t;n_0)\big]$ is strictly increasing in $t$ for $t\le n_0$.
\end{proposition}

\emph{Proof}. See Appendix~\ref{app:proof gap}.

\medskip

Positive rents after a finding are enough to create a learning incentive. The key question here is how the need to deter concealment shapes this incentive. First, because the agent can conceal a finding, the rent after a report must be higher than if he could not, which enlarges the learning incentive. Second, truthful reporting requires the rent after a report today to be no lower than the present value of delaying the report to collect the rent after a report tomorrow. Therefore, the present value of the rent after a finding declines over time, and so does the learning incentive. Both features thus come from the private, dynamic arrival of a finding and emerge in every feasible contract.

\subsection{Comparative Statics}\label{sec:comparative statics}

This section asks how the finding affects the two parties' payoffs. The finding matters in two ways, since it carries information about the state and only the agent observes when it arrives. I separate the two effects, varying first how likely a finding is and then whether it is private.

To study the value of information, I raise the finding probability $q_H$, so that in state $H$ a finding tends to arrive sooner. In the special case of raising $q_H$ from zero, the comparison is between a project in which no finding can arrive and one that can produce a finding. The next result holds whether the finding is contractible or private, so it concerns the information itself rather than its privacy.

\begin{proposition}\label{prop:information}
    Whenever $n_0^{FB}>0$, raising $q_H$ strictly increases first-best total surplus. Whether the finding is contractible or private, raising $q_H$ can increase or decrease either party's payoff, and both payoffs can decrease at once.
\end{proposition}

\emph{Proof}. See Appendix~\ref{app:proof information}.

\medskip

In the first best, a higher $q_H$ makes work more informative: a finding reveals the state sooner, and its absence is stronger evidence that the state is low. More information guides the planner's decisions better, so the first-best surplus rises whenever the first best involves any work.

In the second best with moral hazard and limited liability, raising $q_H$ changes incentives. It makes silence worse news, so an agent who works becomes more pessimistic more quickly. Formally, $P_t$ decreases with $q_H$ for any fixed $t>0$. This faster learning allows shirking to preserve more optimism and can raise the payoff the agent must receive to work for a given uninformed duration. When a given duration becomes more expensive, the principal can respond by shortening it. She does so whenever the reduction in the agent's rent exceeds the loss of surplus. Thus, the cut can reduce total surplus, and raising $q_H$ can leave both parties worse off.\footnote{Let $c=1$, $p_H=2/5$, $p_L=3/20$, $\mu_0(H)=9/10$, $\delta=1$, and $\pi=11/2$. Raising $q_H$ from $7/20$ to $2/5$ keeps $n_0^{FB}=6$ and raises the first-best surplus from $2.6178$ to $2.6215$, but it shortens the uninformed duration at the optimal contract from $3$ periods to $2$, whether the finding is contractible or private, and the total surplus falls from $2.4948$ to $2.3177$. The agent's payoff falls from $0.2480$ to $0.0774$ when the finding is contractible and from $0.2503$ to $0.0774$ when it is private, and the principal's payoff falls from $2.2469$ to $2.2403$ and from $2.2445$ to $2.2403$. The two cases coincide at $q_H=2/5$ because the uninformed duration is then $2$, so privacy costs nothing (Proposition~\ref{prop:privacy}). All numbers are rounded from exact values.}

To isolate the effect of privacy, I hold $q_H$ fixed and compare a private finding with a contractible one. The next result identifies when privacy is costly to the principal.

\begin{proposition}\label{prop:privacy}
    Relative to a contractible finding, a private finding makes the principal weakly worse off, and strictly so if and only if $n_0^{CF}>2$. The agent can be either better or worse off.
\end{proposition}

\emph{Proof}. See Appendix~\ref{app:proof privacy}.

\medskip

Privacy never helps the principal, because any contract feasible when the finding is private is also feasible when it is contractible. It harms her when $n_0^{CF}>2$. In that case, every optimal contract of Section~\ref{sec:cf} pays an agent informed in period $2$ only $\underline b$ in period $3$, less than the uninformed bonus $b_0^{CF}(3)$. When the finding is private, such an agent would conceal it and collect the uninformed bonus, so the principal must concede an additional rent to restore feasibility.

When $n_0^{CF}\le2$, privacy costs nothing. For $n_0^{CF}=1$, the PBS contract coincides with the contractible-finding contract. For $n_0^{CF}=2$, the contractible-finding contract invites concealment when the finding is private, because an agent informed in period $1$ receives $b_{11}^{CF}(2)=\underline b$ in period $2$, less than the uninformed bonus $b_0^{CF}(2)$. By Section~\ref{sec:cf}, the agent's expected payoff in period $1$ from a success and a finding together is pinned down, but its division between the two is not. The bonus $b_0(1)$ and the bonus $b_{11}(2)$ are therefore substitutes. The principal can raise $b_{11}(2)$ to $b_0^{CF}(2)$ and lower $b_0(1)$ accordingly, which closes the bonus gap at no cost. This adjustment yields the PBS contract, which is therefore optimal when the finding is contractible.

Privacy can either benefit or harm the agent. When the uninformed duration is held fixed, the agent gains exactly what the principal loses. But privacy also makes each additional uninformed period weakly more costly for the principal, because of the positive informed payoffs that privacy requires, so she can shorten the uninformed duration. Such a cut lowers the agent's payoff, and this loss can outweigh his gain at a fixed duration.\footnote{At the primitives of the right panel of Figure~\ref{fig:main result}, the uninformed duration falls from $n_0^{CF}=6$ to $n_0^\ast=5$, and the agent's payoff falls from $0.3469$ to $0.2075$, rounded from exact values.}

\subsection{Proof Sketch and Mechanism}\label{sec:proof sketch}

I now sketch the proof of Theorem~\ref{thm:main result} and explain why the value of reporting is exactly zero.

I focus on program (\ref{eq:first step}), which finds the cheapest transfers implementing a given effort schedule. For any finite uninformed duration $n_0$ and infinite informed durations, I show that the PBS contract $\phi^\ast(n_0)$ is a solution to this program. To solve the problem, I relax (IC). The \emph{relaxed problem} retains (LL), effort obedience, and no-concealing, and drops all other incentive constraints, including no-faking. The relaxed problem is a linear program in the transfers, and I apply weak duality to bound its optimal value from below. I subtract from the agent's payoff the slack of each effort obedience and no-concealing constraint, weighted by a nonnegative multiplier. I choose the multipliers so that every transfer cancels except the uninformed wages, each of which keeps a positive coefficient. By (LL), setting these wages to zero yields a lower bound on the agent's payoff. The PBS contract attains this bound, and it is feasible in the original problem, so it solves the first step. The same lower bound on the agent's payoff applies to finite informed durations, while extending informed work increases total surplus. Since, in addition, an infinite uninformed duration is never optimal, the focus on a finite uninformed duration and infinite informed durations entails no loss. The full proof is in Appendix~\ref{app:proof main result}. For a contractible finding, a similar version of the argument, in which no-concealing is replaced by the constraint that informed payoffs be nonnegative, proves Theorem~\ref{thm:cf} (Appendix~\ref{app:proof cf}).

Why is reporting worthless? The answer has two steps, and the first one clarifies the key mechanism. First, the principal optimally satisfies the no-concealing constraints with equality from period $2$ on. Second, a report-free contract can then deliver the payoffs that these binding constraints pin down.

\emph{Step 1: the principal wants no-concealing to bind.} The agent's payoff comes through two channels, the bonus upon a success and the informed payoff upon a finding. Which channel tempts the shirker and the concealer more than it tempts the obedient agent? Two lemmas in Appendix~\ref{app:proof main result} answer this question. The first, the \emph{shirker lemma}, says that a more optimistic agent expects more findings per success. In the main model, this holds because a success can arrive in either state, while a finding arrives only in state $H$. A shirker is more optimistic than the obedient agent, since shirking preserves his belief, so per success he expects more findings than the obedient agent does. The second, the \emph{concealer lemma}, shows that a concealer expects to collect the informed payoff more often per success than the obedient agent does. The concealer already has a finding to report and collects this payoff unless success occurs while he delays. The obedient agent, by contrast, receives it only if his work produces a finding but no success. Therefore, compared to the success bonus, it is the informed payoff upon a finding that makes the deviations more attractive.

It follows that, if no-concealing does not bind, the principal benefits from adjusting the transfers in the following way: she cuts the informed payoff upon a period-$t$ report by one unit and raises the uninformed bonus $b_0(t)$ by $Q_{t-1}/P_{t-1}$, the number of findings the obedient agent expects per success. The obedient agent's payoff is then unchanged. An agent who shirked in one earlier period, in contrast, reaches period $t$ more optimistic, perceiving $P_{t-2}$ and $Q_{t-2}$, and loses $P_{t-2}\big[Q_{t-2}/P_{t-2}-Q_{t-1}/P_{t-1}\big]$, which is positive by the shirker lemma. Moreover, a concealer delaying a period-$(t-1)$ finding works in period $t$ knowing that $\omega=H$ and loses $1-p_H-p_HQ_{t-1}/P_{t-1}$, which is positive by the concealer lemma. Each deviator thus loses more from the cut than he gains from the higher bonus, and this adjustment helps to deter both deviations. This cut, however, tightens no-concealing in period $t$, because a lower informed payoff tempts an agent who obtains a finding in period $t$ to conceal it. Since less attractive deviations let the principal lower the agent's pay elsewhere, she lowers the informed payoff until the no-concealing constraint binds. The multiplier on this constraint measures the value of reducing the informed payoff by one more unit. It equals the losses this adjustment imposes on earlier deviators, weighted by the multipliers on their respective incentive constraints:
\[
\eta_t\,P_{t-2}\Big[\frac{Q_{t-2}}{P_{t-2}}-\frac{Q_{t-1}}{P_{t-1}}\Big]+v_t\Big[1-p_H-p_H\,\frac{Q_{t-1}}{P_{t-1}}\Big],
\]
where $\eta_t$ sums the multipliers on effort obedience in the periods before $t$, each weighted by the discounted probability that its shirker reaches period $t$ uninformed, and $v_t$ is $\delta$ times the multiplier on no-concealing in period $t-1$. From period $2$ on, the multipliers on no-concealing are strictly positive. The exception is period $1$, before which no agent can have deviated, so $\eta_1=v_1=0$, and no-concealing holds with equality under the PBS contract, with a zero multiplier. This step does not depend on the finding being conclusive. In fact, one can show that the constraints against concealing remain a binding family under the condition
\[
\frac{q_L}{q_H}<\frac{p_L(1-p_H)}{p_H(1-p_L)},
\]
which is exactly the condition for the shirker lemma.\footnote{With an inconclusive finding, the constraints against concealing that can matter deter deviations that combine shirking before the finding with concealing after it, each of any length (Section~\ref{sec:inconclusive}). This family contains no-concealing, which deters only a one-period delay of the report.} This condition holds here because $q_L=0<p_L$.

\emph{Step 2: a posted bonus sequence delivers the payoffs that the binding constraints pin down.} When no-concealing holds with equality from period $t$ on, an agent informed in period $t$ is indifferent to any length of delay, so he receives what he would get by concealing the finding and continuing on the uninformed track. Since a finding reveals the state, all informed agents at a given date value the remaining uninformed bonuses in the same way. A single posted sequence that pays the uninformed bonuses during the research stage and $\underline b$ afterward therefore delivers this payoff to every informed agent, so reports add no value once the no-concealing constraints hold with equality.

Wages and lump sums are not needed for optimality. A wage is paid in every period the relationship reaches, so it favors shirking, which keeps the relationship alive, while work may end the relationship with a success. A report lump sum can deter concealment by raising the informed payoff, but the informed-track bonuses do the same job while making faking less attractive. A faker is uninformed and less optimistic than an informed agent, so in each period he works he expects to collect a bonus less often. Replacing informed bonuses with a lump sum of the same value to the informed agent would therefore raise the faker's payoff and make faking harder to deter.

\section{Discussion and Extensions}\label{sec:extension and discussion}

This section discusses the main assumptions of the model and studies four extensions.

\subsection{Discussion of Assumptions}\label{sec:discussions}

\paragraph{General contracts.} Lemma~\ref{lem:direct} justifies the focus on direct mechanisms. Online Appendix~\ref{app:general contracts} proves it for a class of general contracts with finite message spaces. This class allows public and hidden randomization, so the transfers, effort schedule and recommendations can be random, and the effort schedule can skip periods. To prove the lemma, rather than turning each general contract into a direct mechanism, I compare every general contract directly with the PBS contract.

The direct mechanism never skips a period of work and never randomizes. These two restrictions entail no loss. Skipping a period, under common discounting, only postpones work and thus success. As for randomization, note that the agent receives no benefit from shirking. Termination is thus covered by offering no further transfers, so it suffices to consider random transfers. If the agent receives no new information about future transfers as the project continues, risk neutrality means that randomizing those transfers cannot help. If the agent instead learns about future transfers gradually, randomization could in principle help, since agents with different beliefs could be induced to choose different efforts. However, the constraints that matter deter the shirker and the concealer, who are more optimistic than the obedient agent and so work whenever he does. Given the same efforts, only expected transfers matter. In \cite*{grta2016} and \cite*{ftwz2024}, by contrast, the agent receives a benefit from shirking, so termination serves as an incentive tool separate from transfers, and random termination can be optimal.

\paragraph{Conclusive success.} In the main model, a success can arrive in either state. If instead $p_L=0$, a success, like a finding, arrives only in state $H$. The ratio of the finding probability to the success probability then no longer depends on the agent's belief, so uninformed agents with different beliefs trade off payments upon success against payoffs after a finding at the same rate. A shirker therefore values any shift of pay between a success and a finding exactly as the obedient agent does, and one can show that the set of optimal contracts expands. When the finding is contractible, the division of pay between a success and a finding is no longer pinned down in any period, and when it is private, other report-free optimal contracts exist that pay a higher development-stage bonus. Theorems~\ref{thm:cf} and~\ref{thm:main result} still hold (Online Appendix~\ref{app:conclusive}), and when the finding is contractible, the PBS contract is optimal alongside the contractible-finding contract. Privacy therefore costs nothing in this conclusive-success setting, and the principal obtains the same payoff whether the finding is private or contractible.

\paragraph{Other optimal contracts.} I focus on the PBS contract because it is consistent with practice and, as the only report-free optimal contract when $n_0^\ast>1$, would be selected by a small reporting cost (Section~\ref{sec:main result}). Other optimal contracts give the agent the same payoff but deliver it in other forms. None of them pays a wage on the uninformed track. Lump sums are never needed but can appear, because in period $1$ only the total expected pay from a success and a finding is pinned down (Section~\ref{sec:cf}), and because an informed payoff can be divided between a lump sum upon the report and subsequent payments on the informed track. A report lump sum, whether it replaces part of the period-$1$ bonus or part of the informed-track bonuses, makes faking more attractive, since a faker values a lump sum as much as an informed agent does but expects to collect bonuses less often. The principal can offset this higher payoff from faking by spreading the remaining informed payoff over more periods on the informed track, with lower bonuses that still exceed $\underline b$. A faker, who becomes more pessimistic as he works without news, is then willing to work in only a few of these periods.\footnote{For example, with the parameters in Table~\ref{tab:example}, the following modification of the PBS contract is also optimal. It leaves $b_0$ and the period-$1$ informed payoff unchanged, but pays a lump sum $\ell_1\approx0.186$ upon a period-$1$ report. In place of the PBS bonuses $(3.610,\,3.684)$ in periods $2$ and $3$, it pays bonuses of approximately $(3.045,\,3.046,\,3.048,\,3.053,\,3.063,\,3.086,\,3.136)$ in periods $2$ through $8$ of the $1$-informed track and $\underline b$ afterward. A faker works in at most two of these periods.}

\paragraph{Commitment.} The main model assumes that the principal commits to a contract in period $0$. Suppose instead that the principal cannot commit. In each period, the principal offers a bonus for that period only, and the agent chooses effort and, absent a success, sends a message.\footnote{The equilibrium notion is perfect Bayesian equilibrium, with beliefs updated by Bayes' rule wherever possible, including after the principal's own deviations, which do not change her belief about the agent's information.} Once a message reveals a finding, it is sequentially rational for the principal to offer $\underline b$ in every later period, which leaves the informed agent no rent. An informed agent thus conceals his finding as long as some later bonus offered without a report is higher than $\underline b$. Depending on parameters, it is then possible that every equilibrium with pure on-path play is outcome-equivalent to one in which no on-path reporting happens.\footnote{For an example, take $c=1$, $p_H=1/2$, $p_L=1/100$, $q_H=3/5$, $\mu_0(H)=9/10$, $\delta=1/2$, and $\pi=9/2$. An equilibrium with pure on-path play exists, and in every such equilibrium, the uninformed agent works in exactly two periods, the second at the bonus $c/P_1$. An agent who obtains a finding in the first of these two periods conceals it, and every bonus paid after the second is $\underline b$.} In that case, pay does respond to reports, but no reports need to be made on the equilibrium path.

\subsection{Costless Findings}\label{sec:costless}

This section studies a \emph{costless finding}, which can arrive whether or not the agent works. While the agent is uninformed, the finding arrives with probability $q_H$ in each period in state $H$ and never in state $L$, whereas success still requires effort. I show that the PBS contract remains optimal within the class of threshold schedules defined below and that, for a given uninformed duration, it pays exactly the bonuses of the main model. The learning incentive, however, reverses. Working may produce a success that ends the relationship before a costless finding arrives, so the prospect of an informed payoff now makes work harder, rather than easier, to motivate. For a given uninformed duration, the PBS bonus therefore lies above the bonus that would be optimal if the costless finding were contractible, except in the last research-stage period, where the two are equal.

I focus on direct mechanisms with \emph{threshold schedules}. A threshold schedule recommends that an uninformed agent work during the research stage, $t\leq n_0$, and shirk afterward, and that an informed agent work until success. Unlike in the main model, the uninformed track continues after period $n_0$. In state $H$, a finding can still arrive while the agent shirks; he can then report it and work until success. Let $n_0^C$ be the shortest optimal uninformed duration within this class when the costless finding is private. Whether or not the agent reports a finding, the PBS contract $\phi^\ast(n_0)$ pays the PBS bonus $b_0^\ast(t;n_0)$ of Section~\ref{sec:main result} upon a success in each period $t\le n_0$, and $\underline b$ afterward. If the costless finding were contractible, an optimal contract with uninformed duration $n_0$ would be $\tilde\phi^{CF}(n_0)$, which pays $\underline b$ upon a success after a finding or after period $n_0$, and otherwise the bonus $\tilde b_0^{CF}(t;n_0)$ at which effort obedience holds with equality. Online Appendix~\ref{app:proof costless} proves the optimality of this contract and the following result.

\begin{proposition}\label{prop:costless}
    Suppose a finding is costless and private. Within the class of threshold schedules, the principal optimally offers the PBS contract $\phi^\ast(n_0^C)$, where $n_0^C\le n_0^\ast$, and $n_0^C=0$ if $\delta=1$. Moreover, for every uninformed duration $n_0>0$, the bonus satisfies $b_0^\ast(t;n_0)\ge\tilde b_0^{CF}(t;n_0)$ for $t\le n_0$, strictly iff $t<n_0$.
\end{proposition}

For a given uninformed duration, the bonuses of the main model stay optimal, for two reasons. First, the PBS bonuses still satisfy effort obedience with equality. Under the PBS contract, a finding changes neither the agent's bonuses nor his effort during the research stage, and the work it induces after period $n_0$ earns him no rent at the bonus $\underline b$. His payoff in state $H$ therefore does not depend on whether or when the finding arrives, and in state $L$ no finding arrives, so the obedient agent and a shirker each earn exactly what they earn in the main model. Second, when the finding is private, no feasible contract with the same uninformed duration gives the agent less than the PBS contract does. An agent who obtains a finding while he shirks can act as if it had not yet arrived, and he then earns at least what the same deviation earns in the main model, so every constraint of the main model still applies.

Before the last research-stage period, the PBS bonus exceeds $\tilde b_0^{CF}$ because the learning incentive reverses. Under the PBS contract, an agent who obtains a finding before period $n_0$ earns a positive informed payoff, since the later research-stage bonuses exceed $\underline b$, whereas under $\tilde\phi^{CF}(n_0)$ he earns no rent. In the main model, only working can bring a finding, so the prospect of this payoff encourages work. Here, shirking can also bring a finding, while working may bring a success that ends the relationship before the finding arrives, so the prospect of this payoff discourages work, and the PBS bonus must exceed $\tilde b_0^{CF}$.

The optimal uninformed duration is weakly shorter than in the main model. For a given uninformed duration, the principal leaves the agent the same payoff as in the main model, but each research-stage period adds less surplus, because a finding still arrives for free after the research stage ends. At $\delta=1$, there is no research stage at all, because waiting costs nothing. In state $H$, the finding eventually arrives without effort and the agent then works until success, so earlier uninformed work adds no surplus in that state. In state $L$, every period of work generates a surplus loss of $c-p_L\pi>0$. Without a research stage, the agent also earns no rent, so the principal keeps the whole surplus.

\subsection{Private Success}\label{sec:private success}

I now let success be privately observed, so the agent can delay disclosing it. Without discounting, the bonus then cannot rise over time, since an agent with a success would wait for the highest bonus. Yet I show that, for a fixed uninformed duration, a posted bonus sequence that never rises attains the principal's optimal payoff.

Only the agent observes a success, and he chooses when to \emph{disclose} it with verifiable evidence. Upon disclosure, the agent receives the current bonus, the principal receives $\pi$, and the relationship ends. No further success or finding arrives while a success remains undisclosed. The obedient agent discloses a success at once. I focus on $\delta=1$. The PBS contract of Theorem~\ref{thm:main result} is then no longer feasible if its research stage lasts more than one period, because its bonus rises and an agent would delay disclosing an early success until the last research-stage period.

At a fixed uninformed duration $n_0$, one of two posted bonus sequences that never rise is optimal. The \emph{flattened PBS contract} posts the last research-stage bonus of the PBS contract, the myopic bonus $c/P_{n_0-1}$, in every period $t\le n_0$, and $\underline b$ afterward. The \emph{constant-bonus contract} offers in every period the bonus that makes the uninformed agent just willing to work in period $n_0$. As he works without news, he becomes more pessimistic, so this bonus no longer induces him to work after period $n_0$. Online Appendix~\ref{app:proof private success} proves the following result.

\begin{proposition}\label{prop:private success}
    Suppose success is private and $\delta=1$, and fix a finite $n_0$. A posted bonus sequence attains the principal's optimal payoff. If $n_0>1$ and $q_H\le p_L$, the constant-bonus contract does so at every prior. If $n_0>1$ and $q_H>p_L$, the flattened PBS contract does so at priors below a threshold, and the constant-bonus contract at priors above it.
\end{proposition}

To see which contract is optimal, consider incentives in the last research-stage period, $n_0$. The principal can reward success with a higher research-stage bonus or make a finding more valuable by raising the development-stage bonus. The cost of the first option is that the higher bonus must also be paid upon an earlier success, since bonuses cannot rise; the cost of the second is a higher informed payoff from an earlier finding. Which cost is larger depends on how likely each event is to arrive before, rather than in, period $n_0$. If successes arrive faster than findings, raising the development-stage bonus is cheaper, so the constant-bonus contract is optimal: the principal raises the development-stage bonus and lowers the research-stage bonus until they coincide. If findings arrive faster, raising the research-stage bonus is cheaper, so the flattened PBS contract is optimal.

If $q_H\le p_L$, successes arrive in either state at least as fast as findings arrive in state $H$, so the constant-bonus contract is optimal at every prior. If $q_H>p_L$ and the prior is low, the state is relatively likely to be $L$, where successes arrive at rate $p_L$, more slowly than findings arrive in state $H$. The flattened PBS contract is then optimal. A higher prior raises the perceived probability of success, since $p_H>p_L$. In state $H$, moreover, an agent who obtains a finding before period $n_0$ still collects the research-stage bonus if he then succeeds by period $n_0$, which makes raising that bonus costlier still. The constant-bonus contract is therefore optimal at high priors, whatever the value of $q_H$.

\subsection{Inconclusive Findings}\label{sec:inconclusive}

The main model assumes that the finding arrives only in state $H$, so it is conclusive. Suppose instead that $q_H>q_L>0$, so that a finding is still good news but no longer reveals the state. A posted bonus sequence can remain optimal in this setting.\footnote{For example, let $c=1$, $p_H=1/2$, $p_L=1/50$, $q_H=1/4$, $q_L=1/8$, $\mu_0(H)=3/10$, $\delta=1$, and $\pi=25$. The optimal schedule is then $n_0=2$, $n_{11}=2$ and $n_{12}=1$, and posting the single bonus sequence $(10.36,\,10.44,\,10.30)$ attains the principal's optimal payoff. I establish optimality exactly among bonus-only contracts and numerically among all direct mechanisms. The numbers are rounded from exact values.} A general characterization, however, becomes much harder, because an informed agent keeps learning and the principal must choose how long informed work lasts after each report.

Specifically, after a finding, each period of work without success lowers the agent's belief. This means that shirking keeps him more optimistic than working, as it does before a finding. To keep the agent working after a report, the principal must therefore pay bonuses above those that just cover his effort cost in expectation, and the longer informed work lasts, the larger the informed payoff she must leave him. The principal weighs the surplus from longer informed work against this larger informed payoff. With a conclusive finding, by contrast, the high-state bonus $\underline b$ keeps an informed agent working while leaving him zero rent, so the principal keeps him working until success and chooses only the uninformed duration.

With an inconclusive finding, the principal must therefore find the cheapest contract for each combination of uninformed and informed durations, rather than only for each uninformed duration as in the main model. Two features make this task harder. First, which deviations are most attractive can change with the informed durations. When informed work after every report is short, keeping an informed agent working requires only bonuses below those before a report. The most attractive way to conceal a finding then involves no shirking, since shirking gives up chances to collect the higher bonuses before a report. When informed work after a report in some period $k$ is long, keeping the informed agent working can require bonuses above those before a report. The resulting high informed payoff after a period-$k$ report tempts an uninformed agent to shirk for a while to stay optimistic, then work, and, if he obtains a finding, conceal it until period $k$ to collect this payoff. Second, combinations of shirking and concealing, each of any length, now matter, so the principal must deter many more deviations than in the main model, where deterring one-period shirking and one-period concealing suffices (Section~\ref{sec:proof sketch}).

Nevertheless, a posted bonus sequence is close to optimal when the finding nearly reveals the state. Let $q_L=\varepsilon$, and let $\Delta U_P(\varepsilon)\ge0$ be the principal's loss from posting a single bonus sequence, that is, how much less she obtains with the best such sequence than with the best direct mechanism of Section~\ref{sec:model}. By Theorem~\ref{thm:main result}, $\Delta U_P(0)=0$. Online Appendix~\ref{app:proof almost conclusive} gives the precise definition of $\Delta U_P$ and proves the following result.

\begin{proposition}\label{prop:almost conclusive}
    Suppose a finding is inconclusive, with $q_L=\varepsilon$. There exist $C,\hat\varepsilon>0$ and $\alpha\in(\tfrac12,1)$ such that $\Delta U_P(\varepsilon)\le C\varepsilon^{\alpha}$ for every $\varepsilon\in(0,\hat\varepsilon]$. Thus, a posted bonus sequence is nearly optimal when the finding nearly reveals the state.
\end{proposition}

The principal's gain from using reports vanishes as the finding becomes conclusive. So if writing and enforcing a contract that depends on reports entails any positive additional fixed cost, a posted bonus sequence beats every direct mechanism that depends on reports once the finding is close enough to conclusive.

This result does not follow from a direct application of a standard continuity theorem, because the PBS contract keeps informed agents working forever at $\varepsilon=0$ but is infeasible at every $\varepsilon>0$.\footnote{The maximum theorem \citep*{berg1963} needs a compact choice set and a feasible set that varies continuously with $\varepsilon$, but the schedule space is discrete and unbounded, and a PBS contract with $n_0>0$ violates (IC) at every $\varepsilon>0$, since an informed agent no longer works for $\underline b$. Envelope theorems \citep*{mise2002} admit a parameter in the constraints only through a saddle point of the Lagrangian, which the problem of jointly choosing a schedule and its transfers need not have. The stability of linear programs \citep*{dafs1967} covers the first step at one schedule, but not the maximum over infinitely many schedules.} The proof instead compares the principal's payoffs with and without reports to her optimal payoff when the finding is conclusive. With reports, she gains at most a multiple of $\varepsilon$ over this payoff, because a finding still leaves the agent almost certain that $\omega=H$. Without reports, she loses at most a multiple of $\varepsilon^\alpha$ by slightly increasing the posted bonus sequence, so that it pays slightly more in the research stage and slightly above $\underline b$ for finitely many development-stage periods, after which the contract ends. These higher bonuses keep informed agents working as they become more pessimistic after periods without success. The exponent $\alpha$ reflects the trade-off between a longer development stage, which adds surplus, and the larger bonus increases needed to sustain effort.

\subsection{Bad-News Findings}\label{sec:bad news}

This section turns to a finding that is bad news, with $q_H=0$ and $q_L\in(0,1)$. A finding then reveals $\omega=L$, so an informed agent succeeds with probability $p_L$ in a period in which he works and breaks even at the bonus $\bar b:=c/p_L$. Each silent worked period multiplies an uninformed agent's likelihood ratio of $\omega=H$ to $\omega=L$ by the silence factor $\rho:=(1-p_H)/[(1-p_L)(1-q_L)]$. I focus on $\rho\ge1$, which holds when a finding arrives often enough in state $L$. No news is then good news, so the success probability $P_t$ that an uninformed agent perceives after $t$ silent worked periods weakly increases as $t$ grows.

The \emph{myopic contract} pays the myopic bonus $c/P_{t-1}$ upon a success in each period $t\le n_0$, and it recommends no work and pays nothing after a report. Since this bonus lies below $\bar b$, an informed agent never works for it, so the contract is equivalent to posting this bonus sequence. The next proposition maintains Assumption~\ref{ass:efficiency}, so that work is wasteful once the state is known to be low.\footnote{When no news is bad news ($\rho<1$), privacy also plays no role if the finding is verifiable. The principal then attains the first-best surplus and leaves the agent zero payoff by running the first-best schedule and paying $c/p_H$ upon a success, the lump sum $c(p_H-p_L)/[p_Hq_L(1-p_L)]$ upon a report, and nothing after a report. Every worked period then yields no rent in each state: in state $H$, where no finding arrives, the bonus exactly covers the effort cost, and in state $L$, the lump sum covers the rest of the effort cost. Also, the payments do not rise over time, so delaying a report never pays. Hence, neither shirking nor concealment earns more than zero, whatever the agent believes.}

\begin{proposition}\label{prop:bad news}
Suppose that the finding is bad news and that no news is good news. Maintain the assumption that $p_L\pi<c$. A myopic contract, which induces no informed work, attains the first-best surplus and leaves the agent zero payoff. It is optimal whether the finding is private or contractible, so privacy plays no role.
\end{proposition}

After bad news, further work is wasteful, so the principal wants the agent to stop, and the myopic contract recommends exactly that. An agent who obtains a finding gains nothing by concealing it, because the myopic bonus lies below $\bar b$. An uninformed agent gains nothing by faking a finding either, because a report ends all payments. Since no news is good news, a shirker is never more optimistic than the obedient agent, so the myopic bonus, which exactly compensates the obedient agent for his effort cost, leaves a shirker no rent either. The principal thus collects the entire first-best surplus. The contract is simple here for a plain reason: the principal wants an informed agent to stop. By contrast, with a good-news finding, the principal wants both the informed and uninformed agents to work, so eliciting private information can be costly.

I next consider what happens when $p_L\pi>c$, so that work is worthwhile even when the state is known to be low. One might then expect the principal to have every agent, informed or not, work until success. The next result shows that she does so when revenue $\pi$ exceeds a threshold $\pi_c$, but not otherwise. Online Appendix~\ref{app:proof bad news} defines this threshold and proves the two propositions of this section.

\begin{proposition}\label{prop:bad news efficient}
Suppose that the finding is bad news and that no news is good news. Moreover, suppose that $p_L\pi>c$ and $\delta=1$.
\begin{enumerate}[nolistsep,label=(\roman*)]
    \item If $\pi<\pi_c$, the myopic contract with $n_0=\infty$, under which only an uninformed agent works, is optimal.
    \item If $\pi>\pi_c$, posting $\bar b$ in every period, under which every agent works until success, is optimal.
\end{enumerate}
\end{proposition}

Faking is now the main concern. Because an informed agent knows that the state is low, the principal must offer him a bonus of at least $\bar b$ in every period to keep him working. An uninformed agent is more optimistic, so he values these bonuses more and can fake a finding to collect them. Inducing work after a finding therefore requires leaving the agent a positive payoff. At $\delta=1$, I show that this payoff is at least what an uninformed agent can earn by shirking from the start and then faking a finding in the period with the longest informed track. The contract of part (ii) leaves the agent exactly the payoff from this deviation. The threshold $\pi_c$ is the revenue at which the additional surplus from letting informed agents work until success equals the rent from this shirk-then-fake deviation. When $\pi$ is below it, the principal lets only the uninformed agent work, and when $\pi$ is above it, she lets every agent work until success.

\section{Conclusion}\label{sec:conclusion}

This paper studies contracting for experimentation when the agent can privately obtain findings about a project's quality. Such information could guide both how long the project runs and how much the agent is paid. The principal must, however, account for the agent's incentives to conceal or fake a finding. I show that an optimal contract has a simple form: it pays success bonuses tied to time alone. Bonuses rise during a research stage that induces learning, then fall to a low, constant level in a development stage. The finding determines whether the agent continues working in this second stage, but reports need not affect his pay.

More broadly, my analysis shows that dynamically arriving private signals make concealment a key deviation. With moral hazard, an agent's option to conceal changes the rent he earns from learning, and this change shapes contracting in two ways. First, the rent after a signal must be higher than when the agent cannot conceal the signal. This higher rent limits pay differences across report histories and so reduces the value of reports. Second, the agent must be willing to report a signal at once and collect the rent after the report, which implies that a later report is worth less to him. The learning incentive---the prospect of this rent that motivates the agent to work---therefore decreases over time, and the explicit incentive must then rise to sustain effort.

Whether concealment or fabrication matters for incentive design is relevant to research in practice. In a national survey, 19.8 percent of life-science faculty reported delaying the publication of their results by more than six months at least once in the previous three years \citep*{blca1997}. Such delays resemble concealment in my model, where a researcher withholds a finding by delaying its report. In the main model, making findings verifiable rules out fabrication but leaves the option to conceal, and I show that the same contract remains optimal, with the funder's payoff also unchanged.

There are other settings where fabrication can play a more important role. In particular, competition among researchers would raise different reporting issues. With competition, an agent's report can reach his rivals, so he may want to distort or withhold information to prevent others from replicating his work. For example, a biotechnology chief executive suggested that his firm sometimes presents misleading chemical structures in papers and conference posters to avoid revealing the actual structures to competitors \citep*{dech2026}. \citet*{anro2007} document scientists leaving critical details out of papers and talks so that competitors cannot repeat the work. A model with competing researchers could capture these incentives, shedding light on how concealment and faking constraints might both influence optimal contracts for experimentation.

\bibliographystyle{aer}
\pdfbookmark[1]{References}{references}
\bibliography{bib}

\begin{appendices}

\section{Preliminaries}\label{app:preliminaries}

This appendix introduces the notations that the proofs rely on. Let $\mu_t$ be the probability that an uninformed agent who has worked $t$ silent periods assigns to $\omega=H$, and $\theta_t:=\mu_t/(1-\mu_t)$ the likelihood ratio. With a slight abuse of notation, $\mu_0$ also denotes the prior probability of $\omega=H$. Recall that $P_t$ and $Q_t$ denote the probabilities, perceived by an uninformed agent who has worked $t$ silent periods, that a worked period produces a success and a finding but no success, respectively (Sections~\ref{sec:cf} and~\ref{sec:main result}). As functions of the agent's belief $\mu$, these probabilities are $P(\mu):=p_L+(p_H-p_L)\mu$ and $Q(\mu):=\mu q_H(1-p_H)$, so $P_t=P(\mu_t)$ and $Q_t=Q(\mu_t)$. Let $\nu(\mu):=1-P(\mu)-Q(\mu)=\mu\gamma'+(1-\mu)(1-p_L)$ denote the probability of silence, where $\gamma':=(1-p_H)(1-q_H)$, and let $\nu_t:=\nu(\mu_t)$. In this notation, the factor $\rho$ of Section~\ref{sec:model} is $\gamma'/(1-p_L)$. After a silent worked period, Bayes' rule implies
\begin{equation}\label{eq:app bayes}
\nu_t\,\mu_{t+1}=\mu_t\,\gamma',
\qquad
\nu_t\,(1-\mu_{t+1})=(1-\mu_t)(1-p_L),
\qquad
\theta_{t+1}=\rho\,\theta_t .
\end{equation}
Four facts below are used repeatedly.

(Fact 1) $\mu_t$, $P_t$, and $Q_t$ are strictly decreasing in $t$ and interior, and $\nu_t\in(0,1)$. These properties follow immediately from \eqref{eq:app bayes}, $\rho\in(0,1)$, and the fact that $P$ and $Q$ increase in $\mu$.

(Fact 2) $(1+\theta)P(\mu)=\theta p_H+p_L$.

(Fact 3) $P(\mu)Q(\mu')-P(\mu')Q(\mu)=-p_Lq_H(1-p_H)(\mu-\mu')$. Since $p_L>0$, this identity implies that the ratio $Q_t/P_t$ is strictly decreasing in $t$.

(Fact 4) Under the obedient strategy, the discounted probability that the relationship reaches period $t$ on the uninformed track is $\mathcal D_t:=\delta^{t-1}\prod_{s=0}^{t-2}\nu_s>0$, with $\mathcal D_1=1$ and $\mathcal D_t=\delta\nu_{t-2}\mathcal D_{t-1}$.

Define the discounted surplus of an informed track whose informed duration is infinite, entered in the period after the finding,
\[
V^{\mathrm{informed}}:=\sum_{j\ge1}\delta^{j}(1-p_H)^{j-1}(p_H\pi-c)=\frac{\delta(p_H\pi-c)}{1-\delta(1-p_H)}>0,
\]
the value of uninformed work in period $t$,
\[
\Pi_t:=P_{t-1}\pi-c+Q_{t-1}V^{\mathrm{informed}},
\]
and $S(n):=\sum_{t\le n}\mathcal D_t\Pi_t$, which is explained below. Write $\infty\mathbf 1$ for the list of informed durations with every entry infinite, and $\mathbf 0$ for the list with every entry zero. Then $S(n)=S(n,\infty\mathbf 1)$ is the surplus of the schedule with uninformed duration $n$ and infinite informed durations. To see this, group the total surplus under the obedient strategy by the period of the finding: uninformed period $t$ contributes $\mathcal D_t(P_{t-1}\pi-c)$ and, upon a finding there, the discounted surplus $V^{\mathrm{informed}}$ of the $t$-informed track, where the finding carries the weight $\mathcal D_tQ_{t-1}$. Summing over $t\le n$ gives $S(n)$.

Fix a contract and a period $k$. Let $F_k$ be the value to an informed agent of working in every period of the $k$-informed track.
Let $\mathcal W_k$ be his optimal value on that track, the supremum over his effort rules. Both values include the track's wages and severance, leave out the lump sum $\ell_k$, and are discounted to period $k$. Write $Y_k:=\ell_k+F_k$ for the informed payoff. At the PBS contract, which pays no lump sum, $F_k=Y_k$ is the informed payoff of Section~\ref{sec:main result}.

Throughout the appendices, report-free refers to a class larger than that of Section~\ref{sec:main result}, and only the proof of Proposition~\proofref{prop:report free} uses this larger class. A contract is report-free if it pays no report lump sum, every informed track pays the uninformed-track bonus and wage in every period $t\le n_0$ in which both tracks are active, and all informed tracks pay the same bonus and the same wage in every period after $n_0$. Showing Proposition~\ref{prop:report free} for this larger class thus shows a stronger result.

\section{Proofs for Sections \ref*{sec:model} and \ref*{sec:cf}}\label{app:proofs model}

\subsection{Proof of Proposition \ref*{prop:frictionless}}\label{app:proof frictionless}

\emph{Step 1: the planner's problem.} Let $n_0^{FB}:=\max\{t\ge1:\Pi_t>0\}$, with $n_0^{FB}:=0$ if $\Pi_1\le0$. To find the first-best schedule and surplus, which Steps 2 and 3 use, I show that, among all effort rules of Section~\ref{sec:model}, total surplus is maximized by working in the first $n_0^{FB}$ uninformed periods and working until success after a finding, so the maximal surplus is $S(n_0^{FB})$. First, I show that $\Pi_t$ strictly decreases, so that $\Pi_t>0$ exactly when $t\le n_0^{FB}$. Plugging the formulas for $P_{t-1}$ and $Q_{t-1}$ into $\Pi_t$ and rearranging yield
\[
\Pi_t=(p_L\pi-c)+\big[\pi(p_H-p_L)+q_H(1-p_H)V^{\mathrm{informed}}\big]\mu_{t-1},
\]
with a positive coefficient on $\mu_{t-1}$. As each silent worked period multiplies the likelihood ratio by the factor $\rho<1$, the belief $\mu_{t-1}$ strictly decreases to zero, so $\Pi_t$ strictly decreases to $p_L\pi-c$, which is negative by Assumption~\ref{ass:efficiency}. It follows that $n_0^{FB}<\infty$, $\Pi_t>0$ if and only if $t\le n_0^{FB}$, and $S(n)$ is single-peaked in $n$ with smallest maximizer $n_0^{FB}$.

Next, to show that no effort rule generates more surplus than the rule above, I take any effort rule and improve it in three moves.
First, let the agent work in every period after a finding. After a finding, the $i$-th worked period yields $(1-p_H)^{i-1}(p_H\pi-c)>0$ in expectation,
so working in every informed period
cannot lower the surplus. The $s$-th worked uninformed period then yields $\Pi_s$ whenever it occurs, because a shirked uninformed period generates nothing and leaves the belief unchanged. Thus, an effort rule that works in exactly the first $n$ uninformed periods generates at most $S(n)$. Second, drop every worked uninformed period after the $n_0^{FB}$-th, each of which yields $\Pi_s\le0$. Third, move the remaining worked uninformed periods to the beginning, leaving no shirked periods in between. This only reduces discounting of the positive surplus $\Pi_s>0$. The resulting effort rule works in the first $j\le n_0^{FB}$ uninformed periods and generates $S(j)\le S(n_0^{FB})$. Hence the effort rule I started from generates at most $S(n_0^{FB})$, which the effort rule in the previous paragraph attains.

\emph{Step 2: no moral hazard.} Suppose that effort is contractible, with (LL) retained. Let the principal offer the schedule $(n_0^{FB},\infty\mathbf 1)$ and pay $c$ in each period in which the agent is recommended to work and works, and nothing else. A worked recommended period then yields zero while any other worked period yields $-c$, so every strategy is worth at most zero to the agent. Since the obedient strategy is worth exactly zero, it is optimal and the principal collects $S(n_0^{FB})$ under it. No contract does better, since the surplus never exceeds $S(n_0^{FB})$ by Step 1, while shirking in every period guarantees the agent a nonnegative payoff under (LL).

\emph{Step 3: no limited liability.} Suppose that transfers may be negative and that the participation constraint $U_A(\phi,e^\ast,r^\ast)\ge0$ is imposed. The principal sells the project at the first-best schedule: she sets every bonus to $\pi$, charges the price $w_0(1)=-S(n_0^{FB})$, and sets every other transfer to zero. The agent then collects $\pi$ from every success and pays $S(n_0^{FB})$ up front, so his problem is the planner's problem of Step 1, restricted to the effort rules that fit this contract, minus the price $S(n_0^{FB})$. The obedient strategy follows the surplus-maximizing effort rule of Step 1,
so it solves the agent's problem and is worth zero to him. Thus participation holds with equality, and the principal collects $S(n_0^{FB})$. This is the most she can obtain because the surplus never exceeds $S(n_0^{FB})$ and participation keeps $U_A\ge0$. $\hfill\square$

\subsection{Proof of Theorem \ref*{thm:cf}}\label{app:proof cf}

The proof follows the sketch in Section~\ref{sec:proof sketch}. To ease exposition, I assume $n_0<\infty$ and show in Online Appendix~\ref{app:infinite} that $n_0=\infty$ is suboptimal. The proof has five steps. Step 1 defines the objective and the constraints of the relaxed problem and shows that they are linear functions of the transfers. Since checking (IC) requires comparing the obedient strategy with every strategy, Step 1 also writes the first-step problem \eqref{eq:first step} in a recursive form, which makes this check convenient. Step 2 fixes $n_0$ but allows $n_1$ to vary, and applies weak duality to the relaxed problem and concludes that every feasible contract pays the agent at least a constant $\kappa$. Step 3 solves an abstract work-or-shirk problem, which Step 4 uses to show the feasibility of the contractible-finding contract $\phi^{CF}(n_0)$. Step 4 also shows that $\phi^{CF}(n_0)$ attains $\kappa$ at the schedule $(n_0,\infty\mathbf 1)$, so it solves the first step, with the agent's least payoff $\kappa$. Step 5 solves the second step, which yields $n_{1k}=\infty$ and $n_0^{CF}\le n_0^{FB}$.

\emph{Step 1: the first step and the relaxed problem.} To write the objective and (IC) of the first-step problem in terms of the transfers, I use the contract form \eqref{eq:contract}, with each $k$-informed track initiated by a period-$k$ finding. Fix a schedule $(n_0,n_1)$ and a contract at it, with $F_k$, $\mathcal W_k$ and $Y_k$ as in Appendix~\ref{app:preliminaries}.
The uninformed agent's problem is the recursion, on the period $t\le n_0+1$ and the number $d\le t-1$ of silent periods he has worked before period $t$,
\begin{equation}\label{eq:app benchV}
V^{CF}(t,d)=w_0(t)+\max\big\{\delta V^{CF}(t{+}1,d),\ -c+P_d\,b_0(t)+Q_d\big(\ell_t+\mathcal W_t\big)+\delta\nu_d\,V^{CF}(t{+}1,d{+}1)\big\},
\end{equation}
where $V^{CF}(n_0{+}1,\cdot)=w_0(n_0{+}1)$ is the terminal value and $P_d$, $Q_d$ and $\nu_d$ are the probabilities of Appendix~\ref{app:preliminaries} after $d$ silent worked periods. This recursion gives the agent's best-response value.
The continuation payoff $U_t^0$ of the obedient uninformed agent from period $t$ on satisfies
\[
U^0_t=w_0(t)-c+P_{t-1}\,b_0(t)+Q_{t-1}\big(\ell_t+F_t\big)+\delta\nu_{t-1}U^0_{t+1}
\qquad\text{with }U^0_{n_0+1}=w_0(n_0{+}1).
\]
Applying this recursion repeatedly from period $1$ gives
\begin{equation}\label{eq:app U0}
U^0_1=\sum_{t\le n_0}\mathcal D_t\Big[w_0(t)-c+P_{t-1}\,b_0(t)+Q_{t-1}\big(\ell_t+F_t\big)\Big]+\delta\,\mathcal D_{n_0}\nu_{n_0-1}\,w_0(n_0{+}1).
\end{equation}
The first step minimizes $U^0_1=U_A(\phi,e^\ast,r^\ast)$ over contracts satisfying (LL) and (IC), where (IC) requires the obedient strategy to be optimal, that is, $V^{CF}(1,0)=U^0_1$.

The relaxed problem keeps (LL) and two families of constraints that every feasible contract satisfies: nonnegative informed payoffs and effort obedience. I first derive the former and then write the latter as linear constraints. The first family is $Y_k\ge0$ for all $k\le n_0$, which is implied by (IC) and (LL) because upon a finding in period $k$, the agent guarantees a payoff of at least $\ell_k\ge0$ by shirking afterward.

To write the slack of each effort obedience constraint as a linear function of the transfers, let $f_s(d):=w_0(s)-c+P_d\,b_0(s)+Q_d\,Y_s$ be the expected payoff of a period $s\le n_0$ worked at the belief $\mu_d$, so that \eqref{eq:app U0} becomes $U^0_1=\sum_{s\le n_0}\mathcal D_sf_s(s{-}1)+\mathcal D_{n_0+1}w_0(n_0{+}1)$, since $\mathcal D_{n_0+1}=\delta\nu_{n_0-1}\mathcal D_{n_0}$ by (Fact 4).
The shirker who follows $e^t_{\text{Shirk}}$ plays like the obedient agent before period $t$ and, if no finding has arrived by then, collects $w_0(t)$ in period $t$ without working and keeps the belief $\mu_{t-1}$. He then works in every later period $s\le n_0$, which he reaches uninformed with the discounted probability $\delta\mathcal D_{s-1}$ and at the belief $\mu_{s-2}$, collects $Y_s$ upon a finding there, and reaches the severance with the discounted probability $\delta\mathcal D_{n_0}$. His value falls short of $U^0_1$ by $\mathcal D_t\,G^S_t$, where the \emph{effort obedience slack} is
\begin{equation}\label{eq:eo slack}
G^S_t:=f_t(t{-}1)-w_0(t)+\frac1{\mathcal D_t}\Big[\sum_{t<s\le n_0}\big(\mathcal D_sf_s(s{-}1)-\delta\mathcal D_{s-1}f_s(s{-}2)\big)+\big(\mathcal D_{n_0+1}-\delta\mathcal D_{n_0}\big)w_0(n_0{+}1)\Big],
\end{equation}
so effort obedience holds if and only if $G^S_t\ge0$ for all $t\le n_0$.

The \emph{relaxed problem} minimizes $U^0_1$ over the contracts at the schedule $(n_0,n_1)$ subject to (LL), $G^S_t\ge0$ for all $t\le n_0$, and $Y_k\ge0$ for all $k\le n_0$.
The relaxed problem is a linear program in the transfers: $U^0_1$ and every slack $G^S_t$ are linear in the uninformed transfers and the informed payoffs, and each informed payoff $Y_k=\ell_k+F_k$ is linear in the lump sum and the transfers of the $k$-informed track.

\emph{Step 2: the weak-duality bound.} To bound the agent's least payoff in the relaxed problem from below, I apply weak duality. I assign a nonnegative multiplier to each constraint of the relaxed problem other than (LL), subtract the weighted slacks from $U^0_1$, and choose the multipliers so that every transfer cancels except the uninformed wages. Each uninformed wage must then have a positive coefficient in the subtraction. The subtraction is well defined, since every $Y_t$ is finite due to bounded transfers.

I first construct the multiplier $\sigma^S_t$ for each effort obedience slack $G_t^S$, choosing it so that every uninformed bonus cancels. Let
\[
\sigma^S_t:=\mathcal D_t+\delta K_{t-1}\Big(\frac1{P_{t-1}}-\frac{\nu_{t-2}}{P_{t-2}}\Big)\quad(t\ge1),
\qquad\text{where }K_t:=\sum_{j\le t}\delta^{t-j}\mathcal D_jP_{j-1}
\]
and $K_0:=0$, so that $\sigma^S_1=\mathcal D_1=1$. Every term with $P_{-1}$, $Q_{-1}$ or $\nu_{-1}$, here and in Theorem~\proofref{thm:main result}'s proof, has a zero factor and vanishes.
These multipliers do not depend on the duration, and $\sigma^S_t\ge\mathcal D_t>0$, since $P_{t-1}<P_{t-2}$ and $\nu_{t-2}<1$ (Fact 1). Note that the constraints $Y_t\ge0$ contain no uninformed bonus, so their multipliers do not affect the claim below. I choose them at the end of this step, where they make the informed payoffs cancel.

I claim that every uninformed bonus has the coefficient zero in the subtraction $U^0_1-\sum_{t\le n_0}\sigma^S_tG^S_t$. The bonus $b_0(s)$ enters $U^0_1$ through the payoff $f_s(s{-}1)$, with the coefficient $\mathcal D_sP_{s-1}$. It enters $G^S_s$ with the coefficient $P_{s-1}$ and, for $t<s$, it enters $G^S_t$ through the obedient agent's and the shirker's period-$s$ payoffs, with the coefficient $(\mathcal D_sP_{s-1}-\delta\mathcal D_{s-1}P_{s-2})/\mathcal D_t$. Its coefficient in $U^0_1-\sum_t\sigma^S_tG^S_t$ is therefore
\[
\mathcal D_sP_{s-1}\Big(1-\sum_{t\le s}\frac{\sigma^S_t}{\mathcal D_t}\Big)+\delta\mathcal D_{s-1}P_{s-2}\sum_{t<s}\frac{\sigma^S_t}{\mathcal D_t}.
\]
To evaluate the sums, note that $\delta\nu_{t-2}/\mathcal D_t=1/\mathcal D_{t-1}$ by (Fact 4) and $K_t=\mathcal D_tP_{t-1}+\delta K_{t-1}$, so that
\[
\frac{\sigma^S_t}{\mathcal D_t}=1+\frac{\delta K_{t-1}}{\mathcal D_tP_{t-1}}-\frac{\delta\nu_{t-2}K_{t-1}}{\mathcal D_tP_{t-2}}=1+\frac{\delta K_{t-1}}{\mathcal D_tP_{t-1}}-\frac{K_{t-1}}{\mathcal D_{t-1}P_{t-2}}=\frac{K_t}{\mathcal D_tP_{t-1}}-\frac{K_{t-1}}{\mathcal D_{t-1}P_{t-2}} .
\]
Summing over $t\le s$, consecutive terms cancel and $\sum_{t\le s}\sigma^S_t/\mathcal D_t=K_s/(\mathcal D_sP_{s-1})$. Likewise, $\sum_{t<s}\sigma^S_t/\mathcal D_t=K_{s-1}/(\mathcal D_{s-1}P_{s-2})$. The coefficient of $b_0(s)$ therefore equals $\mathcal D_sP_{s-1}-K_s+\delta K_{s-1}$, which is zero by the definition of $K_s$.

Next, I compute the coefficients of the wages $w$ and the informed payoffs $Y$, and the constant terms, in $U^0_1-\sum_{t\le n_0}\sigma^S_tG^S_t$. In $U^0_1$ and in every slack, each term of a period $s\le n_0$ belongs to one of two payoffs of period $s$, the obedient agent's payoff $f_s(s{-}1)$ and the payoff $f_s(s{-}2)$ of the shirkers of the periods $t<s$, except the term $-w_0(s)$ of $G^S_s$.

So I first compute the coefficients of these two payoffs in $U^0_1-\sum_{t\le n_0}\sigma^S_tG^S_t$. To do so, write the terms of period $s$ in the subtraction as
\[
\eta_s\,f_s(s{-}2)-\beta_s\,f_s(s{-}1)+\sigma^S_s\,w_0(s),
\]
so that $\eta_s$ is the coefficient of the shirkers' period-$s$ payoff $f_s(s{-}2)$, $-\beta_s$ is the coefficient of the obedient agent's period-$s$ payoff $f_s(s{-}1)$, and the last term comes from $-w_0(s)$ in $G^S_s$. The payoff $f_s(s{-}1)$ has the coefficient $\mathcal D_s$ in $U^0_1$, $1$ in $G^S_s$ and $\mathcal D_s/\mathcal D_t$ in $G^S_t$ for $t<s$, and the payoff $f_s(s{-}2)$ has the coefficient $-\delta\mathcal D_{s-1}/\mathcal D_t$ in $G^S_t$ for $t<s$. Therefore, $\beta_s=\mathcal D_s\sum_{t\le s}\sigma^S_t/\mathcal D_t-\mathcal D_s$ and $\eta_s=\delta\mathcal D_{s-1}\sum_{t<s}\sigma^S_t/\mathcal D_t$. These coefficients do not depend on the duration and satisfy
\begin{equation}\label{eq:app cumul}
\sum_{t\le s}\frac{\sigma^S_t}{\mathcal D_t}=\frac{\mathcal D_s+\beta_s}{\mathcal D_s}=\frac{\eta_{s+1}}{\delta\mathcal D_s}\qquad(s\ge1).
\end{equation}
By the two sums computed in proving the claim above, we have $\beta_s=\delta K_{s-1}/P_{s-1}$ and $\eta_s=\delta K_{s-1}/P_{s-2}$, so $\beta_1=\eta_1=0$, $\beta_s\ge0$ and $\beta_sP_{s-1}=\eta_sP_{s-2}$. This last relation restates the claim, since the coefficient of $b_0(s)$ displayed there is $\eta_sP_{s-2}-\beta_sP_{s-1}$. By \eqref{eq:app cumul} at $s-1$, also $\eta_s=\delta(\mathcal D_{s-1}+\beta_{s-1})\ge\delta\mathcal D_{s-1}>0$ for $s\ge2$.

The coefficients of the informed payoff and the wage of period $s\le n_0$, and the constant term, follow from $\eta_s$, $-\beta_s$ and the formula for $f_s(d)$. The informed payoff $Y_s$ has the coefficient $Q_{s-2}$ in the shirkers' payoff and $Q_{s-1}$ in the obedient agent's, so its coefficient in $U^0_1-\sum_t\sigma^S_tG^S_t$ is
\[
p^Y_s:=\eta_sQ_{s-2}-\beta_sQ_{s-1}=\eta_sP_{s-2}\Big[\frac{Q_{s-2}}{P_{s-2}}-\frac{Q_{s-1}}{P_{s-1}}\Big],
\]
with $p^Y_1=0$ and $p^Y_s>0$ for $2\le s\le n_0$, since $\eta_s>0$ and the ratio $Q_t/P_t$ strictly decreases in $t$ (Fact 3). The wage $w_0(s)$ has the coefficient $1$ in both payoffs, so its coefficient is $\eta_s-\beta_s+\sigma^S_s$, which equals $p^w_s:=\mathcal D_s+(1-\nu_{s-2})\,\eta_s$, since $\sigma^S_s=\mathcal D_s+\beta_s-\nu_{s-2}\eta_s$ by the formulas for $\beta_s$ and $\eta_s$. The severance $w_0(n_0{+}1)$ has the coefficient $\mathcal D_{n_0+1}$ in $U^0_1$ and $(\mathcal D_{n_0+1}-\delta\mathcal D_{n_0})/\mathcal D_t$ in every $G^S_t$, so its coefficient is $p^w_{n_0+1}$, by \eqref{eq:app cumul} at $s=n_0$ and $\mathcal D_{n_0+1}=\delta\nu_{n_0-1}\mathcal D_{n_0}$ (Fact 4). Thus $p^w_1=1$ and, by $\eta_s=\delta(\mathcal D_{s-1}+\beta_{s-1})$ and (Fact 4),
\[
p^w_s=\delta\big[\mathcal D_{s-1}+(1-\nu_{s-2})\beta_{s-1}\big]>0\qquad(2\le s\le n_0+1).
\]
Lastly, both payoffs contain the cost $-c$, so the constant terms of each period $s\le n_0$ sum to $c\,(\beta_s-\eta_s)$. Hence, for every contract at the schedule $(n_0,n_1)$,
\begin{equation}\label{eq:app cf identity}
U^0_1-\sum_{t\le n_0}\sigma^S_t\,G^S_t=\kappa+\sum_{t=1}^{n_0+1}p^w_t\,w_0(t)+\sum_{t=2}^{n_0}p^Y_t\,Y_t,
\qquad \kappa:=c\sum_{t\le n_0}\big(\beta_t-\eta_t\big).
\end{equation}

I now check the conditions of weak duality and draw its conclusion. Assign the multiplier $p^Y_t\ge0$ to the constraint $Y_t\ge0$ of the relaxed problem. Then \eqref{eq:app cf identity} says $U^0_1-\sum_t\sigma^S_tG^S_t-\sum_tp^Y_tY_t=\kappa+\sum_tp^w_tw_0(t)$, so every transfer cancels except the uninformed wages.
Three conditions make the constant $\kappa$ a lower bound on $U^0_1$ at every feasible contract of the relaxed problem. First, every multiplier is nonnegative: $\sigma^S_t\ge\mathcal D_t>0$ and $p^Y_t\ge0$, as shown above. Second, every weighted slack is nonnegative there, since $G^S_t\ge0$ and $Y_t\ge0$ are constraints of the relaxed problem. Third, every uninformed wage keeps the coefficient $p^w_t>0$ and is nonnegative by (LL).
Hence, by weak duality, the agent's least payoff in the relaxed problem is at least $\kappa$: every feasible contract of the relaxed problem satisfies
\begin{equation}\label{eq:app cf bound}
U^0_1=\kappa+\sum_{t\le n_0}\sigma^S_t\,G^S_t+\sum_{t=1}^{n_0+1}p^w_t\,w_0(t)+\sum_{t=2}^{n_0}p^Y_t\,Y_t\;\ge\;\kappa .
\end{equation}
This value $\kappa$ therefore also bounds the agent's payoff in the first-step problem \eqref{eq:first step} from below. It does not depend on the informed durations, so the same bound holds at every $n_1$ when $n_0$ is fixed.

\emph{Step 3: a work-or-shirk problem.} To show in Step 4 that working in every uninformed period is optimal at $\phi^{CF}(n_0)$, and hence that $\phi^{CF}(n_0)$ satisfies (IC), I study \eqref{eq:app benchV} in an abstract form. For arbitrary sequences $b=(b(t))_{t\le n_0}$ and $y=(y_t)_{t\le n_0}$, where $y$ stands in for the values $\ell_t+\mathcal W_t$ of \eqref{eq:app benchV}, let $\mathrm{DP}(b,y)$ denote the problem
\begin{equation}\label{eq:app dp}
V(t,d)=\max\big\{\delta V(t{+}1,d),\ -c+P_d\,b(t)+Q_d\,y_t+\delta\nu_d\,V(t{+}1,d{+}1)\big\},
\qquad V(n_0{+}1,\cdot)\equiv0,
\end{equation}
on the same pairs $(t,d)$. Next, I define the value at the pair $(t,d)$ of working in every remaining period, denoted $A(t,\mu_d)$. To do so, first define, backward from $A_H(n_0{+}1)=A_L(n_0{+}1)=0$,
\[
A_H(t):=-c+p_H\,b(t)+q_H(1-p_H)y_t+\delta\gamma' A_H(t{+}1),
\qquad
A_L(t):=-c+p_L\,b(t)+\delta(1-p_L)A_L(t{+}1),
\]
set $A(t,\mu):=\mu A_H(t)+(1-\mu)A_L(t)$, and let $\chi_t:=A_H(t)-\delta A_H(t{+}1)$ and $\psi_t:=A_L(t)-\delta A_L(t{+}1)$. Expanding the definitions and applying \eqref{eq:app bayes} yield the recursion
\begin{equation}\label{eq:app onestep}
A(t,\mu_d)=-c+P_d\,b(t)+Q_d\,y_t+\delta\nu_d\,A(t{+}1,\mu_{d+1}).
\end{equation}
At a contract without wages and with $A$ evaluated at the contract's terms $(b,y)=(b_0,Y)$, this value at the on-path pair $(t,t{-}1)$ is the obedient agent's continuation payoff, namely $U^0_t=A(t,\mu_{t-1})$.

I claim that if $\chi_t\ge0$ for $t=2,\dots,n_0$ and
\begin{equation}\label{eq:app bind}
A(t,\mu_{t-1})=\delta A(t{+}1,\mu_{t-1})
\qquad\text{for every }t\le n_0,
\end{equation}
then in \eqref{eq:app dp} the work branch attains the maximum at every pair, so that $V(t,d)=A(t,\mu_d)$ everywhere. To see this, I argue by backward induction on $t$, starting from $V(n_0+1,\cdot)=0=A(n_0+1,\cdot)$, with the hypothesis that $V(t+1,d)=A(t+1,\mu_d)$ for every $d\le t$. At any pair $(t,d)$ the shirk branch equals $\delta A(t{+}1,\mu_d)$ and, by \eqref{eq:app onestep}, the work branch equals $A(t,\mu_d)$. Since $\mu_d=\theta_d/(1+\theta_d)$, the work-minus-shirk gap is
\[
M(t,d):=A(t,\mu_d)-\delta A(t{+}1,\mu_d)=\mu_d\chi_t+(1-\mu_d)\psi_t=\frac{\theta_d\chi_t+\psi_t}{1+\theta_d}.
\]
Condition \eqref{eq:app bind} says $M(t,t{-}1)=0$, that is, $\psi_t=-\theta_{t-1}\chi_t$, so $(1+\theta_d)M(t,d)=(\theta_d-\theta_{t-1})\chi_t$ for every $d\le t-1$. We then have $M(t,d)\ge0$ because $\theta_d\ge\theta_{t-1}$ and $\chi_t\ge0$ (at $t=1$ and $d=0$, $\theta_d=\theta_{t-1}$). Work therefore attains the maximum and $V(t,d)=A(t,\mu_d)$, which closes the induction.

\emph{Step 4: the contract $\phi^{CF}(n_0)$ attains the bound and is feasible.} I first show that $\phi^{CF}(n_0)$ is feasible and then that it attains the bound $\kappa$ of Step 2. At the schedule $(n_0,\infty\mathbf 1)$, the contract $\phi^{CF}(n_0)$ sets $w_0\equiv0$, $w_{1k}\equiv0$, $\ell\equiv0$, $b_{1k}(t)=\underline b$ in every informed-track period, and $b_0=b_0^{CF}$. I check (LL), then (IC) on the informed tracks, and finally (IC) on the uninformed track, through the claim of Step 3. First, backward induction on \eqref{eq:cf bonus} gives $b_0^{CF}(t)-c/P_{t-1}=\delta\,[\,b_0^{CF}(t{+}1)-c/P_{t-1}\,]\ge\delta\,c\,[1/P_t-1/P_{t-1}]\ge0$ for $t<n_0$, by (Fact 1). Hence $b_0^{CF}(t)\ge c/P_{t-1}$ for all $t\le n_0$, so $b_0^{CF}$ is positive and (LL) holds. Second, an informed track paying $\underline b$ in every period is worth zero to an informed agent under every effort rule, since each worked period yields $p_H\underline b-c=0$ and the track pays no other transfer. Thus $F_k=\mathcal W_k=0$ for each $k$, so working in every period of each informed track is optimal for the informed agent, as (IC) requires. The uninformed agent's problem at $\phi^{CF}(n_0)$ is therefore $\mathrm{DP}(b_0^{CF},0)$.

I verify the two conditions for the claim of Step 3 at $(b_0^{CF},0)$. One can show that $\chi_{n_0}=p_Hb_0^{CF}(n_0)-c=c\,[p_H-P_{n_0-1}]/P_{n_0-1}>0$ and $\chi_t=p_H\big[b_0^{CF}(t)-\delta b_0^{CF}(t{+}1)\big]-(1-\delta)c+\delta\gamma'\chi_{t+1}$. Rearranging \eqref{eq:cf bonus} gives $b_0^{CF}(t)-\delta b_0^{CF}(t{+}1)=(1-\delta)c/P_{t-1}$, so the first two terms of $\chi_t$ sum to $(1-\delta)c\,[p_H-P_{t-1}]/P_{t-1}\ge0$. Backward induction then yields $\chi_t\ge(\delta\gamma')^{n_0-t}\chi_{n_0}>0$ for every $t$. For \eqref{eq:app bind}, which by Step 3 holds at period $t$ if and only if $\theta_{t-1}\chi_t+\psi_t=0$, note that $\psi_{n_0}=p_Lb_0^{CF}(n_0)-c$ and $\psi_t=(1-\delta)c\,(p_L-P_{t-1})/P_{t-1}+\delta(1-p_L)\psi_{t+1}$. In the last period, since $(1+\theta)P(\mu)=\theta p_H+p_L$ by (Fact 2),
\[
\theta_{n_0-1}\chi_{n_0}+\psi_{n_0}=(\theta_{n_0-1}p_H+p_L)\,b_0^{CF}(n_0)-(1+\theta_{n_0-1})\,c=(1+\theta_{n_0-1})\big[P_{n_0-1}b_0^{CF}(n_0)-c\big]=0 .
\]
Assuming $\theta_t\chi_{t+1}+\psi_{t+1}=0$ for induction, the recursions give
\[
\theta_{t-1}\chi_t+\psi_t=\frac{(1-\delta)c}{P_{t-1}}\big[\theta_{t-1}p_H+p_L-(1+\theta_{t-1})P_{t-1}\big]+\delta\big[\theta_{t-1}\gamma'\chi_{t+1}+(1-p_L)\psi_{t+1}\big],
\]
where the first bracket is zero by the same identity (Fact 2) and the second equals $(1-p_L)[\theta_t\chi_{t+1}+\psi_{t+1}]=0$ by the likelihood-ratio form of \eqref{eq:app bayes}. By induction, \eqref{eq:app bind} holds in every period, so work is optimal at every pair and $V^{CF}(1,0)=U^0_1$. Hence $\phi^{CF}(n_0)$ is feasible, and the agent's payoff under it is
\[
R^{CF}(n_0):=U^0_1(\phi^{CF}(n_0))=\sum_{t\le n_0}\mathcal D_t\big(P_{t-1}b_0^{CF}(t)-c\big).
\]

It remains to show that $\phi^{CF}(n_0)$ attains the bound $\kappa$ of Step 2. The wages and the informed payoffs are zero at this contract. Also, every effort obedience slack is zero: at a contract without wages, $G^S_t$ is the obedient agent's continuation payoff $A(t,\mu_{t-1})$ from period $t$ on minus the shirker's, $\delta A(t{+}1,\mu_{t-1})$.
So \eqref{eq:app bind} says exactly that every effort obedience constraint holds with equality, and it holds at $(b_0^{CF},0)$, as verified above. The identity \eqref{eq:app cf identity} then implies $\kappa=U^0_1(\phi^{CF}(n_0))=R^{CF}(n_0)$.

Consequently, at the schedule $(n_0,\infty\mathbf 1)$, $\phi^{CF}(n_0)$ solves the relaxed problem and also the first-step problem due to its feasibility there. By Step 2, every feasible contract at a schedule with uninformed duration $n_0$ pays the agent at least $R^{CF}(n_0)$.

As a by-product that the main text cites, $b_0^{CF}(t)>c/P_{t-1}$ in every period $t<n_0$, since $\delta\,c\,[1/P_t-1/P_{t-1}]>0$. Moreover, $b_0^{CF}(t)>\underline b$ in every period, since $c/P_{t-1}>c/p_H=\underline b$ for all $t\le n_0$.
Section~\ref{sec:cf} states both bounds $c/P_{t-1}$ and $\underline b$. Section~\ref{sec:cf} and Section~\ref{sec:comparative statics} use the second bound $b_0^{CF}(t)>\underline b$.

As a second by-product that the main text cites, every multiplier $\sigma^S_t$ is strictly positive, whereas weak duality needs only nonnegative multipliers. By \eqref{eq:app cf identity}, every contract of the relaxed problem that attains the bound has $\sum_t\sigma^S_tG^S_t=0$ with every term nonnegative, so strict positivity forces every effort obedience constraint to bind there. The introduction cites this when it says that effort obedience binds in every period before a finding.

\emph{Step 5: the second step.} To show that the principal optimally offers $\phi^{CF}(n_0^{CF})$, with $n_0^{CF}\le n_0^{FB}$, I first prove that every optimal schedule sets $n_{1k}=\infty$ for each $k$, so that the second step reduces to a choice of the uninformed duration. Fix a schedule with $n_0<\infty$ and $n_{1k}<\infty$ for some $k\le n_0$. Its surplus is strictly below $S(n_0)$, since each dropped informed period is reached with positive probability and yields $p_H\pi-c>0$. Every feasible contract at this schedule pays the agent at least $R^{CF}(n_0)$ by \eqref{eq:app cf bound}, which holds for all $n_1$, and Step 4, which identifies $\kappa=R^{CF}(n_0)$. The principal's payoff of every such contract is therefore strictly below $S(n_0)-R^{CF}(n_0)$, which $\phi^{CF}(n_0)$ attains at the schedule $(n_0,\infty\mathbf 1)$ by Step 4. The second step thus reduces to
\[
\max_{n\in\{0,1,2,\dots\}}\;\big[\,S(n)-R^{CF}(n)\,\big],
\]
where $S(0)=0$ and $R^{CF}(0):=0$ is the agent's least payoff at $n_0=0$.

Second, I show that $R^{CF}$ is nondecreasing, so that the maximum is attained at a duration no longer than $n_0^{FB}$. Since the multipliers of Step 2 do not depend on the duration, $R^{CF}(n)$, which is the constant $\kappa$ of Step 2 at $n_0=n$ by Step 4, is the sum of the first $n$ terms of the fixed sequence $c(\beta_t-\eta_t)$. With $R^{CF}(0)=0$, the agent's marginal payoff is therefore
\begin{equation}\label{eq:app rcf}
\Delta R^{CF}(n):=R^{CF}(n)-R^{CF}(n{-}1)=c\,\big(\beta_n-\eta_n\big)=c\,\eta_n\,\frac{P_{n-2}-P_{n-1}}{P_{n-1}}\ \ge\ 0
\qquad(n>0),
\end{equation}
by $\beta_nP_{n-1}=\eta_nP_{n-2}$ and the monotonicity of $P_t$ (Fact 1). Now, for any $n>n_0^{FB}$,
\[
\big[S(n)-R^{CF}(n)\big]-\big[S(n_0^{FB})-R^{CF}(n_0^{FB})\big]=\underbrace{S(n)-S(n_0^{FB})}_{\le0}-\underbrace{\big[R^{CF}(n)-R^{CF}(n_0^{FB})\big]}_{\ge0}\;\le\;0,
\]
by the single-peakedness of $S$ from Step 1 of Appendix~\ref{app:proof frictionless}. The maximum is therefore attained at some $n\le n_0^{FB}$. The smallest such $n$ is therefore $n_0^{CF}$. Hence the principal optimally offers $\phi^{CF}(n_0^{CF})$.

As a by-product that the main text cites, the agent's payoff $R^{CF}(n)$ strictly increases in $n$ from $n=1$ on, since \eqref{eq:app rcf} gives $\Delta R^{CF}(1)=0$ and $\Delta R^{CF}(n)>0$ for $n>1$, as $\eta_n>0$. Section~\ref{sec:cf} cites this increase to explain why $n_0^{CF}$ can fall short of $n_0^{FB}$. $\hfill\square$

\section{Proofs for Section \ref*{sec:pf}}\label{app:proofs pf}


\subsection{Proof of Theorem \ref*{thm:main result}}\label{app:proof main result}

The proof follows the sketch in Section~\ref{sec:proof sketch} and shares the structure of the proof of Theorem~\proofref{thm:cf}. I assume $n_0<\infty$ and show in Online Appendix~\ref{app:infinite} that $n_0=\infty$ is suboptimal. The proof has five steps. Step 1 defines the objective and the constraints of the relaxed problem and shows that they are linear functions of the transfers. Unlike in the proof of Theorem~\proofref{thm:cf}, no-concealing now takes the place of nonnegative informed payoffs. Step 1 also reformulates the first-step problem \eqref{eq:first step} in a recursive form. Step 2 fixes $n_0$ but allows $n_1$ to vary, and applies weak duality to the relaxed problem and concludes that every feasible contract pays the agent at least a constant $\kappa$. Step 3 shows that the PBS contract $\phi^\ast(n_0)$ is feasible and that it attains $\kappa$ at the schedule $(n_0,\infty\mathbf 1)$, so it solves the first step, with the agent's least payoff $\kappa$. Step 4 proves parts (i) and (ii). Step 5 solves the second step and compares the uninformed duration with $n_0^{CF}$.

The proof sketch in Section~\ref{sec:proof sketch} describes the two lemmas below, which are used in Steps 2 to 4.

\begin{lemma}[Shirker lemma]\label{lem:shirker}
For all beliefs $\mu>\mu'$,
\[
P(\mu)Q(\mu')-P(\mu')Q(\mu)=-p_Lq_H(1-p_H)(\mu-\mu')<0 .
\]
Hence, the ratio $Q(\mu)/P(\mu)$ is strictly increasing in $\mu$.
\end{lemma}

\emph{Proof}. The identity is (Fact 3), and its right side is negative since $p_L>0$, $q_H>0$ and $p_H<1$. $\hfill\square$

\begin{lemma}[Concealer lemma]\label{lem:concealer}
For every belief $\mu>0$,
\[
(1-p_H)P(\mu)-p_HQ(\mu)=(1-p_H)\big[p_L(1-\mu)+p_H\mu(1-q_H)\big]>0.
\]
Hence, $(1-p_H)/p_H>Q(\mu)/P(\mu)$.
\end{lemma}

\emph{Proof}. Substituting the two affine forms gives $(1-p_H)P(\mu)-p_HQ(\mu)=(1-p_H)\big[p_L+(p_H-p_L)\mu-p_Hq_H\mu\big]$, which rearranges to the right side. The bracket is positive since $p_H\mu(1-q_H)>0$. $\hfill\square$

\bigskip

One can show that the shirker lemma holds if and only if $q_L/q_H<p_L(1-p_H)/[p_H(1-p_L)]$, which is satisfied here because $q_L=0$. The concealer lemma, however, needs only interior $p_H$ and $q_H$.

\emph{Step 1: the first step and the relaxed problem.} To write the objective and (IC) of the first-step problem in terms of the transfers, I use the contract form \eqref{eq:contract}. Fix a schedule $(n_0,n_1)$ with $n_0>0$ and a contract at it, with $F_k$, $\mathcal W_k$ and $Y_k$ as in Appendix~\ref{app:preliminaries}. To verify (IC) for the PBS contract in Step 3 and to record what (IC) forces at every contract, I first write the agent's best-response values as recursions. An informed agent's values before sending a message are, for $t\le n_0$,
\[
X_t:=\max\{\ell_t+\mathcal W_t,\ \delta V^1(t{+}1)\}\quad(t<n_0),\qquad
X_{n_0}:=\max\{\ell_{n_0}+\mathcal W_{n_0},\ \delta w_0(n_0{+}1)\},
\]
\[
V^1(t):=w_0(t)+\max\big\{X_t,\ -c+p_Hb_0(t)+(1-p_H)X_t\big\}.
\]
Here, $X_t$ is the value of an informed agent who
chooses between reporting now and delaying, and $V^1(t)$ is his value in period $t$ while unreported. The uninformed agent's best-response value $V^0(t,d)$ depends on the period $t\le n_0+1$ and the number $d\le t-1$ of silent periods he has worked before period $t$:
\begin{equation}\label{eq:app V0}
\begin{aligned}
V^0(t,d)=w_0(t)+\max\Big\{&\max\{Z_t(\mu_d),\ \delta V^0(t{+}1,d)\},\\
&-c+P_db_0(t)+Q_dX_t+\nu_d\max\{Z_t(\mu_{d+1}),\ \delta V^0(t{+}1,d{+}1)\}\Big\},
\end{aligned}
\end{equation}
with the terminal value $V^0(n_0{+}1,\cdot):=w_0(n_0{+}1)$. Here, $\mu_d$, $P_d$, $Q_d$ and $\nu_d$ are the belief and the probabilities of Appendix~\ref{app:preliminaries} after $d$ silent worked periods. The faking value $Z_t(\mu)$ equals $\ell_t$ plus the optimal value of an agent with belief $\mu$ on the $t$-informed track, the supremum over his effort rules there, discounted to period $t$. The outer maximum compares shirking with working, and the inner maxima compare faking with continuing when the agent sends his message in a period that brings neither a success nor a finding. The continuation payoff $U_t^0$ of the obedient uninformed agent from period $t$ on is
\begin{equation}\label{eq:app U0star}
U^0_t=-c+w_0(t)+P_{t-1}b_0(t)+Q_{t-1}Y_t+\delta\nu_{t-1}U^0_{t+1},
\qquad U^0_{n_0+1}:=w_0(n_0{+}1).
\end{equation}
Conditional on the agent's private history, $X_t$, $V^1(t)$ and $V^0(t,d)$ are his best-response values, and $U^0_t$ is his value under the obedient strategy.
In particular, $U^0_1=U_A(\phi,e^\ast,r^\ast)$ and $V^0(1,0)=\sup_{(e,r)}U_A(\phi,e,r)$. The first-step problem therefore minimizes $U^0_1$ over contracts satisfying (LL) and (IC), where (IC) requires the obedient strategy to be optimal, that is, $V^0(1,0)=U^0_1$.

Feasibility then makes working in every period of an informed track optimal for the informed agent, which the proof of Proposition~\proofref{prop:report free} uses: for each $t$-informed track reached on the path, (IC) implies
\begin{equation}\label{eq:app forced}
\mathcal W_t=F_t.
\end{equation}

The relaxed problem keeps (LL) and two families of constraints implied by (IC): effort obedience and no-concealing. I borrow effort obedience from Step 1 of Theorem~\proofref{thm:cf}'s proof and write no-concealing as linear constraints. With the payoff $f_s(d)$ of that step, \eqref{eq:app U0star} gives $U^0_1=\sum_{s\le n_0}\mathcal D_sf_s(s{-}1)+\mathcal D_{n_0+1}w_0(n_0{+}1)$ as there. Thus, the same formula \eqref{eq:eo slack} for the effort obedience slack $G_t^S$ applies here.

The concealer who follows $r^m_{\text{Conceal}}$ works in every recommended period and reports a finding in its arrival period, except a finding in period $m$, which has the discounted probability $\mathcal D_mQ_{m-1}>0$. Rather than telling the truth and collecting $Y_m$, he stays on the uninformed track in period $m{+}1$, where he works, succeeds with probability $p_H$ because he knows that $\omega=H$, and absent a success, reports and collects $Y_{m+1}$. When $m=n_0$, he collects the severance instead. Let $g_s:=w_0(s)-c+p_Hb_0(s)+(1-p_H)Y_s$ be the expected payoff of a period $s\le n_0$ to an agent who holds an unreported finding, works, and reports absent a success, as this concealer does in period $m{+}1$. His payoff falls short of $U^0_1$ by $\mathcal D_mQ_{m-1}\,G^C_m$, where the \emph{no-concealing slack} is
\[
G^C_m:=Y_m-\delta\,g_{m+1}\quad(m<n_0),\qquad G^C_{n_0}:=Y_{n_0}-\delta w_0(n_0{+}1),
\]
so no-concealing holds if and only if $G^C_m\ge0$ for all $m\le n_0$.

The \emph{relaxed problem} minimizes $U^0_1$ over the contracts at the schedule $(n_0,n_1)$ subject to (LL), $G^S_t\ge0$ for all $t\le n_0$, and $G^C_m\ge0$ for all $m\le n_0$. As in Step 1 of Theorem~\proofref{thm:cf}'s proof, the relaxed problem is a linear program in the transfers, since every no-concealing slack $G^C_m$, like $U^0_1$ and every $G^S_t$, is linear in the uninformed transfers and the informed payoffs.

\emph{Step 2: the weak-duality bound.} I next apply weak duality to bound the agent's least payoff in the relaxed problem from below. I assign a nonnegative multiplier to each effort obedience and no-concealing constraint, subtract the weighted slacks from $U^0_1$, and choose the multipliers so that every transfer cancels except the uninformed wages, each of which keeps a positive coefficient in the subtraction. The subtraction is well defined, since every $Y_t$ is finite due to bounded transfers.

\emph{The multipliers.} With a slight abuse of notation, I reuse the symbols of Step 2 of Theorem~\proofref{thm:cf}'s proof: $\sigma^S_t$ is the multiplier on the period-$t$ effort obedience, and $\eta_t$ and $-\beta_t$ are the coefficients of the shirkers' and the obedient agent's period-$t$ payoffs in the subtraction. Their roles remain the same, but their formulas differ because the subtraction now also contains the weighted no-concealing slacks. Two symbols are new: the multiplier $\sigma^C_t$ on no-concealing in period $t$, and $v_t:=\delta\sigma^C_{t-1}$. I define the multipliers by forward recursion
\[
\begin{gathered}
\beta_1:=0,\qquad \eta_1:=0,\qquad v_1:=0,\qquad \sigma^S_1:=1,\qquad \sigma^C_1:=0;\\
\eta_t:=\delta\big(\beta_{t-1}+\mathcal D_{t-1}\big),\qquad
v_t:=\delta\,\sigma^C_{t-1},\qquad
\beta_t:=\frac{\eta_tP_{t-2}+p_Hv_t}{P_{t-1}},\qquad
\sigma^S_t:=\beta_t-\delta\nu_{t-2}\beta_{t-1},\\
\sigma^C_t:=\eta_tP_{t-2}\Big(\frac{Q_{t-2}}{P_{t-2}}-\frac{Q_{t-1}}{P_{t-1}}\Big)
+v_t\Big(1-p_H-p_H\frac{Q_{t-1}}{P_{t-1}}\Big)\qquad(t\ge2).
\end{gathered}
\]
By $\mathcal D_t=\delta\nu_{t-2}\mathcal D_{t-1}$ of (Fact 4), $\sigma^S_t/\mathcal D_t=\beta_t/\mathcal D_t-\beta_{t-1}/\mathcal D_{t-1}$ for $t\ge2$, so, with $\sigma^S_1=\mathcal D_1=1$, $\beta_1=0$ and the definition of $\eta_t$, these multipliers satisfy \eqref{eq:app cumul}.

I now show by induction on $t\ge2$ that $\eta_t\ge\delta\mathcal D_{t-1}>0$, $\beta_t>0$, $\sigma^C_t>0$ and $v_t\ge0$. Suppose that $\beta_{t-1}\ge0$ and $v_t\ge0$, as holds at $t=2$ because $\beta_1=0$ and $v_2=\delta\sigma^C_1=0$. Then $\eta_t=\delta(\beta_{t-1}+\mathcal D_{t-1})\ge\delta\mathcal D_{t-1}>0$, and $\beta_t\ge\eta_tP_{t-2}/P_{t-1}>\eta_t>0$ by (Fact 1). The first term of $\sigma^C_t$ is positive by Lemma~\ref{lem:shirker}, since $\eta_t>0$ and $\mu_{t-2}>\mu_{t-1}$, and its second term is nonnegative by Lemma~\ref{lem:concealer}, since $v_t\ge0$. Hence $\sigma^C_t>0$ and $v_{t+1}=\delta\sigma^C_t>0$, which completes the induction. Finally, by the definitions of $\sigma^S_t$ and $\eta_t$ and (Fact 4), $\sigma^S_t=\mathcal D_t+\beta_t-\nu_{t-2}\eta_t$ for $t\ge2$, which is at least $\mathcal D_t>0$ since $\beta_t>\eta_t\ge\nu_{t-2}\eta_t$. Also, $\sigma^S_1=\mathcal D_1=1$.

The coefficients of $Y_s$ and of the wages below use a second form of $\sigma^C_t$. Substituting $\beta_tP_{t-1}=\eta_tP_{t-2}+p_Hv_t$ into $\sigma^C_t$ gives, for all $t\ge1$,
\begin{equation}\label{eq:app XDalt}
\sigma^C_t=\eta_tQ_{t-2}+(1-p_H)v_t-Q_{t-1}\beta_t .
\end{equation}

\emph{The coefficients.} I then show that the multipliers make every transfer cancel except the uninformed wages. Since they satisfy \eqref{eq:app cumul}, the computation of Step 2 of Theorem~\proofref{thm:cf}'s proof applies to $U^0_1-\sum_t\sigma^S_tG^S_t$: the obedient agent's period-$s$ payoff $f_s(s{-}1)$ has the coefficient $-\beta_s$, the shirkers' period-$s$ payoff $f_s(s{-}2)$ has the coefficient $\eta_s$, and $G^S_s$ contributes $\sigma^S_sw_0(s)$. Since $G^C_m=Y_m-\delta g_{m+1}$ for $m<n_0$, subtracting also $\sum_m\sigma^C_mG^C_m$ adds $-\sigma^C_sY_s$ from $G^C_s$ and $\delta\sigma^C_{s-1}g_s=v_s\,g_s$ from $G^C_{s-1}$. The terms of period $s\le n_0$ in the new subtraction are therefore
\[
\eta_s\,f_s(s{-}2)-\beta_s\,f_s(s{-}1)+v_s\,g_s+\sigma^S_s\,w_0(s)-\sigma^C_s\,Y_s .
\]
The coefficient of $b_0(s)$ is $\eta_sP_{s-2}-\beta_sP_{s-1}+p_Hv_s=0$, by the definition of $\beta_s$, and the coefficient of $Y_s$ is $\eta_sQ_{s-2}-\beta_sQ_{s-1}+(1-p_H)v_s-\sigma^C_s=0$, by \eqref{eq:app XDalt}. The wage $w_0(s)$ has the coefficient $1$ in all three payoffs, so its coefficient is $\eta_s-\beta_s+\sigma^S_s+v_s$, which equals
\[
W_s:=\mathcal D_s+(1-\nu_{s-2})\,\eta_s+v_s,
\]
since $\sigma^S_s=\mathcal D_s+\beta_s-\nu_{s-2}\eta_s$, as shown above. As in Step 2 of the proof of Theorem~\proofref{thm:cf}, \eqref{eq:app cumul} at $s=n_0$, together with the term $v_{n_0+1}w_0(n_0{+}1)$ from $G^C_{n_0}$, gives the severance the coefficient $W_{n_0+1}$, with $\eta_{n_0+1}$ and $v_{n_0+1}$ given by the recursion. To sign this coefficient, I compute it in closed form. The coefficient $W_1$ equals $\mathcal D_1=1$, since $\eta_1=v_1=0$. For each $s\le n_0$,
\[
W_{s+1}=\delta\big[\mathcal D_s+(P_{s-1}+Q_{s-1})\beta_s+\sigma^C_s\big]
=\delta\big[\mathcal D_s+(P_{s-2}+Q_{s-2})\eta_s+v_s\big]
=\delta\,W_s ,
\]
where the first equality follows from the definitions of $\eta_{s+1}$ and $v_{s+1}$, $\mathcal D_{s+1}=\delta\nu_{s-1}\mathcal D_s$ of (Fact 4) and $1-\nu=P+Q$, and the second equality follows from \eqref{eq:app XDalt} and $\beta_sP_{s-1}=\eta_sP_{s-2}+p_Hv_s$. Hence $W_s=\delta^{s-1}$ for all $s\le n_0+1$. Lastly, the three payoffs contain the cost $-c$, so the constant terms of each period $s\le n_0$ sum to $c\,(\beta_s-\eta_s-v_s)$. Thus, for every contract at the schedule $(n_0,n_1)$,
\begin{equation}\label{eq:app identity}
U^0_1-\sum_{t\le n_0}\sigma^S_t\,G^S_t-\sum_{m\le n_0}\sigma^C_m\,G^C_m=\kappa+\sum_{s=1}^{n_0+1}\delta^{s-1}w_0(s),\qquad \kappa:=c\sum_{s\le n_0}\big(\beta_s-\eta_s-v_s\big) .
\end{equation}

\emph{Weak duality.} I now check the conditions of weak duality and draw its conclusion. Three conditions make the constant $\kappa$ a lower bound on $U^0_1$ at every feasible contract of the relaxed problem. First, the multipliers are nonnegative: $\sigma^S_t>0$ and $\sigma^C_m\ge0$. Second, every weighted slack is nonnegative, since $G^S_t\ge0$ and $G^C_m\ge0$ are constraints of the relaxed problem. Third, every uninformed wage keeps the coefficient $\delta^{s-1}>0$ and is nonnegative by (LL). Hence, by weak duality, the agent's least payoff in the relaxed problem is at least $\kappa$: every feasible contract of the relaxed problem satisfies
\begin{equation}\label{eq:app bound}
U^0_1=\kappa+\sum_{t\le n_0}\sigma^S_t\,G^S_t+\sum_{m\le n_0}\sigma^C_m\,G^C_m+\sum_{s=1}^{n_0+1}\delta^{s-1}\,w_0(s)\ \ge\ \kappa .
\end{equation}
This value $\kappa$ therefore also bounds the agent's payoff in the first-step problem \eqref{eq:first step} from below. It does not depend on the informed durations, so the same bound holds at every $n_1$ when $n_0$ is fixed.

As a by-product that the main text cites, every uninformed wage, including the severance, keeps the strictly positive coefficient $\delta^{s-1}$ in \eqref{eq:app identity}, whereas weak duality needs only a nonnegative one. Section~\ref{sec:proof sketch} states this positivity when it summarizes the proof steps, and the proof of Proposition~\proofref{prop:report free} uses it to rule out uninformed wages.

\emph{Step 3: the PBS contract attains the bound and is feasible.} I first show that the PBS contract is feasible and then that it attains the bound $\kappa$ of Step 2. Fix $n_1=\infty\mathbf 1$, and let $b_0\in\mathbb R^{n_0}_+$. The contract $\phi(b_0)$ sets $w_0\equiv0$, $\ell\equiv0$ and $w_{1k}\equiv0$, and for each $k\le n_0$ it sets $b_{1k}(t)=b_0(t)$ for $k<t\le n_0$ and $b_{1k}(t)=\underline b$ for $t>n_0$. The PBS contract $\phi^\ast(n_0)$ is $\phi(b_0^\ast)$, where, as in Section~\ref{sec:main result}, $b_0^\ast=b_0^\ast(\cdot\,;n_0)$ makes every effort obedience constraint hold with equality. Two facts give the informed payoffs of $\phi(b_0)$. First, an informed track paying $\underline b$ from some period on is worth zero from that period on to any agent. Second, the periods of the $k$-informed track from its $(j{+}1)$-th on replicate those of the $(k{+}j)$-informed track exactly, so the value of working in every period from a given period on depends only on that period, not on the track. Such period-dependent values $F_t$ of $\phi(b_0)$ therefore satisfy
\begin{equation}\label{eq:app Frec}
F_t=\delta\big[p_Hb_0(t{+}1)-c+(1-p_H)F_{t+1}\big],\qquad F_{n_0}=0 .
\end{equation}

Below, sub-step (3a) reduces the agent's best-response problem at the contract $\phi(b_0)$ to a work-or-shirk problem when the margins $b_0(t)-\underline b-F_t$ are nonnegative, (3b) shows that these margins are positive at $b_0^\ast$, and (3c) shows that working solves the work-or-shirk problem at $\phi(b_0^\ast)$, so that $\phi^\ast(n_0)$ satisfies (IC).

\emph{(3a) Report-freeness.} The contract $\phi(b_0)$ pays the bonus $b_0(t)$ in every period $t\le n_0$ on every track, so a report changes no transfer before period $n_0{+}1$. After period $n_0$, any agent is indifferent among all tracks because each yields zero payoff to him. Reports therefore do not affect what any agent can earn, so misreporting gains nothing. Moreover, if $b_0(t)\ge\underline b+F_t$ for $2\le t\le n_0$, an informed agent who shirks in a period $t\le n_0$ and works in every later period loses $p_H[b_0(t)-\underline b-F_t]\ge0$.
Since his payoff after period $n_0$ is zero whatever he does, the one-shot deviation principle applied to his single informed track then implies that working in every period is optimal for him, so a finding in period $t$ is worth $F_t$. The recursion \eqref{eq:app V0} of the uninformed agent therefore reduces to the work-or-shirk problem $\mathrm{DP}(b_0,F)$ of Step 3 of Theorem~\proofref{thm:cf}'s proof, whose value at $(1,0)$ is $V^0(1,0)=\sup_{(e,r)}U_A(\phi(b_0),e,r)$.

\emph{(3b) Positive margins.} With the function $A$ of Step 3 of Theorem~\proofref{thm:cf}'s proof evaluated at $(b_0,F)$, the slack $G^S_t$ at $\phi(b_0)$, which pays no wage and no lump sum, is $A(t,\mu_{t-1})-\delta A(t{+}1,\mu_{t-1})$. Since $b_0^\ast$ makes every such slack zero by definition, \eqref{eq:app bind} holds at $(b_0^\ast,F)$. To verify the condition of (3a) and (LL) at $\phi(b_0^\ast)$, I show that the margins are positive. Write $\xi_t(\mu):=-c+P(\mu)b_0^\ast(t)+Q(\mu)F_t$ for the payoff of working in period $t\le n_0$ at the belief $\mu$. For each $t<n_0$, combining \eqref{eq:app bind} at $t$ and $t{+}1$ with \eqref{eq:app onestep} at the pairs $(t,t{-}1)$ and $(t{+}1,t{-}1)$ yields
\begin{equation}\label{eq:app postpone}
\xi_t(\mu_{t-1})=\delta\,\xi_{t+1}(\mu_{t-1}) .
\end{equation}
At $t=n_0$, \eqref{eq:app bind} reads $P_{n_0-1}b_0^\ast(n_0)=c$, since $F_{n_0}=0$ and $A(n_0{+}1,\cdot)=0$, so the margin is $c[1/P_{n_0-1}-1/p_H]>0$, as $P_{n_0-1}<p_H$. Since $\underline b=c/p_H$, \eqref{eq:app Frec} gives $F_t-\delta F_{t+1}=\delta p_H[b_0^\ast(t{+}1)-\underline b-F_{t+1}]$. Rewriting \eqref{eq:app postpone} with this identity in terms of the margins at $t$ and $t{+}1$ and dividing by $P_{t-1}$ gives
\[
b_0^\ast(t)-\underline b-F_t=(1-\delta)\,c\Big[\frac1{P_{t-1}}-\frac1{p_H}\Big]+\delta\,\frac{(1-p_H)P_{t-1}-p_HQ_{t-1}}{P_{t-1}}\,\big[b_0^\ast(t{+}1)-\underline b-F_{t+1}\big].
\]
The intercept is nonnegative because $P_{t-1}<p_H$, and the coefficient of the margin at $t{+}1$ is positive by Lemma~\ref{lem:concealer}. Backward induction gives $b_0^\ast(t)>\underline b+F_t$ for all $t\le n_0$, and the identity for $F_t-\delta F_{t+1}$ then implies $F_t>\delta F_{t+1}$ for all $t<n_0$, so $F_t\ge0$ by backward induction from $F_{n_0}=0$. Hence every bonus $b_0^\ast(t)$ is positive, as $\underline b>0$.

\emph{(3c) Work attains the maximum everywhere.} At $(b_0^\ast,F)$, the two conditions of the claim of Step 3 of the proof of Theorem~\proofref{thm:cf} hold. Condition \eqref{eq:app bind} holds by (3b). For the other, I first show that $\delta A_H(t{+}1)=F_t$ for every $t\le n_0$.
Both sides are zero at $t=n_0$. If $\delta A_H(t{+}1)=F_t$ for some $t\le n_0$, the definition of $A_H$ at $(b_0^\ast,F)$ and $q_H(1-p_H)+\gamma'=1-p_H$ give
\[
A_H(t)=-c+p_H\,b_0^\ast(t)+q_H(1-p_H)F_t+\gamma'F_t=-c+p_H\,b_0^\ast(t)+(1-p_H)F_t ,
\]
whose product with $\delta$ is $F_{t-1}$ by \eqref{eq:app Frec} when $t\ge2$. Backward induction therefore gives $\delta A_H(t{+}1)=F_t$ for every $t\le n_0$. Hence, since $\underline b=c/p_H$, $\chi_t=A_H(t)-F_t=p_H[b_0^\ast(t)-\underline b-F_t]>0$ for $2\le t\le n_0$, by (3b). By the claim, working attains the maximum at every pair $(t,d)$ and the value of $\mathrm{DP}(b_0^\ast,F)$ at $(1,0)$ is $A(1,\mu_0)=U^0_1$. By (3a), $\phi^\ast(n_0)$ therefore satisfies (IC), and it satisfies (LL) by (3b), so the PBS contract is feasible. Let $R^\ast(n_0):=U^0_1(\phi^\ast(n_0))$ denote the agent's payoff under it.

It remains to show that $\phi^\ast(n_0)$ attains the bound $\kappa$ of Step 2. The contract has no wage. Every effort obedience slack is zero by definition, and every no-concealing slack is zero due to report-freeness.
The identity \eqref{eq:app identity} then implies $\kappa=U^0_1(\phi^\ast(n_0))=R^\ast(n_0)$. Consequently, at the schedule $(n_0,\infty\mathbf 1)$, $\phi^\ast(n_0)$ solves the relaxed problem and also the first-step problem due to its feasibility there:
\[
U_A(n_0,\infty\mathbf 1)=R^\ast(n_0)=c\sum_{s\le n_0}\big(\beta_s-\eta_s-v_s\big) .
\]
By Step 2, every feasible contract at a schedule with $n_0$ pays the agent at least $R^\ast(n_0)$.

As a by-product that the main text cites, every research-stage bonus $b_0^\ast(t)$ exceeds $\underline b$. The introduction and Sections~\ref{sec:main result} and~\ref{sec:costless} cite it to explain why a finding before period $n_0$ gives the agent a positive payoff under the PBS contract.

As a second by-product that the main text cites, binding effort obedience pins down $b_0^\ast$ uniquely, so the PBS contract is well defined. Indeed, $A(t{+}1,\mu_{t-1})$ depends only on the bonuses after period $t$, while $A(t,\mu_{t-1})$ contains $b_0(t)$ only through $P_{t-1}b_0(t)$ with $P_{t-1}>0$. Therefore, \eqref{eq:app bind} at $t$ uniquely pins down $b_0^\ast(t)$ given the later bonuses, backward from $b_0^\ast(n_0)=c/P_{n_0-1}$. The introduction and Section~\ref{sec:main result} state this uniqueness when they introduce $b_0^\ast$, and the proof of Proposition~\proofref{prop:report free} uses it to pin down the bonuses of a report-free contract.

As a third by-product that the main text cites, $\sigma^S_t>0$ for $t\ge1$ and $\sigma^C_m>0$ for $m\ge2$. By \eqref{eq:app bound}, every contract of the relaxed problem at the schedule $(n_0,\infty\mathbf 1)$ that attains the bound $\kappa$ has every weighted slack equal to zero, so strict positivity forces effort obedience to bind in every period and no-concealing to bind from period $2$ on. Step 1 of Section~\ref{sec:proof sketch} explains why no-concealing binds from period $2$ on, and the introduction and Section~\ref{sec:main result} cite the result when they speak of the binding constraints.

\emph{Step 4: parts (i) and (ii).} Since $n_0^\ast$ is determined only in Step 5, I fix an arbitrary $n_0>1$. At $n_0=1$ both parts hold trivially, since $b_0^\ast(1;1)=b_0^{CF}(1;1)=c/P_0$. Both parts rest on a recursion for the bonus increments, which I derive first. By \eqref{eq:app postpone} of (3b), for every $t<n_0$,
\begin{equation}\label{eq:app L1}
-c+P_{t-1}b_0^\ast(t)+Q_{t-1}F_t\ =\ \delta\big[-c+P_{t-1}b_0^\ast(t{+}1)+Q_{t-1}F_{t+1}\big].
\end{equation}
Rearranging and dividing by $P_{t-1}$ gives
\begin{equation}\label{eq:app L2}
b_0^\ast(t{+}1)-b_0^\ast(t)=(1-\delta)\Big[b_0^\ast(t{+}1)-\frac{c}{P_{t-1}}\Big]
+\frac{Q_{t-1}}{P_{t-1}}\,\big(F_t-\delta F_{t+1}\big),
\end{equation}
which is the relation for $b_0^\ast(t{+}1)-b_0^\ast(t)$ in Section~\ref{sec:main result}.

\emph{Proof of part (i).} The second term of \eqref{eq:app L2} is strictly positive for all $t<n_0$ by (3b). Since $P_t<P_{t-1}$ (Fact 1), the first term is nonnegative if $P_tb_0^\ast(t{+}1)\ge c$. I show that $P_{t-1}b_0^\ast(t)\ge c$ for every $t\le n_0$ by backward induction from $t=n_0$, where $P_{n_0-1}b_0^\ast(n_0)=c$ by (3b). For $t<n_0$, isolating $P_{t-1}b_0^\ast(t)$ in \eqref{eq:app L1} and substituting \eqref{eq:app Frec} for $F_t$ give
\[
P_{t-1}b_0^\ast(t)-c=\delta\Big[\frac{P_{t-1}-p_HQ_{t-1}}{P_t}\big(P_tb_0^\ast(t{+}1)-c\big)+\frac{c}{P_t}\big(P_{t-1}-P_t-Q_{t-1}(p_H-P_t)\big)+p_HQ_{t-1}F_{t+1}\Big].
\]
The first term in the bracket is nonnegative by the induction hypothesis, since $P_{t-1}-p_HQ_{t-1}>0$ by Lemma~\ref{lem:concealer}, and the last term is nonnegative because $F_{t+1}\ge0$ by (3b). The middle term is positive: with $P(\mu)=p_L+(p_H-p_L)\mu$, $Q(\mu)=q_H(1-p_H)\mu$, and $\mu_{t-1}-\mu_t=(1-\rho)\,\mu_{t-1}(1-\mu_t)$, which follows from $\theta_t=\rho\,\theta_{t-1}$ in \eqref{eq:app bayes},
\[
P_{t-1}-P_t-Q_{t-1}(p_H-P_t)=(p_H-p_L)\,\mu_{t-1}(1-\mu_t)\,\frac{(1-p_L)-(1-p_H)(1-q_Hp_L)}{1-p_L}\ >\ 0,
\]
since $1-p_H<1-p_L$ and $1-q_Hp_L\le1$. This completes the induction. Consequently, $b_0^\ast(t)<b_0^\ast(t{+}1)$ for every $t<n_0$.

\emph{Proof of part (ii).} Solving \eqref{eq:app L2} for $b_0^\ast(t)$ gives the backward recursion $b_0^\ast(t)=(1-\delta)c/P_{t-1}+\delta b_0^\ast(t{+}1)-(Q_{t-1}/P_{t-1})(F_t-\delta F_{t+1})$, which differs from the contractible-finding recursion \eqref{eq:cf bonus} only by its last term. Let $\Delta b(t):=b_0^{CF}(t;n_0)-b_0^\ast(t;n_0)$. Subtracting the two recursions shows that this difference obeys
\begin{equation}\label{eq:app premium}
\Delta b(t)=\delta\,\Delta b(t{+}1)+\frac{Q_{t-1}}{P_{t-1}}\,\big(F_t-\delta F_{t+1}\big),
\qquad \Delta b(n_0)=0,
\end{equation}
where the terminal condition holds because both sequences end with $c/P_{n_0-1}$. Since $F_t>\delta F_{t+1}$ by (3b), backward induction gives $\Delta b(t)>0$ for all $t<n_0$, which together with $\Delta b(n_0)=0$ proves part (ii).

\emph{Step 5: the second step.} To show that the principal optimally offers $\phi^\ast(n_0^\ast)$, with $n_0^\ast\le n_0^{CF}\le n_0^{FB}$, I first follow Step 5 of Theorem~\proofref{thm:cf}'s proof, with \eqref{eq:app bound} and Step 3 in place of \eqref{eq:app cf bound} and Step 4 there. A schedule with $n_{1k}<\infty$ for some $k$ then yields strictly less surplus than $S(n_0)$, while every feasible contract at it still pays the agent at least $R^\ast(n_0)$, so every optimal schedule sets $n_{1k}=\infty$ for each $k$, and the second step reduces to
\[
\max_{n\in\{0,1,2,\dots\}}\ \big[\,S(n)-R^\ast(n)\,\big],
\]
with $S(0)=0$ and $R^\ast(0):=0$, as there.

Second, I show that this maximum is attained and that its smallest maximizer $n_0^\ast$ satisfies $n_0^\ast\le n_0^{CF}$. By Step 5 of Theorem~\proofref{thm:cf}'s proof, $n_0^{CF}$ is the smallest maximizer of $S(n)-R^{CF}(n)$, so it suffices that the agent's marginal payoff is at least as high when the finding is private as when it is contractible. Since the multipliers of Step 2 do not depend on the duration, $R^\ast(n)$, which is the constant $\kappa$ of Step 2 at $n_0=n$ by Step 3, equals the sum of the first $n$ terms of the fixed sequence $c(\beta_t-\eta_t-v_t)$. With $R^\ast(0)=0$ and $\beta_tP_{t-1}=\eta_tP_{t-2}+p_Hv_t$, the agent's marginal payoff is therefore
\begin{equation}\label{eq:app rstar}
\Delta R^\ast(t):=R^\ast(t)-R^\ast(t{-}1)=c\,\big(\beta_t-\eta_t-v_t\big)
=c\,\Big[\eta_t\,\frac{P_{t-2}-P_{t-1}}{P_{t-1}}+v_t\,\frac{p_H-P_{t-1}}{P_{t-1}}\Big]\ \ge\ 0,
\end{equation}
with $\Delta R^\ast(1)=0$. To avoid notation collision, let $\eta^{CF}_t$ denote the $\eta_t$ of Step 2 of Theorem~\proofref{thm:cf}'s proof, so that $\Delta R^{CF}(t)=c\,\eta^{CF}_t(P_{t-2}-P_{t-1})/P_{t-1}$ by \eqref{eq:app rcf}. Subtracting this from \eqref{eq:app rstar} gives
\[
\Delta R^\ast(t)-\Delta R^{CF}(t)=c\,\Big[\big(\eta_t-\eta^{CF}_t\big)\,\frac{P_{t-2}-P_{t-1}}{P_{t-1}}+v_t\,\frac{p_H-P_{t-1}}{P_{t-1}}\Big]\ \ge\ 0,
\]
because $P_{t-2}>P_{t-1}$ and $p_H>P_{t-1}$ (Fact 1), $v_t\ge0$ by Step 2, and $\eta_t\ge\eta^{CF}_t$, which I show next. Let $\beta^{CF}_t$ denote the $\beta_t$ of Step 2 of Theorem~\proofref{thm:cf}'s proof. Both pairs of sequences start from $\eta_1=\beta_1=\eta^{CF}_1=\beta^{CF}_1=0$ and satisfy, for $t\ge2$,
\[
\begin{aligned}
\eta_t&=\delta(\mathcal D_{t-1}+\beta_{t-1}), & \beta_t&=\frac{\eta_tP_{t-2}+p_Hv_t}{P_{t-1}},\\
\eta^{CF}_t&=\delta(\mathcal D_{t-1}+\beta^{CF}_{t-1}), & \beta^{CF}_t&=\frac{\eta^{CF}_tP_{t-2}}{P_{t-1}},
\end{aligned}
\]
so they differ only by the term $p_Hv_t/P_{t-1}\ge0$ in $\beta_t$. Suppose that $\beta_{t-1}\ge\beta^{CF}_{t-1}$, as holds at $t=2$. Then, $\eta_t\ge\eta^{CF}_t$ follows from above, and hence $\beta_t\ge\eta_tP_{t-2}/P_{t-1}\ge\eta^{CF}_tP_{t-2}/P_{t-1}=\beta^{CF}_t$. By induction on $t$, $\eta_t\ge\eta^{CF}_t$ for every $t$. Therefore, for every $n>n_0^{CF}$,
\[
\begin{aligned}
\big[S-R^\ast\big](n)-\big[S-R^\ast\big](n_0^{CF})
&=\big[S-R^{CF}\big](n)-\big[S-R^{CF}\big](n_0^{CF})\\
&\quad-\sum_{t=n_0^{CF}+1}^{n}\big(\Delta R^\ast(t)-\Delta R^{CF}(t)\big)\ \le\ 0,
\end{aligned}
\]
since $n_0^{CF}$ maximizes $S-R^{CF}$ and the last subtracted sum is nonnegative. The maximum of $S-R^\ast$ over all durations is therefore attained on the finite set $\{0,\dots,n_0^{CF}\}$, so its smallest maximizer $n_0^\ast$ exists and satisfies $n_0^\ast\le n_0^{CF}$. Consequently, the principal optimally offers $\phi^\ast(n_0^\ast)$, with $n_0^\ast\le n_0^{CF}\le n_0^{FB}$ by Theorem~\ref{thm:cf}. Parts (i) and (ii) are Step 4 at $n_0=n_0^\ast$ if $n_0^\ast>0$ and are vacuous at $n_0^\ast=0$.

As a by-product that the main text cites, $R^\ast$ strictly increases from $n_0=1$ on, because for $t\ge2$ the first term of \eqref{eq:app rstar} is positive, as $\eta_t>0$ and $P_{t-2}>P_{t-1}$ (Fact 1). The introduction and Section~\ref{sec:comparative statics} use it in discussing the welfare effect of privacy. $\hfill\square$

\subsection{Proof of Proposition \ref*{prop:report free}}\label{app:proof report free}

Fix a finite $n_0>1$. By Step 5 of Theorem~\proofref{thm:main result}'s proof, a feasible contract with uninformed duration $n_0$ and some finite informed duration yields strictly less surplus than $S(n_0)$, and every feasible contract with uninformed duration $n_0$ gives the agent at least $R^\ast(n_0)$ by \eqref{eq:app bound} and Step 3 of that proof. Since $\phi^\ast(n_0)$ gives the principal $S(n_0)-R^\ast(n_0)$, also by Step 3, a contract is optimal among the contracts with uninformed duration $n_0$ if and only if it sets every informed duration infinite and gives the agent $R^\ast(n_0)$. In particular, $\phi^\ast(n_0)$ is optimal among these contracts, and it is report-free in the sense of Section~\ref{sec:main result}, since it is equivalent to posting the bonus sequence $\big(b_0^\ast(1;n_0),\dots,b_0^\ast(n_0;n_0),\underline b,\underline b,\dots\big)$ with no wage. For uniqueness, since the report-free class of Appendix~\ref{app:preliminaries} contains that of Section~\ref{sec:main result}, it suffices to show that every feasible contract $\phi$ in the larger class that sets every informed duration infinite and gives the agent $R^\ast(n_0)$ equals $\phi^\ast(n_0)$.

Fix such a contract $\phi$, with uninformed bonuses $b_0(t)$ and informed payoffs $Y_t$. The proof has three steps. Step 1 shows that the agent's least payoff forces $\phi$ to pay no uninformed wage and makes every effort obedience slack zero. Step 2 uses report-freeness to show that $\phi$ posts its uninformed bonuses followed by $\underline b$, that is, $\phi$ is the contract $\phi(b_0)$ of Step 3 of Theorem~\proofref{thm:main result}'s proof. Step 3 shows that zero effort obedience slacks at $\phi(b_0)$ force $b_0=b_0^\ast(\cdot\,;n_0)$.

\emph{Step 1: what the agent's least payoff forces.} Since $\phi$ is feasible at the schedule $(n_0,\infty\mathbf 1)$, it satisfies the weak-duality bound \eqref{eq:app bound} there, and $U^0_1(\phi)=R^\ast(n_0)=\kappa$ by Step 3 of Theorem~\proofref{thm:main result}'s proof. Each term added to $\kappa$ in \eqref{eq:app bound} is nonnegative, so all of them are zero. Since every wage coefficient $\delta^{s-1}$ is positive, $\phi$ pays no uninformed wage, the severance included. Since $\sigma^S_t>0$ for all $t$ and $\sigma^C_m>0$ for all $m\ge2$ by Step 2 of that proof, every effort obedience slack is zero, and so is every no-concealing slack from period $2$ on. In particular, $n_0>1$ gives $G^C_{n_0}=Y_{n_0}-\delta w_0(n_0{+}1)=0$, so $Y_{n_0}=0$.

\emph{Step 2: report-freeness fixes the informed tracks.} Report-freeness implies $\ell\equiv0$, and together with Step 1 it gives $b_{1k}(t)=b_0(t)$ and $w_{1k}(t)=w_0(t)=0$ in every period $k<t\le n_0$, in which the $k$-informed track and the uninformed track are both active. Report-freeness also gives all informed tracks the same bonus $b_1(t)$ and the same wage $w_1(t)$ in each period $t>n_0$, so it remains to show that $w_1\equiv0$ and $b_1\equiv\underline b$. The $n_0$-informed track consists of periods after $n_0$ only. By \eqref{eq:app forced}, $\ell_{n_0}=0$ and Step 1, no effort rule on this track gives an informed agent more than $Y_{n_0}=0$. Shirking forever collects the wages alone, so $w_1\equiv0$ by (LL). Working only in a period $s>n_0$ then yields $\delta^{s-n_0}[p_Hb_1(s)-c]\le0$, so $b_1(s)\le\underline b$. Working in every period yields
\[
Y_{n_0}=\sum_{s>n_0}\delta^{s-n_0}(1-p_H)^{s-n_0-1}\big[p_Hb_1(s)-c\big]=0,
\]
a sum of nonpositive terms with positive weights, so $b_1(s)=\underline b$ for every $s>n_0$. Consequently, $\phi$ is the contract $\phi(b_0)$.

\emph{Step 3: binding effort obedience fixes the bonuses.} It remains to show that $b_0=b_0^\ast(\cdot\,;n_0)$. By Step 1, every effort obedience slack of $\phi=\phi(b_0)$ is zero, and by the second by-product of Step 3 of Theorem~\proofref{thm:main result}'s proof, $b_0^\ast(\cdot\,;n_0)$ is the only bonus sequence $b_0$ for which $\phi(b_0)$ has this property. Hence $\phi=\phi(b_0^\ast)=\phi^\ast(n_0)$. $\hfill\square$

\subsection{Proof of Proposition \ref*{prop:verifiable}}\label{app:proof verifiable}

Verifiability removes only the reporting rules that send ``$F$'' before a finding arrives. The lower bound on the agent's payoff is unchanged, because the relaxed problem in Theorem~\proofref{thm:main result}'s proof is unchanged: its constraints, like those that rule out an infinite uninformed duration in Online Appendix~\ref{app:infinite}, guard only against shirking and concealing. Every contract that is feasible in the main model remains feasible, since (IC) now has fewer deviations to deter. Lastly, $\phi^\ast(n_0)$ induces the same obedient strategy, so it still attains the bound at every uninformed duration $n_0$. The second step is therefore unchanged, and $\phi^\ast(n_0^\ast)$ remains optimal. $\hfill\square$

\subsection{Proof of Proposition \ref*{prop:gap}}\label{app:proof gap}

For every $t<n_0$, multiplying the recursion \eqref{eq:app premium} for $\Delta b(t)=b_0^{CF}(t;n_0)-b_0^\ast(t;n_0)$ by $\delta^{t-1}$ gives
\[
\delta^{t-1}\Delta b(t)-\delta^{t}\Delta b(t{+}1)=\delta^{t-1}\,\frac{Q_{t-1}}{P_{t-1}}\,\big(F_t-\delta F_{t+1}\big),
\]
which is positive since $F_t>\delta F_{t+1}$ by (3b) of Step 3 of Theorem~\proofref{thm:main result}'s proof. Therefore, $\delta^{t-1}\Delta b(t)$ strictly decreases in $t$, that is, $\delta^{t-1}\big[b_0^\ast(t;n_0)-b_0^{CF}(t;n_0)\big]$ strictly increases in $t$. $\hfill\square$

\subsection{Proof of Proposition \ref*{prop:information}}\label{app:proof information}

\emph{Step 1: first-best surplus.} By Step 1 of Proposition~\proofref{prop:frictionless}'s proof, the first-best surplus is $\max_{n\ge0}S(n)$. To show that this maximum strictly increases in $q_H$ whenever $n_0^{FB}>0$, I first write $S(n)$ in closed form, grouping its terms by the state. Multiplying the Bayes identities \eqref{eq:app bayes} at the indices $0,\dots,t-2$ gives $\mathcal D_t\,\mu_{t-1}=\mu_0(\delta\gamma')^{t-1}$ and $\mathcal D_t\,(1-\mu_{t-1})=(1-\mu_0)\big(\delta(1-p_L)\big)^{t-1}$. Since $P$ and $Q$ are affine in the belief,
\[
\mathcal D_t\Pi_t=\mu_0(\delta\gamma')^{t-1}\big[p_H\pi-c+q_H(1-p_H)V^{\mathrm{informed}}\big]+(1-\mu_0)\big(\delta(1-p_L)\big)^{t-1}(p_L\pi-c).
\]
Let $x:=\delta(1-p_H)\in(0,1)$, $y:=1-q_H$, and $g:=p_H\pi-c>0$, so that $\delta\gamma'=xy$ and $V^{\mathrm{informed}}=\delta g/(1-x)$. The first bracket equals $g\,[1+q_Hx/(1-x)]=g\,(1-xy)/(1-x)$, and summing over $t\le n$ gives
\[
S(n)=\mu_0\,g\,\frac{1-(xy)^n}{1-x}+(1-\mu_0)\,S_L(n),
\qquad
S_L(n):=(p_L\pi-c)\sum_{t\le n}\big(\delta(1-p_L)\big)^{t-1}.
\]
The term $S_L(n)$ does not depend on $q_H$, and for every $n>0$ the first term strictly increases in $q_H$, since $x\in(0,1)$ and $y$ strictly decreases in $q_H$. Now let $q_H<q'_H$, let $\hat n:=n_0^{FB}>0$ at $q_H$, and write $S(\cdot\,;q)$ for the surplus at the finding probability $q$. The first-best surplus at $q'_H$ is at least $S(\hat n;q'_H)>S(\hat n;q_H)$, which is the first-best surplus at $q_H$.

\emph{Step 2: payoffs.} One example shows that raising the finding probability can raise both the agent's payoff and the principal's payoff, or lower both, whether the finding is contractible or private. By Proposition~\ref{prop:infinite} and Step 5 of the proofs of Theorems~\ref{thm:cf} and~\ref{thm:main result}, the principal's optimal payoff is the maximum over the uninformed duration $n$ of $S(n)-R^{CF}(n)$ when the finding is contractible and of $S(n)-R^\ast(n)$ when it is private, attained on $\{0,\dots,n_0^{FB}\}$. Take $c=1$, $p_H=2/5$, $p_L=3/20$, $\mu_0=9/10$, $\delta=1$ and $\pi=11/2$. Exact rational arithmetic gives the values below, rounded to four decimals. Each maximizer over $\{0,\dots,n_0^{FB}\}$ is unique, and every longer duration does strictly worse, since $S(n)\le S(n_0^{FB})$ while $R^{CF}$ and $R^\ast$ strictly increase from $n=1$ on by the by-products of Step 5 of both theorems' proofs. Hence every optimal contract has the tabulated uninformed duration, solves the first step there and so gives the agent the payoff shown.
\begin{center}
\begin{tabular}{ccccccc}
\toprule
& & & \multicolumn{2}{c}{agent's payoff} & \multicolumn{2}{c}{principal's payoff}\\
\cmidrule(lr){4-5}\cmidrule(lr){6-7}
$q_H$ & $n_0^{FB}$ & $n_0^{CF}=n_0^\ast$ & contractible & private & contractible & private\\
\midrule
$3/10$ & $7$ & $3$ & $0.2117$ & $0.2133$ & $2.2432$ & $2.2417$\\
$7/20$ & $6$ & $3$ & $0.2480$ & $0.2503$ & $2.2469$ & $2.2445$\\
$2/5$ & $6$ & $2$ & $0.0774$ & $0.0774$ & $2.2403$ & $2.2403$\\
\bottomrule
\end{tabular}
\end{center}
Raising $q_H$ from $3/10$ to $7/20$ therefore increases both payoffs, and raising it from $7/20$ to $2/5$ decreases both, whether the finding is contractible or private.

As a by-product that the main text cites, raising $q_H$ from $7/20$ to $2/5$ in this example raises the first-best surplus but lowers the total surplus at the optimal contracts:
\begin{center}
\begin{tabular}{ccc}
\toprule
$q_H$ & $S(n_0^{FB})$ & $S(n_0^{CF})=S(n_0^\ast)$\\
\midrule
$3/10$ & $2.6145$ & $2.4549$\\
$7/20$ & $2.6178$ & $2.4948$\\
$2/5$ & $2.6215$ & $2.3177$\\
\bottomrule
\end{tabular}
\end{center}
A footnote in Section~\ref{sec:comparative statics} uses these numbers to show that a shorter uninformed duration can lower the total surplus. $\hfill\square$

\subsection{Proof of Proposition \ref*{prop:privacy}}\label{app:proof privacy}

\emph{Step 1: the agent's payoffs at a common duration.} To compare the principal's optimal payoffs in Step 2, I first show that
\begin{equation}\label{eq:app RvsRCF}
R^\ast(n_0)\ \ge\ R^{CF}(n_0)\ \text{ for all }n_0,
\qquad\text{strictly if and only if }n_0>2 .
\end{equation}
Both payoffs vanish at $n_0=0$, so $R^\ast(n_0)-R^{CF}(n_0)$ is the sum of the differences $\Delta R^\ast(t)-\Delta R^{CF}(t)$ of the agent's marginal payoffs over the periods $t\le n_0$. It therefore suffices to show that each difference is zero for $t\le2$ and strictly positive for $t\ge3$. As shown in Step 5 of Theorem~\proofref{thm:main result}'s proof, where $\eta^{CF}_t$ is defined,
\[
\Delta R^\ast(t)-\Delta R^{CF}(t)=c\,\Big[\big(\eta_t-\eta^{CF}_t\big)\,\frac{P_{t-2}-P_{t-1}}{P_{t-1}}+v_t\,\frac{p_H-P_{t-1}}{P_{t-1}}\Big],
\qquad \eta_t\ge\eta^{CF}_t .
\]
At $t\le2$ both terms vanish, since $v_1=v_2=0$, $\eta_1=\eta^{CF}_1=0$ and $\eta_2=\eta^{CF}_2=\delta$. At $t\ge3$ the first term is nonnegative, since $\eta_t\ge\eta^{CF}_t$ and $P_{t-2}>P_{t-1}$. The second is strictly positive, since $v_t=\delta\,\sigma^C_{t-1}$ with $\sigma^C_{t-1}>0$ by Step 2 of Theorem~\proofref{thm:main result}'s proof, and $P_{t-1}<p_H$. This proves \eqref{eq:app RvsRCF}.

\emph{Step 2: the principal.} By Proposition~\ref{prop:infinite} and Step 5 of the proofs of Theorems~\ref{thm:cf} and~\ref{thm:main result}, the principal's optimal payoffs with a contractible and with a private finding are the maxima of $S-R^{CF}$ and of $S-R^\ast$ over the uninformed duration, attained on $\{0,\dots,n_0^{FB}\}$ at their smallest maximizers $n_0^{CF}$ and $n_0^\ast$. For every duration $n$,
\[
\big[S-R^\ast\big](n)\ \le\ \big[S-R^{CF}\big](n)\ \le\ \big[S-R^{CF}\big](n_0^{CF})
\]
by \eqref{eq:app RvsRCF} and the optimality of $n_0^{CF}$, so the principal is weakly worse off when the finding is private. If $n_0^{CF}\le2$, then $R^\ast(n_0^{CF})=R^{CF}(n_0^{CF})$ by \eqref{eq:app RvsRCF}, so the maximum of $S-R^\ast$ is at least $[S-R^{CF}](n_0^{CF})$, and the two optimal payoffs are equal. If $n_0^{CF}>2$, the chain is strict at every $n$. At $n>2$, its first inequality is strict by \eqref{eq:app RvsRCF}. At $n\le2<n_0^{CF}$, its second inequality is strict because $n_0^{CF}$ is the smallest maximizer of $S-R^{CF}$. In particular, the chain is strict at $n_0^\ast$, so the principal is strictly worse off.

\emph{Step 3: the agent.} Two examples show that the agent can be better or worse off. As in Step 2 of Proposition~\proofref{prop:information}'s proof, the payoffs are computed exactly in rational arithmetic, and every optimal contract solves the first step at its own duration and so gives the agent the payoff shown. At the primitives of Table~\ref{tab:example}, $n_0^{FB}=6$, and on $\{0,\dots,6\}$ both $S-R^{CF}$ and $S-R^\ast$ have the unique maximizer $3$, so $n_0^{CF}=n_0^\ast=3$, and
\[
R^{CF}(3)=\frac{117}{190}\ <\ \frac{21}{34}=R^\ast(3):
\]
the agent is strictly better off when the finding is private. At the primitives of the right panel of Figure~\ref{fig:main result}, $n_0^{FB}=15$, and on $\{0,\dots,15\}$ the unique maximizer of $S-R^{CF}$ is $6$ and that of $S-R^\ast$ is $5$, so $n_0^{CF}=6$ and $n_0^\ast=5$, and
\[
R^{CF}(6)\approx0.3469\ >\ 0.2075\approx R^\ast(5):
\]
the agent is strictly worse off when the finding is private. $\hfill\square$

\newpage

\section{Online Appendix}\label{app:online appendix}
\addtocontents{toc}{\protect\setcounter{tocdepth}{1}}
\setcounter{proposition}{0}\setcounter{lemma}{0}
\renewcommand{\theproposition}{OA.\arabic{proposition}}
\renewcommand{\thelemma}{OA.\arabic{lemma}}
\providecommand{\theHproposition}{}\renewcommand{\theHproposition}{OA.\arabic{proposition}}\providecommand{\theHlemma}{}\renewcommand{\theHlemma}{OA.\arabic{lemma}}

\subsection{Optimal Contracts with Contractible Findings}\label{app:cf all}

Theorem~\ref{thm:cf} exhibits one optimal contract when the finding is contractible. The following proposition describes what every optimal contract shares. It uses the notation of Theorem~\proofref{thm:cf}'s proof, in which $Y_1$ is the informed payoff upon a period-$1$ finding.

\begin{proposition}\label{prop:cf all}
    Suppose the finding is contractible, and let $\phi$ be an optimal contract with $n_0>0$. Then:\begin{enumerate}[nolistsep,label=(\roman*)]
    \item $w_0\equiv0$;
    \item $b_0(t)=b_0^{CF}(t;n_0)$ for $2\le t\le n_0$;
    \item $\ell_k=0$, $w_{1k}\equiv0$ and $b_{1k}\equiv\underline b$ for $2\le k\le n_0$;
    \item $P_0b_0(1)+Q_0Y_1=P_0b_0^{CF}(1;n_0)$. Conversely, $\phi^{CF}(n_0)$ remains optimal if it also pays a lump sum $\ell_1\le P_0b_0^{CF}(1;n_0)/Q_0$ upon a period-$1$ finding and lowers $b_0(1)$ by $Q_0\ell_1/P_0$;
    \item if $n_0>2$, then $b_{12}(3)=\underline b<b_0(3)$; if $n_0\le2$, the PBS contract $\phi^\ast(n_0)$ is also optimal.
    \end{enumerate}
\end{proposition}

\emph{Proof}. Step 1 shows that $\phi$ attains the bound \eqref{eq:app cf bound}, Steps 2 to 4 prove parts (i) to (iv), and Step 5 proves part (v).

\emph{Step 1: $\phi$ solves the first step at its schedule.} By Proposition~\ref{prop:infinite} and Step 5 of Theorem~\proofref{thm:cf}'s proof, $\phi$ has the schedule $(n_0,\infty\mathbf 1)$ with $n_0$ finite. As an optimal contract, $\phi$ solves the first step there, so $U^0_1=\kappa=R^{CF}(n_0)$ by Step 4 of that proof, and the optimal payoff is $S(n_0)-R^{CF}(n_0)$.

\emph{Step 2: every effort obedience slack is zero, $w_0\equiv0$, and $Y_t=0$ from period $2$ on.} Since $\phi$ is feasible, \eqref{eq:app cf bound} holds at $\phi$, and since $U^0_1=\kappa$, each of the nonnegative terms that it adds to $\kappa$ is zero. As $\sigma^S_t>0$ and $p^w_t>0$ for all $t$, and $p^Y_t>0$ for $2\le t\le n_0$, by Step 2 of Theorem~\proofref{thm:cf}'s proof, it follows that $G^S_t=0$ for all $t\le n_0$, that $w_0\equiv0$, which is part (i), and that $Y_t=0$ for $2\le t\le n_0$.

\emph{Step 3: parts (ii) and (iv).} With $w_0\equiv0$, each slack $G^S_t$ is $P_{t-1}b_0(t)$, with $P_{t-1}>0$, plus an affine function of $Y_t$ and of the later bonuses and informed payoffs, with coefficients that depend only on $n_0$ and the primitives. Given $Y_t=0$ at $2\le t\le n_0$, the equalities $G^S_t=0$ at these periods therefore determine $b_0(n_0),\dots,b_0(2)$ backward, one at a time. The contract $\phi^{CF}(n_0)$ satisfies the same equalities, since its wages, informed payoffs and effort obedience slacks vanish by Step 4 of Theorem~\proofref{thm:cf}'s proof. Hence $b_0(t)=b_0^{CF}(t;n_0)$ for every $2\le t\le n_0$, which is part (ii). In period $1$, the slack $G^S_1$ is $P_0b_0(1)+Q_0Y_1$ plus terms that part (ii) and $Y_t=0$ at $t\ge2$ pin down, and it vanishes at $\phi$ and at $\phi^{CF}(n_0)$, where $Y_1=0$. Thus $P_0b_0(1)+Q_0Y_1=P_0b_0^{CF}(1;n_0)$. For the converse in part (iv), the modified contract changes no strategy's payoff: an agent who shirks in period $1$ collects neither $b_0(1)$ nor $\ell_1$, and an agent who works in period $1$ holds the prior and expects the same total $P_0b_0(1)+Q_0\ell_1$. It is therefore feasible with the agent's payoff $R^{CF}(n_0)$, hence optimal by Step 1.

\emph{Step 4: part (iii).} Fix $2\le k\le n_0$. By Step 2, $\ell_k+F_k=Y_k=0$. Since the $k$-informed track is reached with positive probability, (IC) makes working in every period of it optimal, so $F_k=\mathcal W_k\ge0$, as shirking in every period collects nonnegative transfers. With $\ell_k\ge0$, it follows that $\ell_k=0$ and $F_k=\mathcal W_k=0$. The argument for the $n_0$-informed track in Step 2 of Proposition~\proofref{prop:report free}'s proof then shows that the $k$-informed track, which is infinite by Step 1, pays no wage and the bonus $\underline b$ in every period, which is part (iii).

\emph{Step 5: part (v).} Let $n_0>2$. By part (iii) at $k=2$, $b_{12}(3)=\underline b$, while $b_0(3)=b_0^{CF}(3;n_0)>\underline b$ by part (ii) and Step 4 of Theorem~\proofref{thm:cf}'s proof.

Let $n_0\le2$. By Step 1, it suffices to show that, when the finding is contractible, $\phi^\ast(n_0)$ is feasible at the schedule $(n_0,\infty\mathbf 1)$ and gives the agent $R^{CF}(n_0)$. For feasibility, every effort rule $e$ available when the finding is contractible generates the same transfers and efforts, hence the same payoff, as the strategy $(e,r^\ast)$ when the finding is private, since $r^\ast$ reports each finding in its arrival period. As $\phi^\ast(n_0)$ satisfies (LL) and (IC) when the finding is private by Step 3 of Theorem~\proofref{thm:main result}'s proof, it is therefore feasible when the finding is contractible. By the same equivalence, the agent's payoff under it is $R^\ast(n_0)$, which equals $R^{CF}(n_0)$ at $n_0\le2$ by \eqref{eq:app RvsRCF} in Proposition~\proofref{prop:privacy}'s proof. Hence $\phi^\ast(n_0)$ attains the optimal payoff $S(n_0)-R^{CF}(n_0)$. $\hfill\square$

\subsection{Optimal Contracts with Private Findings}\label{app:pf all}

This section states and proves the characterization of optimal contracts that Section~\ref{sec:main result} describes for a private finding. The notation is that of Theorem~\proofref{thm:main result}'s proof: $F_t$ is the informed payoff of the PBS contract $\phi^\ast(n_0)$, and $Y_t$ that of a contract $\phi$.

\begin{proposition}\label{prop:pf all}
Suppose the finding is private, and let $\phi$ be an optimal contract with $n_0>1$. Then $w_0\equiv0$, and $b_0(t)=b_0^\ast(t;n_0)$ and $Y_t=F_t$ for $2\le t\le n_0$. Moreover, $b_0(1)\le b_0^\ast(1;n_0)$, with equality if and only if $Y_1=F_1$.
\end{proposition}

\emph{Proof}. By Proposition~\ref{prop:infinite} and Step 5 of Theorem~\proofref{thm:main result}'s proof, $\phi$ has the schedule $(n_0,\infty\mathbf 1)$ with $n_0<\infty$. As an optimal contract, it solves the first step at this schedule, so $U^0_1(\phi)=R^\ast(n_0)$ by Step 3 of that proof.

\emph{Step 1: zero wages and zero slacks.} The argument of Step 1 of Proposition~\proofref{prop:report free}'s proof, which uses only that $\phi$ is feasible at $(n_0,\infty\mathbf 1)$ and gives the agent $R^\ast(n_0)$, shows that $w_0\equiv0$, that $G^S_t=0$ for all $t\le n_0$ and that $G^C_m=0$ for all $2\le m\le n_0$.

\emph{Step 2: the periods $t\ge2$.} With $w_0\equiv0$, the zero slacks of Step 1 determine $Y_t$ and then $b_0(t)$ at $t=n_0,n_0-1,\dots,2$, in this order. Indeed, $G^C_t=0$ gives $Y_{n_0}=0$ and $Y_t=\delta\big[-c+p_Hb_0(t{+}1)+(1-p_H)Y_{t+1}\big]$ for $t<n_0$, and each slack $G^S_t$ is $P_{t-1}b_0(t)$ plus an affine function of $Y_t$ and of the later bonuses and informed payoffs, with $P_{t-1}>0$. The PBS contract satisfies the same equalities by Step 3 of Theorem~\proofref{thm:main result}'s proof, with its informed payoffs equal to $F_t$, so $b_0(t)=b_0^\ast(t;n_0)$ and $Y_t=F_t$ for every $2\le t\le n_0$.

\emph{Step 3: period $1$.} The slack $G^S_1$, which vanishes by Step 1, is $P_0b_0(1)+Q_0Y_1$ plus terms that Step 2 pins down. The PBS contract satisfies $G^S_1=0$ as well, so $P_0b_0(1)+Q_0Y_1=P_0b_0^\ast(1;n_0)+Q_0F_1$. No-concealing in period $1$, $G^C_1\ge0$, implies $Y_1\ge\delta\big[-c+p_Hb_0^\ast(2;n_0)+(1-p_H)F_2\big]=F_1$ by \eqref{eq:app Frec}. Hence
\[
b_0(1)=b_0^\ast(1;n_0)-\frac{Q_0}{P_0}\,\big(Y_1-F_1\big)\ \le\ b_0^\ast(1;n_0),
\]
with equality if and only if $Y_1=F_1$.

As a by-product that the main text cites, an optimal contract with $n_0=1$ also pays no uninformed wage, since the wage part of Step 1 does not use $n_0>1$. Section~\ref{sec:discussions} states it.

As a second by-product that the main text cites, every optimal contract has a strictly increasing research-stage bonus, $b_0(1)\le b_0^\ast(1;n_0)<b_0^\ast(2;n_0)=b_0(2)<\dots<b_0(n_0)$, since Step 4 of Theorem~\proofref{thm:main result}'s proof proves part (i) of that theorem for every duration $n_0>1$. Section~\ref{sec:main result} states it. $\hfill\square$

\subsection{Infinite Uninformed Durations}\label{app:infinite}

The proofs of Theorems~\ref{thm:cf} and~\ref{thm:main result} assume a finite uninformed duration. The following proposition shows that this is without loss of optimality. Its proof uses the notation of those proofs.

\begin{proposition}\label{prop:infinite}
Fix a feasible contract $\phi$ at a schedule with $n_0=\infty$. For some finite $T$, $U_P(\phi,e^\ast,r^\ast)<S(T)-R^{CF}(T)$ if the finding is contractible, and $U_P(\phi,e^\ast,r^\ast)<S(T)-R^\ast(T)$ if it is private. Hence no optimal contract sets $n_0=\infty$.
\end{proposition}

\emph{Proof}. Step 1 treats a contractible finding and Step 2 a private finding.

\emph{Step 1: a contractible finding.} Throughout this step, $\sigma^S_t$, $\beta_t$, $p^w_t$ and $p^Y_t$ are the symbols of Step 2 of Theorem~\proofref{thm:cf}'s proof. Since the agent's least payoff at $n_0=\infty$ need not be attained, I truncate $\phi$ at a large finite duration $T$ and apply the identity \eqref{eq:app cf identity} of Step 2 of Theorem~\proofref{thm:cf}'s proof to the truncated contract, which need not be feasible, to show that $\phi$ pays the agent at least $R^{CF}(T)$. If $U^0_1=\infty$, the principal's payoff is $-\infty$, since no schedule's surplus exceeds $S(n_0^{FB})$ by Step 1 of Proposition~\proofref{prop:frictionless}'s proof, and the uninformed duration $n_0=0$ dominates. Henceforth $U^0_1<\infty$. By the argument of Step 1 of Theorem~\proofref{thm:cf}'s proof, $Y_t\ge0$ for every $t$, so every $U^0_t$ is finite, since the terms of the earlier periods are bounded below by $-c$ per period.

Because the truncated contract replaces the continuation from period $T{+}1$ on with a severance, the truncation gives $U^0_1\ge R^{CF}(T)$ only up to an error term $E_T$. I first drop the uninformed wages, write the continuation payoffs through the values of Step 3 of Theorem~\proofref{thm:cf}'s proof, and record two inequalities that feasibility implies. I then derive the bound up to $E_T$ and show that $E_T\ge0$ for arbitrarily large $T$. Dropping the uninformed wages is without loss. Under any effort rule the agent reaches the uninformed period $t$ with probability $\Lambda_d:=\prod_{j<d}\nu_j$, where $d$ is the number of periods he has worked before $t$, and $\Lambda_d\ge\Lambda_{t-1}$ since every factor is at most one and $d\le t-1$, with equality for the obedient agent. Every rule therefore collects at least as much in discounted wages as the obedient strategy, so deleting $w_0$ preserves (IC), weakly lowers the agent's payoff, and leaves the surplus unchanged. Let $w_0\equiv0$ from now on. Evaluate $A_H$, $A_L$, and $A$, the objects of Step 3 of Theorem~\proofref{thm:cf}'s proof, at $(b_0,Y)$, where $A_H$ and $A_L$ are now series, and write $\Delta^A(t):=A_H(t)-A_L(t)$. Since the period payoffs of $A_H$ are at least $-c$, $A_H\ge-C_H$ with $C_H:=c/(1-\delta\gamma')$. Since the model bounds $b_0$ at $n_0=\infty$, the period payoffs of $A_L$ are bounded and $A_L$ is finite. Applying \eqref{eq:app bayes} term by term shows that the obedient agent's continuation payoff $U^0_t$ equals $A(t,\mu_{t-1})$, so $A_H(t)$ is finite as well.

Feasibility supplies two inequalities. First, (IC) implies $U^0_t\ge0$ for all $t$, since shirking in every uninformed period from $t$ on, which collects nothing there, is worth $U^0_1-\mathcal D_tU^0_t$. Second, as in Step 1 of Theorem~\proofref{thm:cf}'s proof, the value of the shirker who follows $e^t_{\text{Shirk}}$ falls short of $U^0_1$ by $\mathcal D_tG^S_t$, where the effort obedience slack now reads $G^S_t=U^0_t-\delta A(t{+}1,\mu_{t-1})$. Effort obedience requires $G^S_t\ge0$. With $\xi_t(\mu):=-c+P(\mu)\,b_0(t)+Q(\mu)\,Y_t$, expanding $U^0_t$ by one period and applying \eqref{eq:app bayes} turns $G^S_t\ge0$ into
\begin{equation}\label{eq:app postponeCF}
\xi_t(\mu_{t-1})\ \ge\ \delta\big[\mu_{t-1}(1-\gamma')\,A_H(t{+}1)+(1-\mu_{t-1})\,p_L\,A_L(t{+}1)\big].
\end{equation}

To compare $\phi$ with a finite duration $T>0$, let $\phi_T$ be the contract at the schedule $(T,(n_{1k})_{k\le T})$ that keeps every transfer of $\phi$ in the periods $t\le T$ and on the tracks entered at $k\le T$ and pays the severance $w_0(T{+}1):=U^0_{T+1}\ge0$. The identity \eqref{eq:app cf identity} holds for every contract with a finite duration, feasible or not, so it holds at $\phi_T$ with $\kappa=R^{CF}(T)$ by Step 4 of Theorem~\proofref{thm:cf}'s proof. Since the slacks of $\phi_T$ may be negative, I express them through the effort obedience slacks of $\phi$, which are nonnegative. The obedient agent's payoff in $\phi_T$ is $U^0_1$, since the severance replaces his continuation payoff from period $T{+}1$ one for one. The shirker of a period $t\le T$ collects the same in $\phi_T$ as in $\phi$ except upon reaching period $T{+}1$ uninformed, which he does with the discounted probability $\delta\mathcal D_T$ at the belief $\mu_{T-1}$. There he collects $U^0_{T+1}=A(T{+}1,\mu_T)$ in $\phi_T$, while in $\phi$ he works on, which is worth $A(T{+}1,\mu_{T-1})$. As the affine function $A(T{+}1,\cdot)$ has the slope $\Delta^A(T{+}1)$, each slack of $\phi_T$ exceeds that of $\phi$ by $\delta(\mathcal D_T/\mathcal D_t)(\mu_{T-1}-\mu_T)\Delta^A(T{+}1)$. Weighting by $\sigma^S_t$ and using \eqref{eq:app cumul} at $s=T$, I rewrite the identity at $\phi_T$, whose only wage is the severance, in terms of the effort obedience slacks $G^S_t$ of $\phi$:
\[
U^0_1-\sum_{t\le T}\sigma^S_tG^S_t=R^{CF}(T)+\sum_{t=2}^{T}p^Y_tY_t+p^w_{T+1}U^0_{T+1}+\delta\big(\mathcal D_T+\beta_T\big)(\mu_{T-1}-\mu_T)\,\Delta^A(T{+}1).
\]
Since $\sigma^S_t>0$ and $p^Y_t\ge0$ by Step 2 of Theorem~\proofref{thm:cf}'s proof, every weighted slack and every term $p^Y_tY_t$ is nonnegative. Moreover $U^0_{T+1}\ge0$ and $p^w_{T+1}=\delta[\mathcal D_T+(1-\nu_{T-1})\beta_T]\ge\delta p_L(\mathcal D_T+\beta_T)$, since $1-\nu_{T-1}=P_{T-1}+Q_{T-1}\ge p_L$. It follows that
\begin{equation}\label{eq:app cf close}
U^0_1\ \ge\ R^{CF}(T)+\delta\big(\mathcal D_T+\beta_T\big)\,E_T,
\qquad E_T:=p_L\,U^0_{T+1}+(\mu_{T-1}-\mu_T)\,\Delta^A(T{+}1),
\end{equation}
so $U^0_1\ge R^{CF}(T)$ whenever $E_T\ge0$.

It remains to show that $E_T\ge0$ for arbitrarily large $T$. Suppose instead that $E_T<0$ for all $T\ge T_0$. I show that $A_L$ and then $A_H$ are positive in all late periods, which makes $E_T$ positive. Because $U^0_{T+1}\ge0$ and $\mu_{T-1}>\mu_T$ by (Fact 1), $E_T<0$ forces $\Delta^A(T{+}1)<0$, that is, $A_L(t)>A_H(t)$ at every $t>T_0$, and so $A_L(t)>A(t,\mu_{t-1})=U^0_t\ge0$. Substituting $A_L(t{+}1)\ge0$ and $A_H(t{+}1)\ge-C_H$ into \eqref{eq:app postponeCF} gives $\xi_t(\mu_{t-1})\ge-\mu_{t-1}C_H$ for every $t\ge T_0$, that is, $P_{t-1}b_0(t)\ge c-Q_{t-1}Y_t-\mu_{t-1}C_H$. Plugging this bound into the value of one worked period conditional on $\omega=H$, dropping the term in $Y_t$, and using $P_{t-1}\ge p_L$ give, for all large $t$,
\[
-c+p_Hb_0(t)+q_H(1-p_H)Y_t\ \ge\ \frac{c\,(p_H-P_{t-1})}{P_{t-1}}-\mu_{t-1}\,\frac{p_HC_H}{p_L}\ \ge\ \frac{c\,(p_H-p_L)}{2p_L}\ >\ 0 .
\]
The dropped term is nonnegative, since $Y_t\ge0$ and its coefficient $q_H(1-p_H)-p_HQ_{t-1}/P_{t-1}=q_H(1-p_H)\,p_L(1-\mu_{t-1})/P_{t-1}$ is positive, and the second inequality holds since $P_{t-1}\to p_L$ and $\mu_{t-1}\to0$. Summing the series of $A_H$ gives $A_H(T{+}1)\ge c(p_H-p_L)/\big(2p_L(1-\delta\gamma')\big)>0$ for all large $T$. Writing $U^0_{T+1}=A(T{+}1,\mu_T)$ and dropping from $E_T$ the terms in $A_H(T{+}1)$, whose total coefficient $p_L\mu_T+\mu_{T-1}-\mu_T$ is positive, leaves $E_T\ge\big[p_L(1-\mu_T)-(\mu_{T-1}-\mu_T)\big]A_L(T{+}1)$. The right side is positive for every large $T$, since $A_L(T{+}1)>0$ and $\mu_{T-1}\to0$. This contradicts $E_T<0$.

\emph{The assembly.} It remains to bound the surplus of $\phi$ and compare the principal's payoff under it with that under $\phi^{CF}(T)$. Choose $T>n_0^{FB}$ with $E_T\ge0$, so that $U^0_1\ge R^{CF}(T)$ by \eqref{eq:app cf close}. Grouped by the period of the finding, as in Appendix~\ref{app:preliminaries}, the surplus of $\phi$ is $\sum_t\mathcal D_t\big[P_{t-1}\pi-c+Q_{t-1}S^1_t\big]\le\sum_t\mathcal D_t\Pi_t$, where $S^1_t\le V^{\mathrm{informed}}$ is the surplus of the $t$-informed track. Since $\Pi_t\le0$ at every $t>n_0^{FB}$ and $\Pi_t$ is strictly decreasing by Step 1 of Proposition~\proofref{prop:frictionless}'s proof, $\Pi_t<0$ at every $t>T$, so the surplus of $\phi$ is at most $S(T)+\sum_{t>T}\mathcal D_t\Pi_t<S(T)$. The principal's payoff of $\phi$ is therefore strictly below $S(T)-R^{CF}(T)$, which $\phi^{CF}(T)$ attains at the schedule $(T,\infty\mathbf 1)$ by Step 4 of Theorem~\proofref{thm:cf}'s proof, so no optimal contract sets $n_0=\infty$.

\emph{Step 2: a private finding.} The constructions of Step 1 of Theorem~\proofref{thm:main result}'s proof are stated period by period, and they apply here without their terminal conditions. As in Step 1 of this proof, the uninformed duration $n_0=0$ dominates every feasible contract with an infinite payoff to the agent, so let $U_A(\phi,e^\ast,r^\ast)<\infty$. Shirking forever after a report, or from any period on without one, collects only nonnegative transfers, so (IC) implies $Y_t\ge0$ and $U^0_t\ge0$ for every $t$, since each of these histories is reached with positive probability. Every $U^0_t$ is also finite, since $\mathcal D_tU^0_t$ is $U^0_1$ less the finite expected payoffs of the periods before $t$. The argument truncates $\phi$ at a large finite duration $T$ and applies the bound of Step 2 of Theorem~\proofref{thm:main result}'s proof to the truncated contract. The truncated contract pays the agent's continuation payoff under the obedient strategy as a severance. The truncation therefore needs this payoff to be at least as high in state $H$ as in state $L$, so that a shirker or a concealer, who is more optimistic than the obedient agent, values the severance no more than continuing. I show this ordering first.

To compare the two states, let $A_H(t)$ and $A_L(t)$ be the values, conditional on $\omega=H$ and on $\omega=L$, of the obedient strategy from period $t$ on, as in Step 3 of Theorem~\proofref{thm:cf}'s proof. This strategy works in every uninformed period, reports a finding in its arrival period and works on the informed track thereafter. These values now include the wage, and they are infinite series that satisfy
\[
\begin{aligned}
A_H(t)&=w_0(t)-c+p_Hb_0(t)+q_H(1-p_H)Y_t+\delta\gamma'A_H(t{+}1),\\
A_L(t)&=w_0(t)-c+p_Lb_0(t)+\delta(1-p_L)A_L(t{+}1).
\end{aligned}
\]
The value of this strategy at belief $\mu$ is the affine function $A(t,\mu):=\mu A_H(t)+(1-\mu)A_L(t)$, so $U^0_t=A(t,\mu_{t-1})$. Transfers are nonnegative and, at $n_0=\infty$, $b_0$ and $w_0$ are bounded, so $A_L$ is bounded, as its series converges geometrically at the rate $\delta(1-p_L)<1$. Moreover $A_H(t)\ge-c/(1-\delta\gamma')$, and $A_H(t)<\infty$ for every $t$ since $\mu_{t-1}A_H(t)\le U^0_t+\sup_s|A_L(s)|$.

I claim that the slope $\Delta^A(t):=A_H(t)-A_L(t)$ of $A(t,\cdot)$ is nonnegative for all $t$. The claim rests on two slacks that (IC) keeps nonnegative. First, the shirker who follows $e^t_{\text{Shirk}}$ collects $w_0(t)$ in period $t$ without working and then follows the obedient strategy from period $t{+}1$ on at the belief $\mu_{t-1}$. His value therefore falls short of $U^0_1$ by $\mathcal D_t\,G^S_t(\phi)$, where, by effort obedience \eqref{eq:effort obedience},
\begin{equation}\label{eq:app infpostpone}
G^S_t(\phi):=U^0_t-w_0(t)-\delta A(t{+}1,\mu_{t-1})\ \ge\ 0\qquad\text{for all }t .
\end{equation}
Second, an agent informed in period $t$ can conceal his finding and, from period $t{+}1$ on, work every period, reporting with probability $q_H$ after each worked period without a success and working on the informed track after the report. This strategy, the \emph{mimic}, reproduces exactly the transfer law of the obedient strategy conditional on $\omega=H$, so it is worth $A_H(t{+}1)$ from period $t{+}1$. It is a randomization over pure strategies, against each of which (IC) holds, and the finding arrives in period $t$ with the discounted probability $\mathcal D_tQ_{t-1}>0$, so
\begin{equation}\label{eq:app conceallow}
G^M_t:=Y_t-\delta A_H(t{+}1)\ \ge\ 0\qquad\text{for all }t .
\end{equation}
These two slacks bound the slope $\Delta^A$ from below. In terms of $\hat b_t:=b_0(t)-\delta A_L(t{+}1)$ and $G^M_t$, the two recursions give, by $\gamma'+q_H(1-p_H)=1-p_H$ and the affine form of $P$,
\[
\begin{aligned}
\Delta^A(t)&=(p_H-p_L)\,\hat b_t+q_H(1-p_H)\,G^M_t+\delta(1-p_H)\,\Delta^A(t{+}1),\\
G^S_t(\phi)&=P_{t-1}\,\hat b_t+Q_{t-1}\,G^M_t-c-\delta\mu_{t-1}p_H\,\Delta^A(t{+}1),
\end{aligned}
\]
since the wage cancels in both. Eliminating $\hat b_t$ gives the identity
\begin{equation}\label{eq:app slope}
\Delta^A(t)=\frac{(p_H-p_L)\,c}{P_{t-1}}+\delta\Big(1-\frac{p_Lp_H}{P_{t-1}}\Big)\Delta^A(t{+}1)+\frac{p_H-p_L}{P_{t-1}}\,G^S_t(\phi)+\frac{q_H(1-p_H)\,p_L}{P_{t-1}}\,G^M_t .
\end{equation}
Suppose now that $\Delta^A(t)<0$ at some $t$. Both slacks are nonnegative, and the factor of $\Delta^A(t{+}1)$ lies in $(0,1]$ since $p_L\le P_{t-1}\le p_H<1$. Then \eqref{eq:app slope} first forces $\Delta^A(t{+}1)<0$, since its right side would otherwise be positive, and then implies $\Delta^A(t{+}1)\le\Delta^A(t)-(p_H-p_L)c/p_H<0$. By induction, $A_L-A_H$ then grows without bound, whereas it never exceeds $\sup_s|A_L(s)|+c/(1-\delta\gamma')$. This contradiction proves the claim.

Now pick $T>n_0^{FB}$, as the surplus comparison below requires, and let $\phi_T$ be the truncation of $\phi$ defined in Step 1, which here keeps the wages of $\phi$ in the periods $t\le T$, with the severance $w_0(T{+}1):=U^0_{T+1}\ge0$. I show that $\phi_T$ satisfies the constraints of the relaxed problem of Step 1 of Theorem~\proofref{thm:main result}'s proof. The severance $U^0_{T+1}=A(T{+}1,\mu_T)$ replaces the value of the obedient strategy from period $T{+}1$ on, so $\phi_T$ gives the obedient agent the payoff $U^0_1$ and has the informed payoffs $Y_t$ of $\phi$. As in Step 1 of this proof, the effort obedience slack of $\phi_T$ at a period $t\le T$ is $G^S_t(\phi)+(\delta\mathcal D_T/\mathcal D_t)(\mu_{T-1}-\mu_T)\Delta^A(T{+}1)$. The no-concealing slacks of $\phi_T$ for $m<T$ are those of $\phi$, and the slack for $m=T$ is $Y_T-\delta U^0_{T+1}=G^M_T+\delta(1-\mu_T)\Delta^A(T{+}1)$. Since $\mu_{T-1}>\mu_T$ and $\Delta^A(T{+}1)\ge0$ by the claim, all these slacks are nonnegative by \eqref{eq:app infpostpone}, \eqref{eq:no-concealing} and \eqref{eq:app conceallow}. Thus \eqref{eq:app bound} at the duration $T$, with $\kappa=R^\ast(T)$ by Step 3 of Theorem~\proofref{thm:main result}'s proof, gives $U^0_1\ge R^\ast(T)$. The surplus of $\phi$ is strictly below $S(T)$, as in the assembly of Step 1 of this proof. Therefore the principal's payoff of $\phi$ is strictly below $S(T)-R^\ast(T)$, which $\phi^\ast(T)$ attains at the schedule $(T,\infty\mathbf 1)$ by Step 3 of Theorem~\proofref{thm:main result}'s proof, so no optimal contract sets $n_0=\infty$. $\hfill\square$

\subsection{General Contracts}\label{app:general contracts}

This section defines the general contract class of Section~\ref{sec:model} and proves Lemma~\ref{lem:direct}.

\paragraph{The general class.} Only the contract form differs from Section~\ref{sec:model}. A success, and nothing else, ends the relationship, so the agent decides whether to work in every period before a success. In period $t$, the contract first makes a public announcement $z_t$ from a finite set. The agent then chooses effort $e\in\{0,1\}$ at cost $ce$, and nature draws the outcome, a success being public. If there is no success, the agent sends a message from the finite message space $M_t$. Finally, the contract pays a transfer $\tau_t\ge0$, which the agent observes. A \emph{general contract} is
\[
\mathcal G:=\big((\Omega,\mathcal B,\mathbb P);\,(z_t),\,(M_t),\,(\tau_t)\big),
\]
where the standard Borel space $(\Omega,\mathcal B,\mathbb P)$ carries all of the contract's randomization, whose draw is independent of the state and of nature. The contractible history records calendar time, the announcements and messages so far, and the success and its period, but neither effort nor the finding. The announcement $z_t$ is measurable in the draw and the contractible history through period $t-1$, the message space $M_t$ is a function of the contractible history through $z_t$, and $\tau_t$ is measurable in the draw and the contractible history through period $t$. This definition covers public, gradually revealed and hidden randomization, arbitrary message play, and random work recommendations that need not form intervals.

A strategy $\zeta$ prescribes distributions over effort and messages at every private history. The agent's mixing and the contract's draw cancel in Bayes' rule, so his belief is $\mu_t$ after $t$ silent worked periods and $1$ after a finding. Each worked period ends the relationship with probability at least $p_L>0$, so the expected discounted effort cost is at most $c/p_L$, while the expected discounted transfers lie in $[0,\infty]$ by Tonelli's theorem. Hence the agent's payoff $U_A(\mathcal G,\zeta)$ lies in $[-c/p_L,\infty]$, the surplus $S(\mathcal G,\zeta)$ in $[-c/p_L,\pi]$, and the principal's payoff $U_P(\mathcal G,\zeta):=S(\mathcal G,\zeta)-U_A(\mathcal G,\zeta)$ in $[-\infty,\pi+c/p_L]$. A strategy is \emph{optimal} if it attains $\sup_{\zeta'}U_A(\mathcal G,\zeta')$, which is how (IC) reads in the general class. Among the agent's optimal strategies, the principal is credited with the one she prefers.

\paragraph{The result.} The following lemma is the formal content of Lemma~\ref{lem:direct}, since the PBS contract is a direct mechanism.

\begin{lemma}\label{lem:general}
For every general contract $\mathcal G$ and every optimal strategy $\zeta$,
\begin{equation}\label{eq:oa direct}
U_P(\mathcal G,\zeta)\ \le\ S(n_0^\ast)-R^\ast(n_0^\ast).
\end{equation}
The PBS contract $\phi^\ast(n_0^\ast)$ is a general contract, the obedient strategy is optimal in it, and the principal's payoff under it equals the right side of \eqref{eq:oa direct}.
\end{lemma}

\emph{Proof}. The proof evaluates the payoff bound behind Theorem~\ref{thm:main result} along the play that a general contract induces. Steps 1 to 3 reduce the contract to a tree of recommendations with one report, Step 4 bounds the agent's payoff on the tree, and Step 5 bounds the surplus and verifies attainment. For a direct mechanism, the lower bound on the agent's payoff is a weighted combination of constraints that the severance closes at the last uninformed period (Step 2 of Theorem~\proofref{thm:main result}'s proof). A general contract may pay forever, so Step 4 runs the combination over the first $T$ periods and closes it with the contract's own values in period $T{+}1$. These closing terms are never negative, because at every node of the tree the obedient strategy is worth at least as much to a more optimistic agent (Lemma~\ref{lem:monotone} below).

If $\sup_{\zeta'}U_A(\mathcal G,\zeta')=\infty$, every optimal strategy gives $U_A=\infty$, which implies $U_P=-\infty$. Henceforth $U_A(\mathcal G,\zeta)=\sup_{\zeta'}U_A(\mathcal G,\zeta')<\infty$.

\emph{Step 1: purification.} This step shows that $\zeta$ may be taken pure. Realize the mixing at each private history through the inverse distribution function of an independent uniform draw, one for the effort and one for the message of each period. These draws suffice, since in any play the agent reaches at most one private history at which he chooses effort and one at which he sends a message in each period. The law under $\zeta$ is then the average over its pure realizations $\hat\zeta$. So is the agent's payoff, by Tonelli's theorem and the cost bound $c/p_L$, and so is the surplus, which is bounded. Every realization has $U_A(\mathcal G,\hat\zeta)\le\sup U_A<\infty$ and the average equals $\sup U_A$, so almost every realization is optimal, and at an optimal realization $U_P(\mathcal G,\hat\zeta)=S(\mathcal G,\hat\zeta)-\sup U_A$ is bounded. These values average to $U_P(\mathcal G,\zeta)$, so some optimal pure realization gives the principal at least $U_P(\mathcal G,\zeta)$.

\emph{Step 2: the announcement is the recommendation.} Call the \emph{record} the contractible history without the announcements, and let $m$ be the arrival period of the finding, equal to $\emptyset$ if it has not arrived. Since $\zeta$ is pure, let $\mathcal A_t$ list, for each arrival period $m\in\{\emptyset,1,\dots,t\}$, the period-$t$ effort and message that $\zeta$ prescribes to an agent with arrival period $m$ whose past messages are those of the record and whose past efforts are those that $\zeta$ prescribes to him. The list is defined for every record, takes finitely many values, is measurable in the draw and the record, and is known to the agent in $\mathcal G$ before his period-$t$ effort. Since redefining $\zeta$ at private histories reached with probability zero changes no payoff, $\zeta$ may be taken to play, at each private history, the effort and message that $\mathcal A_t$ lists for the agent's arrival period. Let $\mathcal G^{\mathcal A}$ announce $\mathcal A_t$ before effort, offer the messages listed in $\mathcal A_t$, and pay $\mathbb E[\tau_t\mid\mathcal A_1,\dots,\mathcal A_t,\text{record}]$. Fix a strategy of $\mathcal G$ whose period-$t$ play is measurable in $(\mathcal A_1,\dots,\mathcal A_t)$ and in the agent's own efforts, outcomes, finding and messages. The likelihood of the state, nature, these variables and the record then depends on the draw only through the path of the $\mathcal A$'s, so the two contracts induce the same joint law of the state, nature, these variables, the record and the path, and, given the path and the record, the draw is distributed as $\mathbb P$ conditioned on the positive-probability event that it generates the path. The expected transfers agree period by period by the tower property, and the two contracts yield the same $U_A$, $S$ and $U_P$. In particular, every conditional expected transfer at a path and record of positive probability is finite, since some strategy reaches them with positive probability and would otherwise earn more than $\sup U_A$. Every strategy of $\mathcal G^{\mathcal A}$ is such a strategy of $\mathcal G$, because the agent in $\mathcal G$ knows $\mathcal A_t$ before his period-$t$ effort and every listed message is available in $\mathcal G$. Hence $\sup U_A(\mathcal G^{\mathcal A},\cdot)\le\sup U_A(\mathcal G,\cdot)$, and $\zeta$ remains optimal at the same payoff to the principal. Without loss, then, the only announcement is the recommendation, and the transfers are deterministic functions of the record and the announcement path.

\emph{Step 3: one report, the tree, and the transfers.} The agent's exogenous private information is one stopping time, the arrival period of the finding, since the finding is conclusive, arrives at most once and only with effort, and a success ends the relationship. Let $\phi$ offer the messages $\{\text{``}F\text{''},\emptyset\}$ in every period before a report and $\{\emptyset\}$ after it. Before a report, $\phi$ computes the \emph{virtual messages} that $\zeta$ prescribes to an agent without a finding, and from the period $k$ of a report on, those that $\zeta$ prescribes to an agent whose finding arrived in period $k$. It draws the announcements and pays the transfers of $\mathcal G^{\mathcal A}$ at the \emph{virtual record}, the record with the virtual messages in place of the agent's messages. A report in a period whose announcement recommends shirking to an agent without a finding is treated as the message $\emptyset$, since no finding arrives there under the obedient strategy. Under the obedient strategy $(e^\ast,r^\ast)$, which reports a finding in its arrival period and follows the recommendations, the virtual record coincides with the record of $(\mathcal G^{\mathcal A},\zeta)$, so $(\phi,(e^\ast,r^\ast))$ and $(\mathcal G^{\mathcal A},\zeta)$ yield the same $U_A$, $S$ and $U_P$. Any strategy of $\phi$ is mirrored in $\mathcal G^{\mathcal A}$ by sending the virtual messages, which are listed in $\mathcal A_t$ and collect the same transfers, so the obedient strategy is optimal in $\phi$.

A \emph{node} $h$ is an announcement path without a report and with positive probability $\Prob(h)$. It has a period $t(h)$, a recommendation $\bar e(h)\in\{0,1\}$ to an agent without a finding, and \emph{children} $h'$, its extensions to period $t(h){+}1$ with positive probability. The conditional probabilities $\Prob(h'\mid h)$ of the children sum to one and do not depend on the agent's play, since before a report the virtual record is a function of the announcement path. It is a \emph{work node} if $\bar e(h)=1$ and a \emph{blank node} otherwise. Each period has finitely many nodes. A report at a work node $h$ opens the \emph{$h$-informed track}, which is unrestricted. Write the success and report payments at $h$ as $b_0(h)$ and $\ell_h$, and write the remaining payment at $h$, made with no success and no report and fixed before the next announcement, as the wage $w_0(h'):=\delta^{-1}\times(\text{remaining payment at }h)$ for each child $h'$, with $w_0(h):=0$ at the period-$1$ nodes. These definitions preserve every cumulative discounted payment, hence every value, and (LL) holds transfer by transfer. No strategy used below works at a blank node, so a success bonus there plays no role. It remains to bound $U_P(\phi,e^\ast,r^\ast)$.

\emph{Counts and scales.} Let the \emph{count} $n(h)$ be the number of work nodes on the path up to and including $h$, and let $d(h):=n(h)-\bar e(h)$, the number of work nodes strictly above it. Write $\Lambda_d:=\prod_{s<d}\nu_s$, so that $\mathcal D_t=\delta^{t-1}\Lambda_{t-1}$, and define the \emph{scale} $\varpi(h):=\Prob(h)\,\delta^{\,t(h)-n(h)}$. The obedient agent arrives at $h$ uninformed and unreported, with the belief $\mu_{d(h)}$ and the discounted probability $\mathrm{reach}(h):=\Prob(h)\,\delta^{\,t(h)-1}\Lambda_{d(h)}$, which equals $\varpi(h)\mathcal D_{n(h)}$ at a work node. Write $\mathbb E_h$ for the average over the children of $h$.

\emph{Values on the tree.} Values are discounted to the start of period $t(h)$ and include $w_0(h)$ where the agent collects it. Let $U^0_1:=U_A(\phi,e^\ast,r^\ast)$, and let $U^0(h)$ be the continuation payoff at $h$ of the obedient agent, so that $U^0_1=\sum_{t(h)=1}\Prob(h)U^0(h)$. At a work node $h$, the \emph{informed payoff} $Y(h)$ is $\ell_h$ plus the value to an informed agent of following the recommendations of the $h$-informed track. Let $A_\omega(h)$ be the value of the obedient strategy from $h$ on, conditional on $\omega$, to an agent uninformed at $h$, and set $A(h,\mu):=\mu A_H(h)+(1-\mu)A_L(h)$ and $\Delta^A(h):=A_H(h)-A_L(h)$. Then $U^0(h)=A(h,\mu_{d(h)})$. At a work node,
\begin{equation}\label{eq:oa A}
\begin{aligned}
A_H(h)&=w_0(h)-c+p_Hb_0(h)+q_H(1-p_H)Y(h)+\delta\gamma'\,\mathbb E_hA_H(h'),\\
A_L(h)&=w_0(h)-c+p_Lb_0(h)+\delta(1-p_L)\,\mathbb E_hA_L(h'),
\end{aligned}
\end{equation}
and $A_\omega(h)=w_0(h)+\delta\,\mathbb E_hA_\omega(h')$ at a blank node.

\emph{Step 4a: finiteness and two deviations.} This step shows that the values on the tree are finite and defines, for each work node, the two deviations that Step 4c combines. Each bound below compares the obedient strategy with a strategy of $\phi$ that plays like the obedient agent until an event of positive probability and then deviates, possibly by randomizing. By Step 3, no such strategy is worth more than $U^0_1$, which is finite.

First, $U^0(h)<\infty$ for every node $h$, since $U^0_1$ is $\mathrm{reach}(h)\,U^0(h)$ plus flows that sum to at least $-c/p_L$, and $\mathrm{reach}(h)>0$ because each $\nu_s$ is positive (Fact 1). For the same reason, $w_0(h)$ is finite for every node, and $b_0(h)$ and $Y(h)$ are finite for every work node. A strategy that plays like the obedient agent until it arrives at $h$ uninformed and unreported loses $\mathrm{reach}(h)$ times the excess of $U^0(h)$ over its continuation payoff at $h$, so no continuation payoff at $h$ exceeds $U^0(h)$. Likewise, a finding but no success at a work node $h$ has the discounted probability $\mathrm{reach}(h)Q_{d(h)}>0$, so no continuation payoff after such a finding, from the message at $h$ on, exceeds $Y(h)$. Under (LL), shirking forever without a report is worth at least $0$, so $U^0(h)\ge0$ for all nodes and $Y(h)\ge0$ for all work nodes. Put $C_H:=c/(1-\delta\gamma')$ and $C_L:=c/(1-\delta(1-p_L))$. Before a report, the flows of the obedient strategy from $h$ other than effort costs are nonnegative, and a report at a work node $h''$ yields the informed payoff $Y(h'')\ge0$. The obedient strategy from $h$ reaches its $i$-th work node unreported at least $i-1$ periods after $h$, with probability at most $\gamma'^{\,i-1}$ in state $H$ and $(1-p_L)^{i-1}$ in state $L$. Hence $A_H(h)\ge-C_H$ and $A_L(h)\ge-C_L$, and both are finite since $A(h,\mu_{d(h)})=U^0(h)$ is finite with $\mu_{d(h)}\in(0,1)$.

Now fix a work node $h$ with $j:=n(h)$. The \emph{shirker} of $h$ plays like the obedient agent until he arrives at $h$ uninformed and unreported, shirks at $h$, and then follows the obedient strategy at the belief $\mu_{j-1}$. His payoff falls short of $U^0_1$ by $\mathrm{reach}(h)\,G^S(h)$, where the \emph{effort obedience slack}
\[
G^S(h):=U^0(h)-w_0(h)-\delta\,\mathbb E_hA(h',\mu_{j-1})
\]
is nonnegative, since no continuation payoff at $h$ exceeds $U^0(h)$. An informed, unreported agent at a node $h'$ can play the \emph{mimic}, which works at work nodes, shirks at blank nodes, reports with probability $q_H$ after each worked period without a success, and after a report at $h''$ follows the $h''$-informed track, as in Step 2 of Proposition~\ref{prop:infinite}'s proof. The mimic succeeds, reports, or does neither with the state-$H$ probabilities of the obedient strategy. Since the conditional probabilities of the children do not depend on the agent's play and $Y(h'')$ depends on the contractible history alone, the mimic is worth $A_H(h')$. For a period $T\ge t(h)$, the \emph{concealer} of $h$ with horizon $T$ plays like the obedient agent until he obtains a finding but no success at $h$. He then delays his report and shirks at blank nodes until the first work node $h_1$ strictly below $h$. If $t(h_1)\le T$, he works at $h_1$ and, absent a success, reports and follows the $h_1$-informed track. Otherwise, or if no such node exists, he plays the mimic from the node of period $T{+}1$ that he reaches. His payoff falls short of $U^0_1$ by $\mathrm{reach}(h)Q_{j-1}\,G^C_T(h)$, where the \emph{no-concealing slack}
\[
G^C_T(h):=Y(h)-\sum_{\tilde h}\Prob(\tilde h\mid h)\,\delta^{\,t(\tilde h)-t(h)}\,x_T(\tilde h)
\]
is nonnegative, since no continuation payoff after a finding but no success at $h$ exceeds $Y(h)$. The sum runs over the nodes $\tilde h$ that he visits, those strictly below $h$ with $d(\tilde h)=j$ and $t(\tilde h)\le T+1$, and $\Prob(\tilde h\mid h):=\Prob(\tilde h)/\Prob(h)$. His term at $\tilde h$ is $x_T(\tilde h):=w_0(\tilde h)$ at a blank node and $x_T(\tilde h):=w_0(\tilde h)-c+p_Hb_0(\tilde h)+(1-p_H)Y(\tilde h)$ at a work node, both in the first $T$ periods, and $x_T(\tilde h):=A_H(\tilde h)$ in period $T{+}1$. For a direct mechanism and $T\ge n_0$, the two slacks are those of Step 1 of Theorem~\proofref{thm:main result}'s proof.

\emph{Step 4b: monotonicity.}

\begin{lemma}[Monotone continuation payoff on the tree]\label{lem:monotone}
For every node $h$, $\Delta^A(h)\ge0$.
\end{lemma}

At every work node, deterring shirking and a delayed report forces the slope $\Delta^A$ above a positive amount plus a fraction, smaller than $1-p_L$, of the average slope at the next work nodes. A negative slope could therefore persist only by growing geometrically along the tree, faster than the low-state probability of reaching deeper work nodes falls, which contradicts a finite payoff to the agent.

\emph{Proof}. \emph{(a) One node.} Fix a work node $h$, and let $\mu:=\mu_{d(h)}$, $A'_\omega:=\mathbb E_hA_\omega(h')$ and $\Delta':=A'_H-A'_L=\mathbb E_h\Delta^A(h')$. By Step 4a, these are finite, and the effort obedience slack $G^S(h)=U^0(h)-w_0(h)-\delta\big[\mu A'_H+(1-\mu)A'_L\big]$ and the no-concealing slack at the horizon $t(h)$, $G^M(h):=G^C_{t(h)}(h)=Y(h)-\delta A'_H$, are nonnegative. Now let $\Gamma(\mu):=1-p_Lp_H/P(\mu)$. The recursions \eqref{eq:oa A} at $h$ are those of Step 2 of Proposition~\ref{prop:infinite}'s proof with $A'_\omega$ in place of $A_\omega(t{+}1)$. Thus the derivation of \eqref{eq:app slope} there, with $G^S(h)$ and $G^M(h)$ in place of $G^S_t(\phi)$ and $G^M_t$, gives
\begin{equation}\label{eq:oa one node}
\Delta^A(h)=\frac{(p_H-p_L)\,c}{P(\mu)}+\delta\,\Gamma(\mu)\,\Delta'+\frac{p_H-p_L}{P(\mu)}\,G^S(h)+\frac{q_H(1-p_H)\,p_L}{P(\mu)}\,G^M(h).
\end{equation}
Since the first term of \eqref{eq:oa one node} is positive and both slacks are nonnegative, this identity implies
\begin{equation}\label{eq:oa slope}
\Delta^A(h)\ \ge\ \delta\,\Gamma(\mu_{d(h)})\,\mathbb E_h\Delta^A(h')\qquad\text{for every work node $h$.}
\end{equation}

\emph{(b) Blank stretches.} Fix a node $h$ and draw a branch below $h$ with the conditional probabilities $\Prob(h'\mid h)$. Write $\mathbb E_h[Z;B]$ for the expectation of $Z$ on the event $B$ under this law, which extends the average $\mathbb E_h$ over children, and let $\hat h$ be the first work node at or below $h$, if any. Before $\hat h$ the obedient strategy shirks, so in both states the agent collects the same wages, whose expectation is finite since $A_\omega(h)$ is finite and $A_\omega(\hat h)\ge-C_\omega$. The wages cancel from $\Delta^A$, so $\Delta^A(h)=\mathbb E_h\big[\delta^{\,t(\hat h)-t(h)}\,\Delta^A(\hat h);\ \hat h\text{ exists}\big]$. For a work node $h$, let $h_1$ be the first work node strictly below $h$. Applying this identity at each child in \eqref{eq:oa slope} gives
\begin{equation}\label{eq:oa skip}
\Delta^A(h)\ \ge\ \Gamma(\mu_{d(h)})\,\mathbb E_h\big[\delta^{\,t(h_1)-t(h)}\,\Delta^A(h_1);\ h_1\text{ exists}\big].
\end{equation}

\emph{(c) Growth.} Suppose that $\Delta^A(h)<0$ at a work node $h$. Let $h_0:=h$, and let $h_{i+1}$ be the first work node strictly below $h_i$ whenever $h_i$ exists. Each $h_i$ is reached at a stopping time of the branch, with $d(h_i)=d(h)+i$. In state $L$, given the branch, the obedient strategy from $h$ arrives at $h_i$ with probability $(1-p_L)^i$, and its transfers are nonnegative. The obedient strategy's expected discounted costs total at most $C_L$, and $A_L(h_i)^-$ is at most its expected discounted costs from $h_i$ on. Hence, for every $i\ge0$,
\begin{equation}\label{eq:oa growth}
A_L(h)\ \ge\ -C_L+(1-p_L)^i\,E_i,\qquad E_i:=\mathbb E_h\big[\delta^{\,t(h_i)-t(h)}A_L(h_i)^+;\ h_i\text{ exists}\big].
\end{equation}
The same argument in state $H$, where the obedient strategy arrives at $h_i$ unreported with probability $\gamma'^{\,i}$ and its flows other than costs are nonnegative, gives $\mathbb E_h[\delta^{\,t(h_i)-t(h)}A_H(h_i)^+;\,h_i\text{ exists}]\le\gamma'^{-i}(A_H(h)+C_H)$, so $\delta^{\,t(h_i)-t(h)}\Delta^A(h_i)$ is integrable for each $i$. Moreover, $0<\Gamma(\mu_d)\le\bar\Gamma:=\Gamma(\mu_0)$ for every $d\ge0$, because $p_Lp_H/P(\mu)\le p_H<1$ and $\Gamma$ increases in $\mu$ while $\mu_d\le\mu_0$. In addition, $\bar\Gamma-(1-p_L)=p_L(P_0-p_H)/P_0<0$. Iterating \eqref{eq:oa skip} $K$ times and averaging at each $h_i$ by the Markov property gives
\[
\Delta^A(h)\ \ge\ \Big(\prod_{i<K}\Gamma(\mu_{d(h)+i})\Big)\,\mathbb E_h\big[\delta^{\,t(h_K)-t(h)}\,\Delta^A(h_K);\ h_K\text{ exists}\big].
\]
Bound $\Delta^A(h_K)$ below by $-C_H-A_L(h_K)^+\le0$ before bounding the product by $\bar\Gamma^K$. This gives $\Delta^A(h)\ge-\bar\Gamma^K(C_H+E_K)$, and with \eqref{eq:oa growth} at $i=K$,
\[
A_L(h)\ \ge\ -C_L-(1-p_L)^KC_H+\big|\Delta^A(h)\big|\,\Big(\frac{1-p_L}{\bar\Gamma}\Big)^{K},
\]
whose right side grows without bound in $K$ since $\bar\Gamma<1-p_L$. This contradicts the finiteness of $A_L(h)$. Hence $\Delta^A\ge0$ at every work node, and at every node by part (b). $\hfill\square$

\emph{Step 4c: the combination through period $T$.} This step proves that the agent's payoff is at least the sum over the work nodes of the agent's scaled marginal payoffs $\varpi(h)\,\Delta R^\ast(n(h))$, the bound \eqref{eq:oa bound} below. To obtain it, I combine the slacks of Step 4a over the work nodes of the first $T$ periods into an exact identity, with the multipliers of Step 2 of Theorem~\proofref{thm:main result}'s proof times the scales. Lemma~\ref{lem:monotone} then signs the closing terms of this identity in period $T{+}1$.

The forward recursion of Step 2 of Theorem~\proofref{thm:main result}'s proof defines $\beta_j$, $\eta_j$, $\sigma^S_j$, $v_j$ and $\sigma^C_j$ for every count $j\ge1$ and does not depend on the duration. The signs ($\sigma^S_j>0$ and $\beta_j,\eta_j,v_j,\sigma^C_j\ge0$) and the cumulation \eqref{eq:app cumul} therefore hold for every count. So do the identities that cancel the bonuses and informed payoffs (the definition of $\beta_j$ and \eqref{eq:app XDalt}), the wage coefficients $W_j=\mathcal D_j+(1-\nu_{j-2})\eta_j+v_j=\delta^{j-1}$ and the agent's marginal payoffs \eqref{eq:app rstar}. By the definitions of $\eta_{j+1}$ and $v_{j+1}$ and the recursion $\mathcal D_{j+1}=\delta\nu_{j-1}\mathcal D_j$ of (Fact 4), $W_{j+1}=\delta\big[\mathcal D_j+(1-\nu_{j-1})\beta_j+\sigma^C_j\big]$, so
\begin{equation}\label{eq:oa key}
\mathcal D_j+(1-\nu_{j-1})\beta_j+\sigma^C_j=\delta^{j-1}\qquad(j\ge1).
\end{equation}

Fix $T\ge1$, and let $\mathcal H_T:=\{h:\,t(h)\le T\}$ and $\mathcal C_{T+1}:=\{h':\,t(h')=T+1\}$, both finite. Set $\lambda_0:=0$ and $\lambda_j:=\delta^{1-j}\big[(\mathcal D_j+\beta_j)(\mu_{j-1}-\mu_j)+\sigma^C_j(1-\mu_j)\big]$ for $j\ge1$, which is nonnegative by the signs above and the inequalities $1>\mu_{j-1}>\mu_j$ of (Fact 1). Then, identically,
\begin{equation}\label{eq:oa cut}
\begin{aligned}
&U^0_1-\sum_{h\in\mathcal H_T:\,\bar e(h)=1}\varpi(h)\big[\sigma^S_{n(h)}\,G^S(h)+\sigma^C_{n(h)}\,G^C_T(h)\big]\\
&\quad=\sum_{h\in\mathcal H_T:\,\bar e(h)=1}\varpi(h)\,\Delta R^\ast(n(h))+\sum_{h\in\mathcal H_T}\Prob(h)\,\delta^{\,t(h)-1}w_0(h)\\
&\qquad+\sum_{h'\in\mathcal C_{T+1}}\Prob(h')\,\delta^{T}\big[U^0(h')+\lambda_{d(h')}\,\Delta^A(h')\big].
\end{aligned}
\end{equation}
The first two sums on the right are what the identity \eqref{eq:app identity} of Step 2 of Theorem~\proofref{thm:main result}'s proof delivers on a finite horizon. The third sum closes the combination.

To verify \eqref{eq:oa cut}, unroll every payoff through period $T$, close it with the continuation payoffs in period $T{+}1$, and collect coefficients node by node. The loss of the shirker of $h$ is the obedient agent's terms at and below $h$, the wage at $h$ excepted, minus the shirker's own terms strictly below $h$. This loss enters the combination with the factor $\sigma^S_{n(h)}/\mathcal D_{n(h)}$. By \eqref{eq:app cumul}, a term common to the losses of the shirkers of counts $1,\dots,i$ carries the total factor $(\mathcal D_i+\beta_i)/\mathcal D_i$. Now fix a node $\tilde h$ of $\mathcal H_T\cup\mathcal C_{T+1}$ with $d:=d(\tilde h)\ge1$, and write $\bar\varpi(\tilde h):=\Prob(\tilde h)\,\delta^{\,t(\tilde h)-d}$. The shirker of each of the $d$ work nodes strictly above $\tilde h$ reaches $\tilde h$ at the belief $\mu_{d-1}$ with the same discounted probability $\Prob(\tilde h)\,\delta^{\,t(\tilde h)-1}\Lambda_{d-1}=\bar\varpi(\tilde h)\mathcal D_d$. Only the concealer of the last of these nodes, $h$, visits $\tilde h$, and by the definition of $\varpi$ his term $x_T(\tilde h)$ enters with the weight $\varpi(h)\sigma^C_d\Prob(\tilde h\mid h)\,\delta^{\,t(\tilde h)-t(h)}=\bar\varpi(\tilde h)\sigma^C_d$.

At a work node $\tilde h\in\mathcal H_T$ of count $j=d+1$, where $\bar\varpi(\tilde h)=\delta\varpi(\tilde h)$ and $\mathrm{reach}(\tilde h)=\varpi(\tilde h)\mathcal D_j$, the obedient flow $w_0(\tilde h)-c+P_{j-1}b_0(\tilde h)+Q_{j-1}Y(\tilde h)$ enters the losses of the shirkers of counts $1,\dots,j$, and the shirkers from above collect the same flow at the belief $\mu_{j-2}$. The node's own shirker and concealer add $\varpi(\tilde h)\sigma^S_j$ to the weight of $w_0(\tilde h)$ and $-\varpi(\tilde h)\sigma^C_j$ to that of $Y(\tilde h)$. By the definitions of $\eta_j$ and $v_j$, the obedient agent, the shirkers and the concealer from above carry the net weights $\varpi(\tilde h)$ times $-\beta_j$, $\eta_j$ and $v_j$. These are the weights of the period-$j$ transfers in the identity \eqref{eq:app identity}, scaled by $\varpi(\tilde h)$, also at count $1$, where none arrives from above and $\eta_1=v_1=0$. Hence, as there, $b_0(\tilde h)$ and $Y(\tilde h)$ cancel, $w_0(\tilde h)$ keeps $\varpi(\tilde h)W_j=\Prob(\tilde h)\,\delta^{\,t(\tilde h)-1}$, and the constants give $\varpi(\tilde h)\,\Delta R^\ast(j)$.

At a blank node $\tilde h\in\mathcal H_T$ and at $\tilde h\in\mathcal C_{T+1}$, the obedient agent arrives with $\mathrm{reach}(\tilde h)=\bar\varpi(\tilde h)\nu_{d-1}\mathcal D_d$, and his terms enter the losses of the shirkers of counts $1,\dots,d$. Thus the obedient agent, the shirkers and the concealer carry the net weights $\bar\varpi(\tilde h)$ times $-\nu_{d-1}\beta_d$, $\mathcal D_d+\beta_d$ and $\sigma^C_d$. These sum to $\bar\varpi(\tilde h)\delta^{d-1}=\Prob(\tilde h)\,\delta^{\,t(\tilde h)-1}$ by \eqref{eq:oa key}. At a blank node all three collect the wage alone. At $\tilde h\in\mathcal C_{T+1}$ they hold the values $A(\tilde h,\mu_d)=U^0(\tilde h)$, $A(\tilde h,\mu_{d-1})$ and $A_H(\tilde h)=A(\tilde h,1)$, and since $A(\tilde h,\cdot)$ is affine with the slope $\Delta^A(\tilde h)$, their terms sum to $\Prob(\tilde h)\,\delta^T\big[U^0(\tilde h)+\lambda_d\Delta^A(\tilde h)\big]$. At nodes of either kind with $d(\tilde h)=0$, only the obedient agent arrives, with $\mathrm{reach}(\tilde h)=\Prob(\tilde h)\,\delta^{\,t(\tilde h)-1}$. This proves \eqref{eq:oa cut}.

Now sign \eqref{eq:oa cut}. On the left, every slack and every multiplier is nonnegative, so the left side is at most $U^0_1$. On the right, the wages are nonnegative by (LL), $U^0(h')\ge0$ by Step 4a, and $\lambda_{d(h')}\Delta^A(h')\ge0$ by Lemma~\ref{lem:monotone}. So \eqref{eq:oa cut} implies $U^0_1\ge\sum_{h\in\mathcal H_T:\,\bar e(h)=1}\varpi(h)\,\Delta R^\ast(n(h))$ for every $T\ge1$. These terms are nonnegative by \eqref{eq:app rstar}, so letting $T$ grow gives, by monotone convergence,
\begin{equation}\label{eq:oa bound}
U^0_1\ \ge\ \sum_{h:\,\bar e(h)=1}\varpi(h)\,\Delta R^\ast(n(h)).
\end{equation}

\emph{Step 5: the surplus bound and the assembly.} Blank nodes generate no surplus under the obedient strategy. At a work node $h$ with $j:=n(h)$, the surplus under the obedient strategy is $\mathrm{reach}(h)\big[P_{j-1}\pi-c+Q_{j-1}S^1(h)\big]$, where $S^1(h)$ is the surplus of the $h$-informed track when its recommendations are followed, discounted to period $t(h)$. The $i$-th worked period of any effort rule on the track falls at least $i$ periods after $t(h)$ and is reached with probability at most $(1-p_H)^{i-1}$, and $p_H\pi-c>0$ by Assumption~\ref{ass:efficiency}, so $S^1(h)\le V^{\mathrm{informed}}$. Hence $S(\phi,e^\ast,r^\ast)\le\sum_{\bar e(h)=1}\varpi(h)\mathcal D_{n(h)}\Pi_{n(h)}$. This series converges absolutely. Work nodes of one count $j$ never share a branch, so their probabilities sum to at most one, and each has $\varpi(h)\mathcal D_j=\mathrm{reach}(h)\le\Prob(h)(1-p_L)^{j-1}$, since $\nu(\mu)\le1-p_L$. So $\sum_{\bar e(h)=1}\varpi(h)\mathcal D_{n(h)}\le1/p_L$, while $\Pi_j$ is bounded. The right side of \eqref{eq:oa bound} is at most $U^0_1<\infty$, and subtracting \eqref{eq:oa bound} gives
\[
U_P(\phi,e^\ast,r^\ast)\ \le\ \sum_{h:\,\bar e(h)=1}\varpi(h)\,u_{n(h)},\qquad u_j:=\mathcal D_j\Pi_j-\Delta R^\ast(j),
\]
an absolutely convergent series. Draw a branch with the probabilities $\Prob$, let $t_j$ be the period of its $j$-th work node, and let $\psi_j:=\delta^{\,t_j-j}$, with $\psi_j:=0$ if the branch has fewer than $j$ work nodes. A work node $h$ has $\varpi(h)=\Prob(h)\psi_{n(h)}$ on every branch through it, so by Fubini's theorem the series equals $\mathbb E\big[\sum_j\psi_ju_j\big]$. On every branch $\psi_j$ is nonincreasing, since $t_{j+1}\ge t_j+1$, and $\psi_1\le1$. By \eqref{eq:app rstar}, $\sum_{i\le j}u_i=S(j)-R^\ast(j)$. By Step 5 of Theorem~\proofref{thm:main result}'s proof, the maximum of $S(j)-R^\ast(j)$ over all finite $j$ is $S(n_0^\ast)-R^\ast(n_0^\ast)$, which is nonnegative since $S(0)=R^\ast(0)=0$. Summation by parts gives, on every branch and for every $N$,
\[
\begin{aligned}
\sum_{j\le N}\psi_ju_j&=\sum_{1\le j<N}(\psi_j-\psi_{j+1})\big[S(j)-R^\ast(j)\big]+\psi_N\big[S(N)-R^\ast(N)\big]\\
&\le\ \psi_1\big[S(n_0^\ast)-R^\ast(n_0^\ast)\big]\ \le\ S(n_0^\ast)-R^\ast(n_0^\ast).
\end{aligned}
\]
Letting $N$ grow on almost every branch and taking expectations gives $U_P(\phi,e^\ast,r^\ast)\le S(n_0^\ast)-R^\ast(n_0^\ast)$, which is \eqref{eq:oa direct} by Steps 1 to 3.

For attainment, the PBS contract $\phi^\ast(n_0^\ast)$ is a general contract, with trivial announcements, the message space $\{\text{``}F\text{''},\emptyset\}$ before a report in the periods $t\le n_0^\ast$, the message space $\{\emptyset\}$ otherwise, and deterministic, bounded transfers. Relative to the direct mechanism, the class adds only the option to work after period $n_0^\ast$ without a report. Such effort costs $c$, pays nothing and changes no later transfer or message space, so every strategy of the class does weakly worse than the strategy of the direct mechanism that shirks in these periods instead. The obedient strategy is optimal by Step 3 of Theorem~\proofref{thm:main result}'s proof, and the principal's payoff under it is $S(n_0^\ast)-R^\ast(n_0^\ast)$ by Steps 3 and 5 there, trivially so at $n_0^\ast=0$. By \eqref{eq:oa direct}, no other optimal strategy gives the principal more. $\hfill\square$

\subsection{Conclusive Success}\label{app:conclusive}

This section shows that the main characterizations extend to a conclusive success, $p_L=0$, with the other assumptions of Section~\ref{sec:model} unchanged. A worked period in state $L$ then never ends the relationship. When $\delta<1$, effort costs still sum to at most $c/(1-\delta)$, and every payoff is well defined as in Section~\ref{sec:model}. At $\delta=1$, effort costs may sum to $+\infty$, and a payoff sum may take the form $\infty-\infty$. There, a sum that converges absolutely has its usual value, and any other sum is read as the limit superior of its partial sums through period $T$ as $T$ grows. Shirking forever remains available and is worth at least zero under (LL), so every feasible contract still gives the agent a nonnegative payoff.

\begin{proposition}\label{prop:conclusive}
Let $p_L=0$. Then Theorems~\ref{thm:cf} and~\ref{thm:main result} hold. Moreover, $R^\ast(n_0)=R^{CF}(n_0)$ for every finite $n_0$, so $n_0^\ast=n_0^{CF}$, the principal obtains the same payoff whether the finding is private or contractible, and $\phi^\ast(n_0^\ast)$ is optimal when the finding is contractible as well.
\end{proposition}

\emph{Proof}. \emph{Step 1: what carries over.} At $p_L=0$ the beliefs stay interior, and $P_t=p_H\mu_t$, $Q_t$ and $\nu_t$ keep the properties (Fact 1), (Fact 2) and (Fact 4). The first best is unchanged, with $\Pi_t$ decreasing strictly to $-c$. (Fact 3) and Lemma~\ref{lem:shirker} hold with equality, that is, $Q(\mu)/P(\mu)=q_H(1-p_H)/p_H$ for every belief $\mu>0$. The proofs of the two theorems use them only in weak form: through the coefficient $p^Y_t\ge0$ in Step 2 of Theorem~\proofref{thm:cf}'s proof, and through the multiplier $\sigma^C_t\ge0$ in Step 2 of Theorem~\proofref{thm:main result}'s proof. Lemma~\ref{lem:concealer} holds at every belief $\mu>0$, its bracket being $p_H(1-q_H)\mu$, and the proofs apply both lemmas only at the interior beliefs $\mu_t$. It follows that every step at a finite uninformed duration carries over verbatim. These are Steps 1 to 5 of Theorem~\proofref{thm:cf}'s proof and Steps 1 to 5 of Theorem~\proofref{thm:main result}'s proof, including the strict monotonicity in part (i) and the strict inequality in part (ii) of Theorem~\ref{thm:main result}. The two strict results rest on $F_t>\delta F_{t+1}$, Lemma~\ref{lem:concealer} and the positivity of $(1-p_L)-(1-p_H)(1-q_Hp_L)$ in Step~4 of Theorem~\proofref{thm:main result}'s proof (part (i)), the last reducing to $p_H>0$ at $p_L=0$. The exceptions are the strict signs $p^Y_t>0$ in Step 2 of Theorem~\proofref{thm:cf}'s proof and $\sigma^C_t>0$ in Step 2 of Theorem~\proofref{thm:main result}'s proof, with the third by-product of Step 3 there, and the proofs of the two theorems use neither. Since the ratio $Q(\mu)/P(\mu)$ is then constant, every $p^Y_t$ and the first term of every $\sigma^C_t$ vanish. Induction on $t$ from $v_2=\delta\,\sigma^C_1=0$ then shows that every $v_t$ and every $\sigma^C_t$ is zero, that is, the multipliers on no-concealing all vanish. Apart from these signs, only the two steps of Proposition~\ref{prop:infinite}'s proof, which rule out an infinite uninformed duration, use $p_L>0$. Steps 2 and 3 treat them at $p_L=0$.

\emph{Step 2: an infinite uninformed duration at $\delta=1$.} Here $\mathcal D_t=\mu_0\gamma'^{\,t-1}+1-\mu_0$ decreases to $1-\mu_0>0$, while $\mathcal D_tP_{t-1}=p_H\mu_0\gamma'^{\,t-1}$ and $\mathcal D_tQ_{t-1}=q_H(1-p_H)\mu_0\gamma'^{\,t-1}$ vanish geometrically. On a schedule with $n_0=\infty$, the obedient effort cost $c\sum_t\mathcal D_t$ is therefore infinite, while the expected revenue and the surplus of the informed tracks are finite. Every feasible contract gives the agent a nonnegative payoff, so its transfers sum to $+\infty$, and the principal's payoff, the revenue less the transfers, is $-\infty$. The uninformed duration $n_0=0$ therefore dominates the schedule.

\emph{Step 3: an infinite uninformed duration at $\delta<1$.} In Step 1 of Proposition~\ref{prop:infinite}'s proof, everything up to the bound \eqref{eq:app cf close} applies as written. At $w_0\equiv0$, which that step imposes without loss, the obedient strategy collects nothing in state $L$, so $A_L(t)=-c/(1-\delta)<0$, and $U^0_t\ge0$ implies $\mu_{t-1}A_H(t)=U^0_t-(1-\mu_{t-1})A_L(t)>0$. Hence $\Delta^A(T{+}1)>0$ and $E_T>0$ for every $T$, which replaces the argument by contradiction there. The assembly then goes through unchanged. Step 2 of that proof applies as written, since $p_L>0$ enters it only through the rate $\delta(1-p_L)<1$ at which the series of $A_L$ converges, and this rate is $\delta<1$ at $p_L=0$.

\emph{Step 4: the payoff comparison.} I show that $R^\ast(n)=R^{CF}(n)$ for every finite $n$. By Step 1, every $v_t$ is zero at $p_L=0$. The recursion of Step 2 of Theorem~\proofref{thm:main result}'s proof and the relations of Step 2 of Theorem~\proofref{thm:cf}'s proof, compared in Step 5 of Theorem~\proofref{thm:main result}'s proof, differ only by the term $p_Hv_t/P_{t-1}$ in $\beta_t$, so they give the same sequences, with $\eta_t=\eta^{CF}_t$, and the display of $\Delta R^\ast(t)-\Delta R^{CF}(t)$ in that step gives $\Delta R^\ast(t)=\Delta R^{CF}(t)$ at every $t$. Summing from $R^\ast(0)=R^{CF}(0)=0$ gives the claim. Hence $S(n)-R^\ast(n)=S(n)-R^{CF}(n)$ for all $n$, the two smallest maximizers coincide, $n_0^\ast=n_0^{CF}$, and the principal's optimal payoffs are equal. By the argument of Step 5 of Proposition~\ref{prop:cf all}'s proof, when the finding is contractible, $\phi^\ast(n_0^\ast)$ is feasible and gives the principal the same payoff, so it is also optimal. $\hfill\square$

\subsection{Proof of Proposition \ref*{prop:almost conclusive}}\label{app:proof almost conclusive}

Let $U_P^{\mathrm{unr}}(\varepsilon)$ and $U_P^{\mathrm{PBS}}(\varepsilon)$ be the principal's optimal payoffs with and without reports, defined precisely below, and write $\Delta U_P(\varepsilon):=U_P^{\mathrm{unr}}(\varepsilon)-U_P^{\mathrm{PBS}}(\varepsilon)$ for her loss from posting a single bonus sequence. The proof bounds both payoffs against the principal's optimal payoff at $\varepsilon=0$, which is $U_P^\ast:=S(n_0^\ast)-R^\ast(n_0^\ast)=\max_n\,[S(n)-R^\ast(n)]$ by Step 5 of Theorem~\proofref{thm:main result}'s proof. Part 1 shows that allowing reports gains the principal at most a multiple of $\varepsilon$, in the sense that there are $\varepsilon_1\in(0,1)$ and $\bar C>0$ such that
\begin{equation}\label{eq:oa ac upper}
U_P^{\mathrm{unr}}(\varepsilon)\ \le\ U_P^\ast+\bar C\,\varepsilon\qquad\text{for every }\varepsilon\in[0,\varepsilon_1].
\end{equation}
Part 2 shows that posting a single bonus sequence loses at most a multiple of $\varepsilon^\alpha$, for the exponent $\alpha\in(\tfrac12,1)$ of Step 6 there, in the sense that there are $\varepsilon_2\in(0,1)$ and $\underline C>0$ such that
\begin{equation}\label{eq:oa ac lower}
U_P^{\mathrm{PBS}}(\varepsilon)\ \ge\ U_P^\ast-\underline C\,\varepsilon^\alpha\qquad\text{for every }\varepsilon\in[0,\varepsilon_2].
\end{equation}
As $\varepsilon\le\varepsilon^\alpha$ on $[0,1]$, the proposition then holds with $\hat\varepsilon:=\min\{\varepsilon_1,\varepsilon_2\}$ and $C:=\bar C+\underline C$. The notation of Appendices~\ref{app:preliminaries} and~\ref{app:proof main result} is maintained.

\emph{The perturbed environment.} For $\varepsilon\ge0$, state $L$ yields a finding with probability $q_L:=\varepsilon$ in a worked period, and everything else in Section~\ref{sec:model} is kept. In particular, contracts are the direct mechanisms \eqref{eq:contract}, since Lemma~\ref{lem:direct} is proved only for a conclusive finding. In a worked period, an uninformed agent with belief $\mu$ perceives a success with probability $P(\mu)$, a finding but no success with probability $Q_\varepsilon(\mu):=\mu q_H(1-p_H)+(1-\mu)\varepsilon(1-p_L)$, and silence with probability $\nu_\varepsilon(\mu):=1-P(\mu)-Q_\varepsilon(\mu)$. Note that the success probability does not involve $\varepsilon$. By Bayes' rule, silence multiplies his likelihood ratio by $\rho_\varepsilon:=\rho/(1-\varepsilon)$, and a finding multiplies it by $I_\varepsilon:=q_H(1-p_H)/[\varepsilon(1-p_L)]$, read as $+\infty$ at $\varepsilon=0$. Since no second finding can arrive, each later worked period without a success multiplies the likelihood ratio of an informed agent by $\rho_2:=(1-p_H)/(1-p_L)\in(0,1)$. Write $\theta^\varepsilon_d:=\theta_0\rho_\varepsilon^{\,d}$ and $\mu^\varepsilon_d:=\theta^\varepsilon_d/(1+\theta^\varepsilon_d)$ for the likelihood ratio and the belief of an uninformed agent after $d$ silent worked periods, and $P^\varepsilon_d:=P(\mu^\varepsilon_d)$, $Q^\varepsilon_d:=Q_\varepsilon(\mu^\varepsilon_d)$, $\nu^\varepsilon_d:=\nu_\varepsilon(\mu^\varepsilon_d)$ and $\mathcal D^\varepsilon_t:=\delta^{t-1}\prod_{s\le t-2}\nu^\varepsilon_s$. At $\varepsilon=0$ these are the objects of Appendix~\ref{app:preliminaries}. Transfers are nonnegative and each worked period ends the relationship with probability at least $p_L$, so every strategy of the agent is worth at least $-c/p_L$ conditional on either state, discounted to its first period. A contract $\phi$ is \emph{$\varepsilon$-feasible} if it satisfies (LL) and (IC) in this environment. At such a contract, $U^0_t$ is the obedient agent's continuation payoff from the uninformed period $t$, and $S_\varepsilon(\phi)$ is the total surplus. Then $U_P^{\mathrm{unr}}(\varepsilon)$ is the supremum of $S_\varepsilon(\phi)-U^0_1$ over all schedules and all $\varepsilon$-feasible contracts. Under a posted bonus sequence $b\ge0$, bounded as the model requires of an infinite track, the agent only chooses when to work. The optimal payoff $U_P^{\mathrm{PBS}}(\varepsilon)$ is the supremum of the principal's payoff over the sequences and the best responses to them that \emph{follow} a schedule, working in exactly the periods that the schedule recommends given the period of the finding. Such a best response is obedient, at the same payoffs, in the direct mechanism that, at the schedule this best response follows, pays only $b(t)$ upon a success in each recommended period $t$. Indeed, every strategy in this mechanism is worth at most what its own effort rule is worth under the posted bonus sequence, with equality for the best response, which works only in recommended periods. Hence $U_P^{\mathrm{PBS}}(\varepsilon)\le U_P^{\mathrm{unr}}(\varepsilon)$, that is, $\Delta U_P(\varepsilon)\ge0$. Every constant $C_i$ and threshold $\varepsilon_i$ below depends only on the primitives $(c,p_H,p_L,q_H,\mu_0,\delta,\pi)$, and so do $n_0^\ast$, $\bar N:=n_0^{FB}$ and the objects of Theorem~\proofref{thm:main result}'s proof at a given duration.

\emph{Part 1: reports gain at most a multiple of $\varepsilon$.} The proof applies the weak-duality identity of Step 2 of Theorem~\proofref{thm:main result}'s proof to an $\varepsilon$-feasible contract. In the perturbed environment, every probability moves by a multiple of $\varepsilon$, and a deviator who reports evaluates the report at his own belief, since an informed agent no longer knows the state. Once feasibility bounds the transfers (Step 1), the change in the surplus is a multiple of $\varepsilon$ (Step 2), and so are both effects on the combination (Step 4). Step 3 ranks the continuation payoffs so that a long uninformed track can be truncated. Throughout, $\varepsilon\le(1-\rho)/2$ and $\varepsilon\le q_H(1-p_H)/(1-p_L)$, so that $\rho_\varepsilon<1\le I_\varepsilon$, and $\varepsilon_1$ is the least of these two thresholds and the one of Step 3.

Fix an $\varepsilon$-feasible contract $\phi$ at a schedule $(n_0,n_1)$, with $n_0=\infty$ and finite or infinite informed durations allowed. For a report period $s\le n_0$, let $Y^H_s$ and $Y^L_s$ be the values, conditional on each state and discounted to period $s$, of reporting in period $s$ and then working every recommended period of the $s$-informed track, the lump sum included. Since this strategy is fixed, an agent with belief $\mu$ values a period-$s$ report at the \emph{report value} $\mu Y^H_s+(1-\mu)Y^L_s$. The informed payoff $Y_s$ is the report value at the belief $\mu^{\mathrm a}_s$ of an agent informed in period $s$ on path, whose likelihood ratio is $\theta^\varepsilon_{s-1}I_\varepsilon$. The values $A_H(t)$ and $A_L(t)$ of the obedient strategy from the uninformed period $t\le n_0+1$, conditional on each state, and $A(t,\mu)$ and $\Delta^A(t)$, are those of Step 2 of Proposition~\ref{prop:infinite}'s proof, now with the recursions
\[
\begin{aligned}
A_H(t)&=w_0(t)-c+p_Hb_0(t)+q_H(1-p_H)Y^H_t+\delta\gamma'A_H(t{+}1),\\
A_L(t)&=w_0(t)-c+p_Lb_0(t)+\varepsilon(1-p_L)Y^L_t+\delta(1-\varepsilon)(1-p_L)A_L(t{+}1),
\end{aligned}
\]
read as series if $n_0=\infty$ and ended by $A_H(n_0{+}1)=A_L(n_0{+}1)=w_0(n_0{+}1)$ otherwise, so that $U^0_t=A(t,\mu^\varepsilon_{t-1})$.

\emph{Step 1: bounds.} The residuals of Steps 3 and 4 are multiples of $\varepsilon$ times transfers or report values, so I first bound the transfers and report values discounted to period $1$. A contract with $U^0_1>\pi$ gives the principal a negative payoff, since the surplus never exceeds $\pi$, whereas $U_P^\ast\ge S(0)-R^\ast(0)=0$. So let $U^0_1\le\pi$. By (IC), $U^0_1$ is at least the value of each of three strategies, all of which collect only nonnegative transfers in the periods not mentioned. The first never works and never reports, so it collects every uninformed wage. The second works only in the uninformed period $t$, at the prior belief, and never reports. The third works in no uninformed period, reports ``$F$'' in a period $s\le n_0$, and then works in at most one period $r$ of the $s$-informed track. Hence, with $M_b:=(\pi+c)/p_L$,
\[
\sum_t\delta^{t-1}w_0(t)\le\pi,\qquad \delta^{t-1}b_0(t)\le M_b,\qquad \delta^{s-1}\ell_s+\sum_r\delta^{r-1}w_{1s}(r)\le\pi,\qquad \delta^{r-1}b_{1s}(r)\le M_b,
\]
the severances included among the wages. The strategies that define $Y^\omega_s$ and $A_\omega(t)$ collect wages and at most one bonus or one report value, since a success ends the relationship and a report ends the uninformed track. Summing their flows and using the lower bound $-c/p_L$ on the value of every strategy therefore gives, for all $s\le n_0$, $t\le n_0+1$ and $\omega\in\{H,L\}$,
\begin{equation}\label{eq:oa ac bounds}
\big|\delta^{s-1}Y^\omega_s\big|\ \le\ M_Y:=\pi+M_b+c/p_L,\qquad \big|\delta^{t-1}A_\omega(t)\big|\ \le\ M_A:=\pi+M_Y .
\end{equation}
One more bound is used twice. An agent informed in a period $t<n_0$ can conceal the finding and play the mimic of Step 2 of Proposition~\ref{prop:infinite}'s proof. Conditional on $\omega=H$, the mimic still reproduces the law of the obedient strategy, but conditional on $\omega=L$ it does not. The mimic is therefore worth $\mu^{\mathrm a}_tA_H(t{+}1)+(1-\mu^{\mathrm a}_t)A^{\mathrm m}_L$ from period $t{+}1$, where $A^{\mathrm m}_L\ge-c/p_L$ is its value conditional on $\omega=L$. As in \eqref{eq:app conceallow}, (IC) implies
\begin{equation}\label{eq:oa ac mimic}
Y_t\ \ge\ \delta\big[\mu^{\mathrm a}_tA_H(t{+}1)+(1-\mu^{\mathrm a}_t)A^{\mathrm m}_L\big]\qquad(t<n_0).
\end{equation}
Finally, to control the probabilities, fix a horizon $N_1:=\bar N+1+\min\{k\ge1:\delta^{\bar N}(p_H-p_L)\,c\,(1-p_L)^{-k}>4p_HM_A\}$, chosen so that Step 3 can rank the continuation payoffs. Since $\theta^\varepsilon_d=\theta_d(1-\varepsilon)^{-d}$ with $d\le N_1$ and $\varepsilon\le1/2$, the mean value theorem gives the first part of the display below for every $d\le N_1$ and $t\le N_1+1$,
\begin{equation}\label{eq:oa ac drift}
|P^\varepsilon_d-P_d|,\ \ |Q^\varepsilon_d-Q_d|,\ \ |\nu^\varepsilon_d-\nu_d|,\ \ \delta^{1-t}\,|\mathcal D^\varepsilon_t-\mathcal D_t|\ \le\ C_1\varepsilon,
\qquad
1-\mu\ \le\ \frac1\theta\ \le\ C_1\varepsilon.
\end{equation}
The second part holds for the likelihood ratio $\theta$ and the belief $\mu$ of any agent who has obtained a finding no later than period $N_1$, on path or after shirking an earlier period, both upon the finding and after one further worked period without a success. Indeed, $\theta^\varepsilon_d\ge\theta_d$ makes this likelihood ratio at least $\theta_{N_1}\rho_2I_\varepsilon$.

\emph{Step 2: surplus.} To bound the surplus by its conclusive counterpart, couple the environments at $\varepsilon$ and at $0$ by a common state and one common uniform draw per period. This draw determines the outcome of a worked period so that the two environments differ only in that, at $\varepsilon$, a finding can arrive in state $L$. The realized paths under the obedient strategy then differ only after a finding in state $L$, which has probability at most $\varepsilon/p_L$, since the agent works at most $1/p_L$ periods in expectation in state $L$. From that period on, the continuation surplus of either path lies in $[-c/p_L,0]$ conditional on $\omega=L$, as $p_L\pi<c$. Hence $S_\varepsilon(\phi)\le S(n_0,n_1)+C_2\varepsilon$ with $C_2:=c/p_L^2$. Grouping by the period of the finding, as in Appendix~\ref{app:preliminaries}, gives $S(n_0,n_1)\le S(n_0)$ for finite $n_0$, since each informed track yields at most $V^{\mathrm{informed}}$. By Step 1 of Proposition~\proofref{prop:frictionless}'s proof, $S(n_0,n_1)\le S(\bar N)$ always. Therefore
\begin{equation}\label{eq:oa ac surplus}
S_\varepsilon(\phi)\ \le\ S(n)+C_2\,\varepsilon,
\qquad\text{where }n:=n_0\text{ if }n_0\le N_1\text{ and }n:=\bar N\text{ if }n_0>N_1 .
\end{equation}

\emph{Step 3: ranking.} To truncate a long uninformed track at $\bar N$, I show that $\Delta^A(\bar N{+}1)\ge0$ whenever $n_0>N_1$. Fix a period $t$ with $\bar N<t<N_1$. As in \eqref{eq:app infpostpone}, the shirker who follows $e^t_{\text{Shirk}}$ plays the obedient strategy at the belief $\mu^\varepsilon_{t-1}$ from period $t{+}1$, so effort obedience implies $G^S_t(\phi):=U^0_t-w_0(t)-\delta A(t{+}1,\mu^\varepsilon_{t-1})\ge0$. Write $G^\omega_t:=Y^\omega_t-\delta A_\omega(t{+}1)$ for $\omega\in\{H,L\}$. Eliminating $b_0(t)-\delta A_L(t{+}1)$ between the two recursions and $G^S_t(\phi)$, as in Step 2 of Proposition~\ref{prop:infinite}'s proof, gives, with $P:=P^\varepsilon_{t-1}$,
\begin{equation}\label{eq:oa ac slope}
\begin{aligned}
\Delta^A(t)={}&\frac{(p_H-p_L)\,c}{P}+\delta\Big(1-\frac{p_Lp_H}{P}\Big)\Delta^A(t{+}1)\\
&+\frac{(p_H-p_L)\,G^S_t(\phi)+p_Lq_H(1-p_H)\,G^H_t-\varepsilon p_H(1-p_L)\,G^L_t}{P},
\end{aligned}
\end{equation}
which is \eqref{eq:app slope} at $\varepsilon=0$, where $G^H_t=G^M_t$. Since $Y_t=\mu^{\mathrm a}_tY^H_t+(1-\mu^{\mathrm a}_t)Y^L_t$, the mimic bound \eqref{eq:oa ac mimic} can be written as $G^H_t\ge-(Y^L_t-\delta A^{\mathrm m}_L)/\theta$ at the likelihood ratio $\theta:=\theta^\varepsilon_{t-1}I_\varepsilon$. Thus $\delta^{t-1}G^H_t\ge-2M_YC_1\varepsilon$ by \eqref{eq:oa ac bounds} and \eqref{eq:oa ac drift}, while $\delta^{t-1}|G^L_t|\le M_Y+M_A$. Write $u_t:=\delta^{t-1}\Delta^A(t)$ and $h_t:=1-p_Lp_H/P^\varepsilon_{t-1}$, which lies in $[1-p_H,1-p_L]$. Multiplying \eqref{eq:oa ac slope} by $\delta^{t-1}$ and dropping $G^S_t(\phi)\ge0$ gives $u_t\ge\delta^{t-1}(p_H-p_L)c/p_H+h_tu_{t+1}-C_3\varepsilon$, and hence, at $\varepsilon\le\delta^{N_1}(p_H-p_L)c/(2p_HC_3)$, the third threshold of $\varepsilon_1$,
\[
u_t\ \ge\ \delta^{t-1}\,\frac{(p_H-p_L)\,c}{2p_H}+h_t\,u_{t+1}\qquad(\bar N<t<N_1).
\]
Suppose now that $u_{\bar N+1}<0$. Whenever $u_t<0$, the display implies $h_tu_{t+1}<u_t$, so $u_{t+1}<u_t/(1-p_L)<0$, and at $t=\bar N+1$ it gives $u_{\bar N+2}<-\delta^{\bar N}(p_H-p_L)c/[2p_H(1-p_L)]$. By induction, $u_{N_1}<-\delta^{\bar N}(p_H-p_L)c\,(1-p_L)^{\bar N+1-N_1}/(2p_H)<-2M_A$ by the choice of $N_1$, whereas $|u_{N_1}|\le2M_A$ by \eqref{eq:oa ac bounds}. This contradiction shows that $\Delta^A(\bar N{+}1)\ge0$.

\emph{Step 4: the combination.} To bound the agent's payoff from below, I show that, at the duration $n$ of \eqref{eq:oa ac surplus},
\begin{equation}\label{eq:oa ac R}
U^0_1\ \ge\ R^\ast(n)-C_5\,\varepsilon .
\end{equation}
If $n=0$, which happens when $n_0=0$, or when $n_0>N_1$ and $\bar N=0$, the bound $U^0_1\ge0=R^\ast(0)$ holds by (IC), so let $n>0$. The identity \eqref{eq:app identity} of Step 2 of Theorem~\proofref{thm:main result}'s proof is linear in the numbers $b_0(s)$, $w_0(s)$ and $Y_s$, and its derivation collects their coefficients one by one, so it holds for any real values of these numbers. I apply the identity at the duration $n$, with the bonuses and wages of $\phi$ in the periods $s\le n$ and the informed payoffs $Y_s$ of $\phi$. In the slot of the severance $w_0(n{+}1)$, I use the severance of $\phi$ if $n=n_0$ and $U^0_{n+1}$ if $n<n_0$, where $U^0_{n+1}\ge0$ by (IC), as in Step 2 of Proposition~\ref{prop:infinite}'s proof. Write $\tilde U^0_1$, $\tilde G^S_t$ and $\tilde G^C_m$ for the obedient agent's payoff and the slacks of the identity at these numbers. These expressions carry the conclusive coefficients $P_d$, $Q_d$, $\mathcal D_t$ and $p_H$. I claim that, for all $t,m\le n$,
\begin{equation}\label{eq:oa ac tilde}
\big|U^0_1-\tilde U^0_1\big|\ \le\ C_4\varepsilon,\qquad \mathcal D_t\,\tilde G^S_t\ \ge\ -2C_4\varepsilon,\qquad \delta^{m-1}\,\tilde G^C_m\ \ge\ -C_4\varepsilon .
\end{equation}
Given \eqref{eq:oa ac tilde}, the identity implies \eqref{eq:oa ac R}, since its constant $\kappa$ is $R^\ast(n)$ by \eqref{eq:app rstar}, its multipliers $\sigma^S_t$ and $\sigma^C_m$ are nonnegative, and so are its wage terms. The constant $C_5$ collects $C_4$ and the sums of $\sigma^S_t/\mathcal D_t$ and $\delta^{1-m}\sigma^C_m$ over $t,m\le N_1$, which depend only on the primitives because the multipliers are those of the conclusive environment.

To prove \eqref{eq:oa ac tilde}, I compare each expression with the value of the corresponding strategy in $\phi$. By the derivations of $G^S_t$ in Step 1 of Theorem~\proofref{thm:cf}'s proof and of $G^C_m$ in Step 1 of Theorem~\proofref{thm:main result}'s proof, $\tilde U^0_1$ and $\tilde U^0_1-\mathcal D_t\tilde G^S_t$ are the values of the obedient strategy and of $e^t_{\text{Shirk}}$, and, at $m<n$, $\tilde G^C_m$ is the no-concealing slack against $r^m_{\text{Conceal}}$, all computed with the conclusive coefficients. The values of the same strategies in $\phi$ carry the $\varepsilon$-coefficients instead, the concealer succeeds with probability $P(\mu^{\mathrm a}_m)$ instead of $p_H$, and each deviator who reports collects the report value at his own belief instead of $Y_s$. Moreover, when $n<n_0$, the slot $U^0_{n+1}=A(n{+}1,\mu^\varepsilon_n)$ is the obedient agent's continuation payoff, whereas the shirker continues with $A(n{+}1,\mu^\varepsilon_{n-1})$. This value is at least $U^0_{n+1}$ by Step 3, since $n=\bar N$ and $\mu^\varepsilon_{n-1}\ge\mu^\varepsilon_n$, so the slot understates the shirker's value. By \eqref{eq:oa ac drift}, the changes other than the slot move each coefficient by at most a multiple of $\delta^{s-1}\varepsilon$, where $s$ is the period of the entry it multiplies. These changes move each report value by at most $C_1\varepsilon\,|Y^H_s-Y^L_s|$, since both the deviator's belief and $\mu^{\mathrm a}_s$ are within $C_1\varepsilon$ of one by the second part of \eqref{eq:oa ac drift}. As each expression has at most $4n+2$ terms, \eqref{eq:oa ac bounds} bounds the total effect of these changes by $C_4\varepsilon$, discounted to period $1$. This proves the first bound of \eqref{eq:oa ac tilde}. Since $U^0_1$ is at least the value of $e^t_{\text{Shirk}}$ in $\phi$ by (IC), both directions of the comparison give the second. Since the no-concealing slack against $r^m_{\text{Conceal}}$ is nonnegative in $\phi$, the same comparison gives the third at $m<n$. At $m=n=n_0$, $\tilde G^C_n=Y_n-\delta w_0(n{+}1)$ is the no-concealing slack of $\phi$ itself, hence nonnegative. At $m=n<n_0$, $\tilde G^C_n=Y_n-\delta A(n{+}1,\mu^\varepsilon_n)$. The mimic bound \eqref{eq:oa ac mimic} at $t=n$, together with $\mu^{\mathrm a}_n\ge\mu^\varepsilon_n$ and Step 3, gives $\tilde G^C_n\ge-\delta(1-\mu^{\mathrm a}_n)\big[A_L(n{+}1)+c/p_L\big]$, which is at least $-C_4\varepsilon\,\delta^{1-n}$ by \eqref{eq:oa ac bounds} and \eqref{eq:oa ac drift}.

By \eqref{eq:oa ac surplus} and \eqref{eq:oa ac R}, the principal's payoff at $\phi$ is at most $S(n)-R^\ast(n)+(C_2+C_5)\,\varepsilon\le U_P^\ast+\bar C\varepsilon$, with $\bar C:=C_2+C_5$. Since Step 1 dismissed the contracts with $U^0_1>\pi$, and $\phi$ and its schedule are arbitrary, \eqref{eq:oa ac upper} follows.

\emph{Part 2: posting a single bonus sequence loses at most a multiple of $\varepsilon^\alpha$.} If $n_0^\ast=0$, then $U_P^\ast=0$, which posting no bonus attains at every $\varepsilon$, since the agent then never works. At $\varepsilon=0$, posting the PBS bonus sequence $b_0^\ast(\cdot\,;n_0^\ast)$, followed by $\underline b$, attains $U_P^\ast$. Indeed, an effort rule is worth as much under this sequence as in $\phi^\ast(n_0^\ast)$ combined with a report of ``$F$'' in period $1$, so the strategy that follows the schedule $(n_0^\ast,\infty\mathbf 1)$ is a best response by Step 3 of Theorem~\proofref{thm:main result}'s proof. So let $\varepsilon>0$ and $n:=n_0^\ast>0$. It suffices to exhibit one sequence $b$ and one best response $\zeta$ to it that follows a schedule.

\emph{Step 1: the sequence.} To keep both stages running at $\varepsilon>0$, the sequence raises the PBS research-stage bonuses by a small addition, pays slightly above $\underline b$ for finitely many development-stage periods, and then stops. With $\lambda_H:=\delta(1-p_H)$ and $\lambda_L:=\delta(1-p_L)$, for $y\in(0,1]$ and $\tau\ge1$, let $T:=n+\tau$, $x:=2y/(p_L\lambda_H^{\,n-1})$, and
\[
b(t):=b_0^\ast(t;n)+x\quad(t\le n),
\qquad
b(t):=\underline b+y\quad(n<t\le T),
\qquad
b(t):=0\quad(t>T),
\]
which satisfies (LL), since $b_0^\ast(\cdot\,;n)>0$ by (3b) in Step 3 of Theorem~\proofref{thm:main result}'s proof. The strategy $\zeta$ works in every period $t\le n$ while uninformed, shirks afterward while uninformed, and works in every period through $T$ once informed, so it follows the schedule $(n,(T-k)_{k\le n})$. In Step 5, the research-stage addition $x$ offsets two small effects that can make postponing work attractive to an uninformed agent: the drift due to $\varepsilon$ and the development-stage addition $y$. Every bonus is at most $B:=\max_{t\le n}b_0^\ast(t;n)+2/(p_L\lambda_H^{\,n-1})+\underline b+1$, and a success pays once, so every strategy of the agent is worth at most $K:=B+c/p_L$ in absolute value conditional on either state.

\emph{Step 2: the agent's problem.} To show that $\zeta$ is a best response, I compare its values with the agent's best-response values. At any private history, the agent maximizes $a_H\mathbb E_H[\cdot]+a_L\mathbb E_L[\cdot]$ over continuation strategies, where $a_\omega$ is the probability of the history and the state $\omega$. The supremum is positively homogeneous in $(a_H,a_L)$, so, divided by $a_H+a_L$, it depends on the history only through the belief $a_H/(a_H+a_L)$. Hence the best-response value depends only on the period, the belief, and whether the agent is informed. In this part, write $V^1(t,\theta)$ for this value at the start of period $t$ for an informed agent with likelihood ratio $\theta$ and belief $\mu:=\theta/(1+\theta)$, and $V^0(t,d)$ for an uninformed agent after $d$ silent worked periods. These values satisfy
\[
\begin{aligned}
V^1(t,\theta)&=\max\big\{\delta V^1(t{+}1,\theta),\ -c+P(\mu)\,b(t)+\delta\big(1-P(\mu)\big)V^1(t{+}1,\rho_2\theta)\big\},\\
V^0(t,d)&=\max\big\{\delta V^0(t{+}1,d),\ -c+P^\varepsilon_d\,b(t)+\delta Q^\varepsilon_d\,V^1\big(t{+}1,\theta^\varepsilon_dI_\varepsilon\big)+\delta\nu^\varepsilon_d\,V^0(t{+}1,d{+}1)\big\},
\end{aligned}
\]
with $V^1(T{+}1,\cdot)=V^0(T{+}1,\cdot)=0$, as nothing is paid after $T$. Let $J^1(t,\theta)$ and $J^0(t,d)$ be the values of $\zeta$ from the same pairs. Given the state, $\zeta$ plays the same after a finding whatever the belief, so $J^1(t,\theta)=\mu A^1_H(t)+(1-\mu)A^1_L(t)$, where $A^1_\omega(t):=\sum_{s=t}^{T}[\delta(1-p_\omega)]^{s-t}\big(p_\omega b(s)-c\big)$. Steps 3 to 5 show that $V^0(1,0)=J^0(1,0)$ for small $\varepsilon$.

\emph{Step 3: uninformed agents stop after period $n$.} I show that $V^0(t,d)=0$ for all $t>n$, provided $\varepsilon\le1-\rho$ and $y\le c(p_H-p_L)/[2p_H(\theta_0+p_L)]$, so that $c':=c-p_L(\underline b+y)>\theta_0y$. Fix an effort rule from $(t,d)$ that first works in period $s$. In state $H$, each worked period earns at most $p_Hy$ in expectation and the $i$-th is reached with probability at most $(1-p_H)^{i-1}$, so the effort rule is worth at most $\delta^{s-t}y$. In state $L$, the first worked period is reached for sure and each loses at least $c'$, so the effort rule is worth at most $-\delta^{s-t}c'$. As $\rho_\varepsilon\le1$, the likelihood ratio is $\theta^\varepsilon_d\le\theta_0$, so the effort rule is worth at most $\delta^{s-t}(1-\mu^\varepsilon_d)(\theta_0y-c')<0$, and shirking forever does better.

\emph{Step 4: informed agents work through $T$.} With $\bar\theta_t:=c\rho_2^{\,t-T}/(p_Hy)$, I show that
\begin{equation}\label{eq:oa ac informed}
V^1(t,\theta)=J^1(t,\theta)\qquad\text{for every }2\le t\le T+1\text{ and }\theta\ge\bar\theta_t ,
\end{equation}
that is, an informed agent who is optimistic enough works in every period through $T$. The claim holds provided $y\le c\,m_\ast/(2K)$, where $m_\ast$ is the least margin $b_0^\ast(t;n)-\underline b-F_t$ over $2\le t\le n$, positive by (3b), and $m_\ast:=1$ if $n=1$. At $t=T+1$ both sides vanish. In a development-stage period $n<t\le T$, an informed agent observes nothing until success, so an effort rule is a set of worked periods $s_1<s_2<\cdots$. Such an effort rule is worth $\sum_{i:\,s_i\le T}\delta^{s_i-t}z_i$ plus negative terms from the periods after $T$, with $z_i:=\mu(1-p_H)^{i-1}p_Hy-(1-\mu)(1-p_L)^{i-1}c'$. For $\theta\ge\bar\theta_t$ and $i\le T-t+1$, $\theta\rho_2^{\,i-1}p_Hy\ge c\ge c'$, so $z_i\ge0$. As at most $T-t+1$ worked periods fall in $[t,T]$ and $s_i\ge t+i-1$, the effort rule is worth at most $\sum_{i\le T-t+1}\delta^{i-1}z_i=J^1(t,\theta)$, so working in every period through $T$ is optimal.

In a research-stage period $2\le t\le n$, suppose that \eqref{eq:oa ac informed} holds at $t+1$. As $\bar\theta_{t+1}=\rho_2\bar\theta_t$, both continuation payoffs in the recursion for $V^1(t,\theta)$ are values of working in every period through $T$, so by Bayes' rule the work branch is $J^1(t,\theta)$, the shirk branch is $\delta J^1(t{+}1,\theta)$, and
\[
(1+\theta)\big[J^1(t,\theta)-\delta J^1(t{+}1,\theta)\big]
=\theta\,p_H\big[b(t)-\underline b-\delta A^1_H(t{+}1)\big]+A^1_L(t)-\delta A^1_L(t{+}1).
\]
The last two terms are at least $-2K$. In the bracket, $\delta A^1_H(t{+}1)$ is the PBS informed payoff $F_t$ of \eqref{eq:app Frec} plus the value of the additions after period $t$, as an informed agent earns zero at $\underline b$. This value of the additions is at most $x$, since $x>y$ and a success pays once. Since $b(t)$ includes $x$, the bracket is at least $b_0^\ast(t;n)-\underline b-F_t\ge m_\ast$, and $\theta\ge\bar\theta_t\ge c/(p_Hy)$ implies that the right side is nonnegative, as $c\,m_\ast/y\ge2K$. So work attains the maximum, and $V^1(t,\theta)=J^1(t,\theta)$.

\emph{Step 5: uninformed agents work through period $n$.} I show that $V^0(t,d)=J^0(t,d)$ for all $t\le n+1$ and $d\le t-1$, provided, besides the bounds of Steps 3 and 4, that $4Kn\varepsilon\le y$ and that the development stage is short enough, in the sense that
\begin{equation}\label{eq:oa ac tail}
\rho_2^{\,T-2}\ \ge\ C_6\,\frac{\varepsilon}{y},
\qquad
C_6:=\frac{c\,(1-p_L)}{p_H\,\theta_{n-1}\,q_H(1-p_H)} .
\end{equation}
At $t=n+1$ both sides vanish by Step 3. At $t\le n$, suppose that the claim holds at $t+1$. A finding in period $t$ leaves the agent with the likelihood ratio $\theta^\varepsilon_dI_\varepsilon\ge\theta_{n-1}I_\varepsilon\ge\bar\theta_2\ge\bar\theta_{t+1}$, the middle inequality by \eqref{eq:oa ac tail}, so he then works through $T$ by \eqref{eq:oa ac informed}. The work branch at $(t,d)$ is therefore $J^0(t,d)$ and the shirk branch is $\delta J^0(t{+}1,d)$, and it remains to show that the gap $M(t,d):=J^0(t,d)-\delta J^0(t{+}1,d)$ is nonnegative. Let $A_H(t)$ and $A^\varepsilon_L(t)$ now denote the values of $\zeta$ from the uninformed period $t$ conditional on each state, which vanish at $t=n+1$. The state-$H$ law of $\zeta$ does not involve $\varepsilon$, so
\[
M(t,d)=\mu^\varepsilon_d\,\chi_t+(1-\mu^\varepsilon_d)\,\psi^\varepsilon_t,
\qquad
\chi_t:=A_H(t)-\delta A_H(t{+}1),\quad \psi^\varepsilon_t:=A^\varepsilon_L(t)-\delta A^\varepsilon_L(t{+}1).
\]
I compare $M(t,d)$ with the gap at $\varepsilon=0$, $M^0(\mu):=\mu\chi_t+(1-\mu)\psi^0_t$, evaluated at $\mu=\mu^\varepsilon_d$, which lies in $[\mu_{t-1},\mu_0]$ since $d\le t-1$ and $\rho\le\rho_\varepsilon\le1$.

First, $M^0(\mu)\ge y$ on $[\mu_{t-1},\mu_0]$. The values of $\zeta$ are affine in the bonuses, so $M^0$ splits into the PBS part, due to $b_0^\ast(\cdot\,;n)$ followed by $\underline b$ through period $T$, and the parts due to the two additions. At $\varepsilon=0$ only agents in state $H$ are informed, and they earn zero at $\underline b$. The PBS part is therefore an affine function of $\mu$ that equals the nonnegative work-minus-shirk gap of (3c) in Step 3 of Theorem~\proofref{thm:main result}'s proof at each $\mu_d$ with $d\le t-1$, which implies that this part is nonnegative on $[\mu_{t-1},\mu_0]$. The addition $x$ is paid upon a success in a period $s\le n$, which $\zeta$ works in either state, so it contributes $\mu\chi^x_t+(1-\mu)\psi^x_t$, where
\[
\chi^x_t=p_Hx\,\frac{(1-\delta)+\delta p_H\lambda_H^{\,n-t}}{1-\lambda_H}\ \ge\ p_Hx\,\lambda_H^{\,n-t}
\]
and $\psi^x_t\ge p_Lx\,\lambda_L^{\,n-t}$ by the same expression with $p_L$ and $\lambda_L$ in place of $p_H$ and $\lambda_H$, so the contribution is at least $p_L\lambda_H^{\,n-1}x=2y$. The addition $y$ reaches only informed agents, so it contributes $\mu[A^y(t)-\delta A^y(t{+}1)]\ge-y$, where $A^y(t)\in[0,y]$ is its value in state $H$.

Second, $|\psi^\varepsilon_t-\psi^0_t|\le4Kn\varepsilon$, so that $M(t,d)\ge M^0(\mu^\varepsilon_d)-4Kn\varepsilon\ge y-4Kn\varepsilon\ge0$. Under the coupling of Step 2 of Part 1, the realized paths under $\zeta$ differ only if a finding arrives in state $L$ in one of the at most $n$ periods in which $\zeta$ works uninformed, an event of probability at most $n\varepsilon$. From that period on, the agent's continuation payoff on each path is at most $K$ in absolute value conditional on $\omega=L$. Hence $|A^\varepsilon_L(t)-A^0_L(t)|\le2Kn\varepsilon$ for every $t$.

\emph{Step 6: the principal's loss.} Three terms separate the principal's payoff at $\zeta$ from $U_P^\ast$. First, her realized payoff lies in $[-B,\pi]$, so the coupling of Step 5 moves her expected payoff by at most $(\pi+B)n\varepsilon$. Second, at $\varepsilon=0$, the surplus of $\zeta$ falls short of $S(n)$ by at most $\lambda_H^{\,\tau}V^{\mathrm{informed}}$, since the track entered after a finding in period $k\le n$ stops after $T\ge k+\tau$, which forgoes $\lambda_H^{\,T-k}V^{\mathrm{informed}}$, and tracks are entered with total discounted probability at most one. Third, at $\varepsilon=0$, by (3c) in Step 3 of Theorem~\proofref{thm:main result}'s proof and since an informed agent earns zero at $\underline b$, the agent's payoff at $\zeta$ is $R^\ast(n)$ plus the value of the additions, which is at most $x$. Combining the three terms gives
\[
U_P^{\mathrm{PBS}}(\varepsilon)\ \ge\ U_P^\ast-\frac{2y}{p_L\lambda_H^{\,n-1}}-\lambda_H^{\,\tau}V^{\mathrm{informed}}-(\pi+B)\,n\,\varepsilon .
\]
The exponent $\alpha$ balances the addition $y$ against the surplus lost by ending the development stage, of order $\lambda_H^{\,\tau}$. By \eqref{eq:oa ac tail}, $\rho_2^{\,\tau}$ can be of order $\varepsilon/y$, and then $\lambda_H^{\,\tau}$ is of order $(\varepsilon/y)^{\alpha/(1-\alpha)}$, where
\[
\alpha:=\frac{\log(1/\lambda_H)}{\log\big(1/(\lambda_H\rho_2)\big)},
\qquad\text{so that}\qquad
\lambda_H=\rho_2^{\,\alpha/(1-\alpha)} .
\]
Here $\alpha<1$ since $\rho_2<1$, and $\alpha>\tfrac12$ since $\lambda_H/\rho_2=\lambda_L<1$ by $p_L>0$. So let $y:=\varepsilon^\alpha$, and let $\tau$ be the largest integer with $\rho_2^{\,n+\tau-2}\ge C_6\varepsilon^{1-\alpha}$, which exists once $C_6\varepsilon^{1-\alpha}\le\rho_2^{\,n-1}$ and gives \eqref{eq:oa ac tail}. By its maximality, $\rho_2^{\,\tau}<C_6\rho_2^{\,1-n}\varepsilon^{1-\alpha}$, hence $\lambda_H^{\,\tau}=(\rho_2^{\,\tau})^{\alpha/(1-\alpha)}$ is at most $(C_6\rho_2^{\,1-n})^{\alpha/(1-\alpha)}\varepsilon^\alpha$. Let $\varepsilon_2\le1-\rho$ be small enough that, on $(0,\varepsilon_2]$, $y=\varepsilon^\alpha$ meets the bounds of Steps 1, 3 and 4, and both $4Kn\varepsilon\le\varepsilon^\alpha$ and $C_6\varepsilon^{1-\alpha}\le\rho_2^{\,n-1}$ hold, which is possible as $\alpha<1$. As $\varepsilon\le\varepsilon^\alpha$, the bound on $U_P^{\mathrm{PBS}}(\varepsilon)$ above and the bound on $\lambda_H^{\,\tau}$ give \eqref{eq:oa ac lower}, with a constant $\underline C$ that depends only on the primitives. $\hfill\square$

\subsection{Proof of Proposition \ref*{prop:costless}}\label{app:proof costless}

Steps 1 to 4 prove the claims of the proposition on the optimal contract and its uninformed duration, and Step 5 proves the comparison of bonuses, together with the optimality of $\tilde\phi^{CF}(n_0)$ when the costless finding is contractible. The notation of Appendix~\ref{app:preliminaries} is maintained, together with the agent's payoff $R^\ast(n_0)$ of Step 3 of Theorem~\proofref{thm:main result}'s proof, with $R^\ast(0):=0$. The likelihood ratio of a belief $\mu$ is $\theta:=\mu/(1-\mu)$.

\emph{Step 1: the model with a costless finding.} As in Section~\ref{sec:costless}, in every period in which the agent is uninformed, the finding arrives with probability $q_H$ in state $H$ and never in state $L$, whatever his effort and independently of the success. A worked period thus has the probabilities $P(\mu)$, $Q(\mu)$, and $\nu(\mu)$ of the main model, and a shirked period brings no finding with probability $\nu^{s}(\mu):=1-\mu q_H$. After a silent worked period or a shirked period without a finding, the belief moves to $\mu^{w}$ or $\mu^{s}$, with likelihood ratio $\theta\rho$ or $\theta(1-q_H)$, and Bayes' rule implies, for every $\mu\in(0,1)$,
\begin{equation}\label{eq:oa costless bayes}
\begin{aligned}
\nu(\mu)\,\mu^{w}&=\mu\gamma',&\qquad\nu(\mu)\,(1-\mu^{w})&=(1-\mu)(1-p_L),\\
\nu^{s}(\mu)\,\mu^{s}&=\mu(1-q_H),&\qquad\nu^{s}(\mu)\,(1-\mu^{s})&=1-\mu .
\end{aligned}
\end{equation}
An uninformed agent who reaches period $t$ after working $j\le t-1$ periods, all silent, holds the likelihood ratio $\theta_0\rho^{j}(1-q_H)^{t-1-j}\ge\theta_{t-1}$, because $\rho<1-q_H$ by $p_H>p_L$. So every belief that an uninformed agent can hold in period $t$ is at least $\mu_{t-1}$.

Fix an uninformed duration $n_0\in\{0,1,2,\dots\}$. On the uninformed track, a contract pays a bonus $b_0(t)$ upon a success in period $t$ and a wage $w_0(t)$ in every period $t\ge1$ that the track reaches. For every $k\ge1$, it pays on the $k$-informed track a lump sum $\ell_k$ and the bonuses $b_{1k}(t)$ and wages $w_{1k}(t)$, $t>k$. The $k$-informed track is entered upon a report of ``$F$'' in period $k$ when the finding is private, and upon the finding in period $k$, worked or shirked, when it is contractible. All transfers are nonnegative by (LL) and bounded over $t$ on every track. While the agent is uninformed, the obedient strategy works in the periods $t\le n_0$ and shirks afterward. It reports the finding in its arrival period and works in every period of the entered track. Strategies, payoffs, (IC), and the first step are those of Section~\ref{sec:model}, with the effort rule and the reporting rule defined in every period (the reporting rule is absent when the finding is contractible), and with reach probabilities computed for the costless finding. Write $U$ for the agent's payoff under the obedient strategy.

For the $k$-informed track, take $F_k$, $\mathcal W_k$ and the informed payoff $Y_k=\ell_k+F_k$ as in Appendix~\ref{app:preliminaries}. The obedient strategy reaches every track with positive probability, a track $k\le n_0$ after a finding but no success in the worked period $k$, and a track $k>n_0$ after a finding in the shirked period $k$ when $\omega=H$. At a feasible contract, (IC) therefore forces $\mathcal W_k=F_k$, and $F_k\ge0$ since shirking in every period is an effort rule. Hence $Y_k\ge0$ by (LL). Conditional on $\omega=L$ and on $\omega=H$, the obedient strategy is worth, from period $n_0+1$ on the uninformed track,
\[
A_L(n_0{+}1):=\sum_{t>n_0}\delta^{t-n_0-1}w_0(t),\qquad
A_H(n_0{+}1):=\sum_{j\ge0}(1-q_H)^jq_H\Big[\sum_{t=n_0+1}^{n_0+1+j}\delta^{t-n_0-1}w_0(t)+\delta^{j}Y_{n_0+1+j}\Big],
\]
since under $\omega=H$ the finding arrives in period $n_0+1+j$ with probability $(1-q_H)^jq_H$. For $t\le n_0$, the corresponding values from period $t$ are defined backward from these terminal values by
\begin{equation}\label{eq:oa costless A}
\begin{aligned}
A_H(t)&=w_0(t)-c+p_Hb_0(t)+q_H(1-p_H)Y_t+\delta\gamma'A_H(t+1),\\
A_L(t)&=w_0(t)-c+p_Lb_0(t)+\delta(1-p_L)A_L(t+1),
\end{aligned}
\end{equation}
as in Step 2 of Proposition~\ref{prop:infinite}'s proof. Write $A(t,\mu):=\mu A_H(t)+(1-\mu)A_L(t)$, so that the obedient agent's continuation payoff in period $t$ is $A(t,\mu_{t-1})$. By \eqref{eq:app bayes} and the recursion $\mathcal D_t=\delta\nu_{t-2}\mathcal D_{t-1}$ of (Fact 4), the discounted probability of reaching period $t$ on the obedient path jointly with $\omega=H$ is $\mathcal D_t\mu_{t-1}=\delta^{t-1}\mu_0\gamma'^{\,t-1}$, and jointly with $\omega=L$ it is $\mathcal D_t(1-\mu_{t-1})=\delta^{t-1}(1-\mu_0)(1-p_L)^{t-1}$, for all $t\le n_0+1$. Unrolling \eqref{eq:oa costless A} in $U=A(1,\mu_0)$ with these weights gives
\begin{equation}\label{eq:oa costless U}
U=\sum_{t=1}^{n_0}\mathcal D_t\big[w_0(t)-c+P_{t-1}b_0(t)+Q_{t-1}Y_t\big]+\mathcal D_{n_0+1}A(n_0{+}1,\mu_{n_0}).
\end{equation}

For $t\le n_0$, the effort rule $e_{\text{Shirk}}^t$ of \eqref{eq:effort obedience} shirks in period $t$ when no report has been made (if the finding is contractible, when no finding has arrived) and works in every other recommended period. When the finding is private, this effort rule is played with the obedient reporting rule. The resulting strategy coincides with the obedient strategy until period $t$ is reached on the uninformed track. From there, the shirker collects $w_0(t)+q_HY_t+\delta(1-q_H)A_H(t+1)$ conditional on $\omega=H$, since the finding still arrives with probability $q_H$ in the shirked period, and $w_0(t)+\delta A_L(t+1)$ conditional on $\omega=L$. Subtracting these from $A_H(t)$ and $A_L(t)$ and weighting by $\mathcal D_t\mu_{t-1}$ and $\mathcal D_t(1-\mu_{t-1})$ shows that the shirker's value falls short of $U$ by $\mathcal D_t$ times the difference between the two sides of
\begin{equation}\label{eq:oa costless eo}
P_{t-1}b_0(t)-c\ \ge\ \mu_{t-1}\,p_H\big[q_HY_t+\delta(1-q_H)A_H(t+1)\big]+(1-\mu_{t-1})\,\delta\,p_L\,A_L(t+1),
\end{equation}
so effort obedience in period $t$ is \eqref{eq:oa costless eo}.

Finally, I give a sufficient condition for (IC). Let $n_0>0$, and consider a contract without wages and lump sums under which the payoff of a strategy depends only on its effort rule. Suppose that an agent whose finding arrives in a period $t\le n_0$ optimally works in every period of the $t$-informed track, so that his optimal value, discounted to period $t$, is $Y_t$. Suppose also that from period $n_0+1$ on, every agent's optimal continuation payoff is zero and is attained by the obedient strategy, so that $A_H(n_0{+}1)=A_L(n_0{+}1)=0$. Let $\chi^{Y}_t:=A_H(t)-q_HY_t-\delta(1-q_H)A_H(t+1)$ and $\psi_t:=A_L(t)-\delta A_L(t+1)$. By \eqref{eq:oa costless A}, the difference between the two sides of \eqref{eq:oa costless eo} is $\mu_{t-1}\chi^{Y}_t+(1-\mu_{t-1})\psi_t$. For $t\le n_0$, the optimal value $V(t,\mu)$ of an uninformed agent in period $t$ at a belief $\mu$ that he can hold satisfies, by the positive-homogeneity argument of Step 2 of Part 2 of Proposition~\proofref{prop:almost conclusive}'s proof,
\begin{equation}\label{eq:oa costless V}
V(t,\mu)=\max\big\{\mu q_HY_t+\nu^{s}(\mu)\,\delta V(t+1,\mu^{s}),\ -c+P(\mu)b_0(t)+Q(\mu)Y_t+\nu(\mu)\,\delta V(t+1,\mu^{w})\big\}
\end{equation}
with $V(n_0+1,\cdot):=0$, the first branch being shirking. If $V(t+1,\cdot)=A(t+1,\cdot)$, then, by \eqref{eq:oa costless bayes}, the work branch equals $A(t,\mu)$ and the shirk branch equals $A(t,\mu)-\big[\mu\chi^{Y}_t+(1-\mu)\psi_t\big]$. Suppose now that \eqref{eq:oa costless eo} holds with equality for every $t\le n_0$, that is, $\psi_t=-\theta_{t-1}\chi^{Y}_t$, and that $\chi^{Y}_t\ge0$ for $2\le t\le n_0$. Then, since $\mu=\theta/(1+\theta)$, for every belief $\mu$ with likelihood ratio $\theta$ that the agent can hold in period $t$, $(1+\theta)\big[\mu\chi^{Y}_t+(1-\mu)\psi_t\big]=\theta\chi^{Y}_t+\psi_t=(\theta-\theta_{t-1})\chi^{Y}_t\ge0$, since $\theta\ge\theta_{t-1}$ and $\chi^{Y}_t\ge0$ for $t\ge2$, while at $t=1$ the only belief is $\mu_0$. Backward induction on $t$, with $\mu^{w}$ and $\mu^{s}$ beliefs that he can hold in period $t+1$, gives $V(t,\cdot)=A(t,\cdot)$, with work attaining the maximum. The obedient strategy thus attains the optimal value in every period and at every belief, so (IC) holds, with the agent's payoff $A(1,\mu_0)$.

\emph{Step 2: the PBS contract.} Recall that, with a costless finding, $\phi^\ast(n_0)$ denotes the contract that posts the bonus $b^{\mathrm{PBS}}(t):=b_0^\ast(t;n_0)$ in every period $t\le n_0$ and $b^{\mathrm{PBS}}(t):=\underline b$ in every period $t>n_0$, on the uninformed track and on every informed track, and pays no wage and no lump sum. Its transfers depend neither on the finding nor on the report, so its feasibility does not depend on whether the finding is contractible, and the payoff of a strategy depends only on its effort rule. (LL) holds since $b_0^\ast>0$ by (3b) in Step 3 of Theorem~\proofref{thm:main result}'s proof.

In a period $t>n_0$, working gives an agent with belief $\mu'$ the expected payoff $P(\mu')\underline b-c\le0$, which is zero only if he is informed. So every strategy is worth at most zero from period $n_0+1$, which shirking attains, and the obedient strategy is optimal and worth zero for every agent. At $n_0=0$ this argument applies from period $1$, so $\phi^\ast(0)$ is feasible with the agent's payoff $0=R^\ast(0)$. Let $n_0>0$, so that $A_H(n_0{+}1)=A_L(n_0{+}1)=0$. The informed tracks of $\phi^\ast(n_0)$ are those of the PBS contract of the main model, so their values $F_t$ satisfy \eqref{eq:app Frec} at $b_0=b_0^\ast$. Backward induction from $F_{n_0}=0=\delta A_H(n_0{+}1)$, using \eqref{eq:app Frec} and $q_H(1-p_H)+\gamma'=1-p_H$, gives $Y_t=F_t=\delta A_H(t+1)$ for all $t\le n_0$, so that $A_H(t)=-c+p_Hb^{\mathrm{PBS}}(t)+\delta(1-p_H)A_H(t+1)$ by \eqref{eq:oa costless A}. Hence $\chi^{Y}_t=A_H(t)-\delta A_H(t+1)=p_Hb^{\mathrm{PBS}}(t)-c-p_HF_t=p_H[b^{\mathrm{PBS}}(t)-\underline b-F_t]$, which is positive for $2\le t\le n_0$ by (3b). By backward induction, working in every period is therefore optimal for an informed agent, worth $A_H(t)$ at the start of period $t$ and $\delta A_H(t+1)=Y_t$ upon a finding in period $t$. In the main model at $\phi^\ast(n_0)$, where a worked period has the same probabilities, the obedient strategy collects $A_H(t)$ and $A_L(t)$ from period $t$ conditional on the state, and $e_{\text{Shirk}}^t$ collects $\delta A_H(t+1)$ and $\delta A_L(t+1)$. Effort obedience in period $t$ there thus requires $\mu_{t-1}\chi^{Y}_t+(1-\mu_{t-1})\psi_t\ge0$, and it holds with equality by (3b) and (3c) in Step 3 of Theorem~\proofref{thm:main result}'s proof. So \eqref{eq:oa costless eo} holds with equality at every $t\le n_0$, the sufficient condition of Step 1 implies (IC), and the agent's payoff $A(1,\mu_0)$ equals $R^\ast(n_0)$, the agent's payoff of $\phi^\ast(n_0)$ in the main model.

\emph{Step 3: when the finding is private, no feasible contract pays less.} At $n_0=0$, every feasible contract pays at least $0=R^\ast(0)$, since the agent can shirk in every period and never report, collecting only nonnegative transfers. Fix $n_0>0$ and a contract $\phi$ that is feasible when the finding is private, and suppose that $U<\infty$, as there is nothing to show otherwise. Then $A(n_0{+}1,\mu_{n_0})$ is finite, since it enters \eqref{eq:oa costless U} with the positive weight $\mathcal D_{n_0+1}$ and every other term is bounded below. Since $\mu_{n_0}\in(0,1)$ and $A_H(n_0{+}1),A_L(n_0{+}1)\ge0$ by (LL) and $Y\ge0$, both are finite as well.

First, $A_H(n_0{+}1)\ge A_L(n_0{+}1)$. In every period $\tau>n_0$, the obedient strategy reaches with positive probability an informed agent who has not reported, namely when the finding arrives in period $\tau$. There, reporting is worth $\ell_\tau+\mathcal W_\tau=Y_\tau$, and concealing forever while shirking collects only the later wages, so (IC) requires $Y_\tau\ge\sum_{t>\tau}\delta^{t-\tau}w_0(t)$. At $\tau=n_0+1+j$, multiplying this inequality by $\delta^j$ gives $\delta^jY_{n_0+1+j}\ge\sum_{t>n_0+1+j}\delta^{t-n_0-1}w_0(t)$, so every bracket of $A_H(n_0{+}1)$ is at least $A_L(n_0{+}1)$, and $A_H(n_0{+}1)\ge\sum_{j\ge0}(1-q_H)^jq_HA_L(n_0{+}1)=A_L(n_0{+}1)$.

Second, let $\hat\phi$ be the contract of the main model at the schedule $(n_0,\infty\mathbf 1)$ that copies $b_0(t)$ and $w_0(t)$ for $t\le n_0$ and the lump sums and informed tracks of the periods $k\le n_0$, and pays the severance $w_0(n_0{+}1):=A(n_0{+}1,\mu_{n_0})$. It satisfies (LL), since the severance is nonnegative, and the agent's payoff under it is $U$. Indeed, through period $n_0$ and on the tracks entered by then, the obedient strategies of the two models generate the same law of transfers, since a worked period has the same probabilities in both models. The two strategies also reach period $n_0+1$ on the uninformed track with the same probability, and there $\hat\phi$ pays what the obedient strategy under $\phi$ is worth from that period.

Third, $\hat\phi$ satisfies (IC). Fix a strategy $(e,r)$ of the main model, and let $(\tilde e,\tilde r)$ be the strategy, in the model with a costless finding under $\phi$, that keeps a simulated history of the main model, in which the finding can occur only in worked periods, and plays the effort and reporting rules of $(e,r)$ at the simulated history. In a worked period in which the simulated history is uninformed, the simulated finding is the true finding if the agent is uninformed, and the draw of a private coin with probability $q_H$ if he is already informed. This case arises only if $\omega=H$ and the finding arrived in an earlier shirked period. Conditional on the state, successes are drawn as in the main model, and in every worked period in which the simulated history is uninformed, the simulated finding arrives with probability $q_H$ if $\omega=H$ and never if $\omega=L$, independently of the success. It follows that, through period $n_0$ and on the tracks entered by then, the joint law of efforts, successes, the simulated finding, and reports under $(\tilde e,\tilde r)$ equals that of efforts, successes, the finding, and reports under $(e,r)$ in the main model, and $\phi$ and $\hat\phi$ pay the same transfers there. If period $n_0+1$ is reached on the uninformed track, $(e,r)$ collects the severance $A(n_0{+}1,\mu_{n_0})$, while $(\tilde e,\tilde r)$ switches to the obedient strategy, except that an already informed agent reports at the first heads of a private coin with probability $q_H$ per period rather than at the finding. An informed agent then collects $A_H(n_0{+}1)$, because $\omega=H$ and his reports follow the law of the finding under $\omega=H$ while he shirks. An uninformed agent collects $A(n_0{+}1,\mu')$, where his belief satisfies $\mu'\ge\mu_{n_0}$ by Step 1. Both amounts are at least $A(n_0{+}1,\mu_{n_0})$ because $A_H(n_0{+}1)\ge A_L(n_0{+}1)$, so $U_A(\phi,\tilde e,\tilde r)\ge U_A(\hat\phi,e,r)$. For every realization of the coins, $(\tilde e,\tilde r)$ is a strategy of the model with a costless finding, so $U_A(\phi,\tilde e,\tilde r)$, an average over the coins, is at most $U$ by (IC) at $\phi$. Hence $U_A(\hat\phi,e,r)\le U$, the agent's payoff of $\hat\phi$ under the obedient strategy.

So $\hat\phi$ is a feasible contract of the main model with uninformed duration $n_0$ and the agent's payoff $U$, and Steps 2 and 3 of Theorem~\proofref{thm:main result}'s proof give $U\ge R^\ast(n_0)$.

\emph{Step 4: the uninformed duration.} At the uninformed duration $n$, the obedient strategy generates, under every contract, the surplus $S^C(n):=S(n)+\Xi(n)$, where $\Xi(n):=\mu_0q_HV^{\mathrm{informed}}(\delta\gamma')^{n}/[1-\delta(1-q_H)]$. Through period $n$, the obedient path is that of the main model at the schedule $(n,\infty\mathbf 1)$, whose surplus is $S(n)$ by Appendix~\ref{app:preliminaries}. Afterward the uninformed agent shirks, and surplus accrues only on an informed track entered upon a finding in a period $n+1+j$. Such a finding has the probability $\mu_0\gamma'^{\,n}(1-q_H)^jq_H$ jointly with $\omega=H$, and the track yields $V^{\mathrm{informed}}$ in that period, discounted by $\delta^{n+j}$. Summing over $j\ge0$ gives $\Xi(n)$. The principal's payoff at a feasible contract is $S^C(n)-U$, so by Steps 2 and 3 her optimal payoff at the uninformed duration $n$ is $S^C(n)-R^\ast(n)$, attained by $\phi^\ast(n)$, and the second step is $\max_{n\ge0}\,[S^C(n)-R^\ast(n)]$. By Step 5 of Theorem~\proofref{thm:main result}'s proof, $S-R^\ast$ attains its maximum at $n_0^\ast$, and $\Xi$ is strictly decreasing since $\delta\gamma'<1$. These two observations imply $S^C(n)-R^\ast(n)<S^C(n_0^\ast)-R^\ast(n_0^\ast)$ for every $n>n_0^\ast$. The maximum is therefore attained on $\{0,\dots,n_0^\ast\}$, its smallest maximizer $n_0^C$ satisfies $n_0^C\le n_0^\ast$, and $\phi^\ast(n_0^C)$ is an optimal contract.

Finally, the uninformed duration $n_0=\infty$ is not optimal. At this duration the uninformed agent works in every period on the obedient path, so the obedient path and its surplus are those of the main model at the schedule $(\infty,\infty\mathbf 1)$. Let $\phi$ be a contract with $n_0=\infty$, with the definitions of Step 1 read at this duration, that is feasible when the finding is private, and let $\hat\phi$ be the contract of the main model at this schedule that copies all of its transfers. It satisfies (LL) and the bounds of the model, and, as in Step 3, it gives the agent and the principal the same payoffs as $\phi$. The argument that shows (IC) for $\hat\phi$ in Step 3 applies in every period, since the period $n_0+1$, where that argument uses the severance and $A_H(n_0{+}1)\ge A_L(n_0{+}1)$, never arrives. So $\hat\phi$ is feasible, and by Proposition~\ref{prop:infinite} the principal's payoff under it is strictly below $S(T)-R^\ast(T)$ for some finite $T$. This bound is at most $S^C(T)-R^\ast(T)$, which $\phi^\ast(T)$ attains.

Let $\delta=1$. Conditional on $\omega=H$, an agent who is still uninformed after period $n$ obtains the finding with probability one, and informed work follows. So the obedient agent works until success, apart from shirked periods that cost nothing, and the surplus is $\pi-c/p_H$ at every $n$. Conditional on $\omega=L$, the finding never arrives, and the surplus is $\sum_{t\le n}(1-p_L)^{t-1}(p_L\pi-c)$. Hence $S^C(n)-S^C(n-1)=(1-\mu_0)(1-p_L)^{n-1}(p_L\pi-c)<0$ for all $n>0$, by Assumption~\ref{ass:efficiency}. Since $R^\ast$ is nondecreasing with $R^\ast(0)=0$ by \eqref{eq:app rstar}, $S^C(n)-R^\ast(n)<S^C(0)-R^\ast(0)$ for every $n>0$. So $n_0^C=0$, and $\phi^\ast(0)$ is optimal.

\emph{Step 5: a contractible finding.} At $n_0=0$, the contract $\phi^\ast(0)$ pays $\underline b$ upon every success, so it coincides with $\tilde\phi^{CF}(0)$, and the arguments of Steps 2 and 3 show that it is optimal when the finding is contractible. Fix $n_0>0$. The contract $\tilde\phi^{CF}(n_0)$ pays no wage and no lump sum, and it pays $\underline b$ upon every success except on the uninformed track in a period $t\le n_0$, where it pays $b(t):=\tilde b_0^{CF}(t;n_0)$. Every informed track is worth zero to an informed agent, since a worked period yields $p_H\underline b-c=0$, so $Y\equiv0$ and the terminal values are zero. Write $\tilde A_H$ and $\tilde A_L$ for the solution of \eqref{eq:oa costless A} at this contract. By \eqref{eq:oa costless eo}, effort obedience in period $t$ holds with equality if and only if
\begin{equation}\label{eq:oa costless cf bind}
P_{t-1}b(t)-c=\delta\big[\mu_{t-1}(1-q_H)\,p_H\,\tilde A_H(t+1)+(1-\mu_{t-1})\,p_L\,\tilde A_L(t+1)\big].
\end{equation}
Given $b(t+1),\dots,b(n_0)$, this condition is affine in $b(t)$ with slope $P_{t-1}>0$, so imposing it backward from $t=n_0$ determines $b$ uniquely, with $b(n_0)=c/P_{n_0-1}=b_0^\ast(n_0;n_0)$. By \eqref{eq:oa costless U}, the agent's payoff is $\tilde R^{CF}(n_0):=\sum_{t\le n_0}\mathcal D_t\big(P_{t-1}b(t)-c\big)$.

\emph{(a) The comparison.} Let $\phi$ have the uninformed duration $n_0$ and be feasible when the finding is contractible, and suppose that $U<\infty$, as there is nothing to show otherwise. Its wages are nonnegative by (LL) and its informed payoffs by Step 1, hence so are its terminal values. By \eqref{eq:oa costless A}, $A_H(t+1)$ and $A_L(t+1)$ are nondecreasing in the bonuses $b_0(\tau)$ with $\tau>t$, in the wages, in the informed payoffs, and in the terminal values. Thus the same holds for the right side of \eqref{eq:oa costless eo}, in which $Y_t$ also enters with the coefficient $\mu_{t-1}p_Hq_H$. At $\tilde\phi^{CF}(n_0)$ this right side equals $P_{t-1}b(t)-c$ by \eqref{eq:oa costless cf bind}. So if $b_0(\tau)\ge b(\tau)$ for every $\tau>t$, then \eqref{eq:oa costless eo} at $\phi$ implies
\[
P_{t-1}b_0(t)-c\ \ge\ P_{t-1}b(t)-c+\mu_{t-1}p_Hq_HY_t ,
\]
and backward induction gives $b_0(t)\ge b(t)$ for every $t\le n_0$. Two consequences follow. First, $U\ge\tilde R^{CF}(n_0)$ by \eqref{eq:oa costless U}, whose coefficients are positive and whose other terms are nonnegative, so $\tilde\phi^{CF}(n_0)$ is optimal at the uninformed duration $n_0$ once it is feasible, which (b) shows. Second, by Step 2, $\phi^\ast(n_0)$ is feasible when the finding is contractible, and its informed payoff is $Y_t=F_t>0$ for all $t<n_0$, since $F_t>\delta F_{t+1}\ge0$ by (3b) in Step 3 of Theorem~\proofref{thm:main result}'s proof. Hence $b_0^\ast(t;n_0)>b(t)$ for every $t<n_0$, while $b_0^\ast(n_0;n_0)=b(n_0)$, which is the proposition's comparison of bonuses.

\emph{(b) Feasibility of $\tilde\phi^{CF}(n_0)$.} To relate the bonuses of consecutive periods, fix $t<n_0$ and compare two effort rules from period $t$ that follow the obedient strategy after period $n_0$. Effort rule (i) works in period $t$, shirks in period $t+1$, and works in the periods $t+2,\dots,n_0$. Effort rule (ii) shirks in period $t$ and works in the periods $t+1,\dots,n_0$. By \eqref{eq:oa costless cf bind} in periods $t+1$ and $t$, each is worth as much as the obedient strategy from period $t$. Conditional on each state, the two effort rules reach period $t+2$ on the uninformed track with the same probability, $\gamma'(1-q_H)$ if $\omega=H$ and $1-p_L$ if $\omega=L$, and continue identically, while a finding leads to a track worth zero. Equating their flows in periods $t$ and $t+1$ gives
\begin{equation}\label{eq:oa costless law}
P_{t-1}b(t)-c=\delta\big[P^{s}_{t-1}\,b(t+1)-c\,\nu^{s}_{t-1}\big],
\end{equation}
where $P^{s}_{t-1}:=\mu_{t-1}(1-q_H)p_H+(1-\mu_{t-1})p_L$ and $\nu^{s}_{t-1}:=\nu^{s}(\mu_{t-1})$. By \eqref{eq:oa costless bayes}, $P^{s}_{t-1}=\nu^{s}_{t-1}P(\mu^{s}_{t-1})$, and $P(\mu^{s}_{t-1})>P_t$ because $\mu^{s}_{t-1}$ has the likelihood ratio $\theta_{t-1}(1-q_H)>\theta_{t-1}\rho=\theta_t$. So \eqref{eq:oa costless law} can be rewritten as $P_{t-1}b(t)-c=\delta\nu^{s}_{t-1}\big[P(\mu^{s}_{t-1})\,b(t+1)-c\big]$, whose right side is positive whenever $P_tb(t+1)\ge c$, since then $b(t+1)>0$. Backward induction from $P_{n_0-1}b(n_0)=c$ thus gives $P_{t-1}b(t)>c$ for every $t<n_0$, so $b>0$, and $f_t:=p_Hb(t)-c>0$ for all $t\le n_0$ since $P_{t-1}<p_H$. At this contract, where $Y\equiv0$, the quantity $\chi^{Y}_t$ of Step 1 is $\tilde\chi_t:=\tilde A_H(t)-\delta(1-q_H)\tilde A_H(t+1)$. Since $\tilde A_H(t+1)-\delta\gamma'\tilde A_H(t+2)=f_{t+1}$, the sequence $\tilde\chi$ satisfies the recursion $\tilde\chi_t=\big[f_t-\delta(1-q_H)f_{t+1}\big]+\delta\gamma'\tilde\chi_{t+1}$, with $f_{n_0+1}:=\tilde\chi_{n_0+1}:=0$. The bracket is $f_{n_0}>0$ at $t=n_0$, and at $t<n_0$, with $\mu:=\mu_{t-1}$, substituting \eqref{eq:oa costless law} for $P_{t-1}b(t)$ gives
\[
P_{t-1}\big[f_t-\delta(1-q_H)f_{t+1}\big]=c\,(p_H-P_{t-1})-\delta c\big[p_H\nu^{s}_{t-1}-(1-q_H)P_{t-1}\big]+\delta p_Hb(t+1)\big[P^{s}_{t-1}-(1-q_H)P_{t-1}\big],
\]
where $p_H-P_{t-1}$ and the two brackets equal $(1-\mu)(p_H-p_L)$, $(1-\mu)(p_H-p_L+q_Hp_L)$ and $(1-\mu)q_Hp_L$, respectively, so the right side is $(1-\mu)\big[(1-\delta)c\,(p_H-p_L)+\delta q_Hp_Lf_{t+1}\big]>0$. Since the bracket $f_t-\delta(1-q_H)f_{t+1}$ is positive for every $t\le n_0$, backward induction on the recursion yields $\tilde\chi_t>0$ for all $t\le n_0$. At $\tilde\phi^{CF}(n_0)$, the payoff of a strategy depends only on its effort rule. An informed agent is indifferent between working and shirking in each period, so working in every period is optimal for him with value $Y\equiv0$. After period $n_0$, the argument of Step 2 shows that the obedient strategy is optimal and worth zero for every agent. As \eqref{eq:oa costless cf bind} holds at every $t\le n_0$ and $\tilde\chi_t>0$, the sufficient condition of Step 1 implies (IC), and $\tilde\phi^{CF}(n_0)$ satisfies (LL) since $b>0$, so it is feasible. $\hfill\square$

\subsection{Proof of Proposition \ref*{prop:private success}}\label{app:proof private success}

Throughout, $\delta=1$, and the notation of Appendices~\ref{app:preliminaries} and~\ref{app:proof main result} is maintained. Write $\gamma:=1-p_H$, so that $\gamma'=\gamma(1-q_H)$. Multiplying the identities \eqref{eq:app bayes} at the indices $0,\dots,t-2$ gives the reach identities $\mathcal D_t\,\mu_{t-1}=\mu_0\gamma'^{\,t-1}$ and $\mathcal D_t\,(1-\mu_{t-1})=(1-\mu_0)(1-p_L)^{t-1}$. Fix the uninformed duration $n_0$ of the proposition. If $n_0\le1$, the PBS contract $\phi^\ast(n_0)$ posts a bonus sequence that never rises, since its research-stage bonus, if any, is $c/P_0>\underline b$ and is followed by $\underline b$. An agent who holds a success therefore gains nothing by delaying its disclosure, so $\phi^\ast(n_0)$ remains feasible when success is private. Since it is optimal among contracts with this uninformed duration when success is contractible, it attains the principal's optimal payoff here. Let $n_0>1$ from now on. Fix arbitrary informed durations $n_1$. The proof runs as follows. Steps 1 to 3 bound the agent's payoff of every feasible contract at the schedule $(n_0,n_1)$ from below by $\min\{R_0,R_\infty\}$, with $R_0$ and $R_\infty$ defined in Step 3. Step 4 shows that the flattened PBS contract attains $R_0$ and the constant-bonus contract attains $R_\infty$. Step 5 signs $R_\infty-R_0$, and Step 6 assembles the proposition.

For the $t$-informed track, with $n:=n_{1t}$ periods, the value $F_t$ of Appendix~\ref{app:preliminaries} reads, at $\delta=1$,
\[
F_t=\sum_{j=1}^{n}\gamma^{\,j-1}\big(w_{1t}(t{+}j)+p_Hb_{1t}(t{+}j)-c\big)+\gamma^{\,n}\,w_{1t}(t{+}n{+}1),
\]
where the severance term is absent at $n=\infty$. For an infinite track, the series converges absolutely, since $\gamma<1$ and the model bounds the transfers of an infinite track. The informed payoff $Y_t=\ell_t+F_t$ is therefore finite.

\emph{Step 1: holding a success.} I first show that no bonus sequence that the agent faces can rise, since the agent would otherwise hold a success for one period. Steps 1 to 3 use (IC) only against deviations that depart from the obedient strategy after an event of positive probability. The obedient agent's payoff and his continuation payoff after such an event are finite, since $n_0<\infty$ and every $Y_t$ is finite, so (IC) requires the deviator's continuation payoff to be at most the obedient agent's. A success occurs in period $t\le n_0$ with probability $\mathcal D_tP_{t-1}>0$, and in period $j$ of the $k$-informed track with probability $\mathcal D_kQ_{k-1}\gamma^{\,j-k-1}p_H>0$. Rather than disclosing it at once, the agent can hold it, collect the next period's wage and disclose in that period, or collect the severance if his track ends. After a success in period $t\le n_0$, he can also report ``$F$'', collect $\ell_t+w_{1t}(t{+}1)$, and disclose in period $t{+}1$ on the $t$-informed track, or keep only these payments if that track is empty. These deviations give
\[
b_0(t)\ge\ell_t+w_{1t}(t{+}1)+b_{1t}(t{+}1),
\qquad
b_0(t)\ge w_0(t{+}1)+b_0(t{+}1)\ \ (t<n_0),
\qquad
b_0(n_0)\ge w_0(n_0{+}1),
\]
the first inequality whenever the $t$-informed track is nonempty, and $b_0(t)\ge\ell_t+w_{1t}(t{+}1)=Y_t$ when it is empty. Holding a success on an informed track likewise gives $b_{1k}(j)\ge w_{1k}(j{+}1)+b_{1k}(j{+}1)$ for each period $j$ of every track, where the right side becomes the track severance at the track's last period. Every bonus sequence that the agent faces is therefore nonincreasing, each bonus covering the next period's wage and bonus. Call these inequalities the \emph{holding constraints}.

\emph{Step 2: a necessary system.} For every feasible contract, the tuple $(b_0,w_0,Y,w_0(n_0{+}1))\ge0$ satisfies
\[
\begin{gathered}
(\mathrm{N1})\ \ Y_t\ \le\ b_0(t)-\underline b,
\\
(\mathrm{N2})\ \ Y_t\ \ge\ w_0(t{+}1)+\max\big\{Y_{t+1},\ -c+p_Hb_0(t{+}1)+\gamma\,Y_{t+1}\big\}\ \ (t<n_0),
\qquad
Y_{n_0}\ \ge\ w_0(n_0{+}1),
\end{gathered}
\]
\[
(\mathrm{N3})\ \ -c+P_{n_0-1}\big(b_0(n_0)-w_0(n_0{+}1)\big)+Q_{n_0-1}\big(Y_{n_0}-w_0(n_0{+}1)\big)\ \ge\ 0,
\]
besides the holding constraints, where (N1) holds in every period whose informed track is nonempty. The agent's payoff is
\[
U^0_1=\sum_{t\le n_0}\mathcal D_t\big[-c+w_0(t)+P_{t-1}\,b_0(t)+Q_{t-1}\,Y_t\big]+\mathcal D_{n_0}\nu_{n_0-1}\,w_0(n_0{+}1).
\]
(N2) is (IC) against concealing a period-$t$ finding for one period, where the finding occurs with probability $\mathcal D_tQ_{t-1}>0$. In period $t{+}1$ the concealer shirks or works, discloses a success at once, and otherwise reports, while at $t=n_0$ he collects the severance. Since (LL) makes every wage nonnegative, (N2) also implies $Y_t\ge Y_{n_0}\ge0$ for all $t\le n_0$. (N3) is (IC) against the strategy that shirks in period $n_0$ and takes the severance. This strategy differs from the obedient strategy only at the on-path pair $(n_0,n_0{-}1)$. Since $P_{n_0-1}+Q_{n_0-1}=1-\nu_{n_0-1}$, the obedient agent's continuation payoff there exceeds the severance by the left side of (N3). For (N1), fix a period $t$ whose informed track is nonempty. Write $\Upsilon_j:=\sum_{i\ge j}\gamma^{\,i-j}\big(w_{1t}(t{+}i)+p_Hb_{1t}(t{+}i)-c\big)-\big(w_{1t}(t{+}j)+b_{1t}(t{+}j)-\underline b\big)$, where, for a finite track of $n$ periods, the sum stops at $i=n$ and adds the severance term $\gamma^{\,n+1-j}w_{1t}(t{+}n{+}1)$, as in the formula for $F_t$. The holding constraints and $p_H\underline b=c$ give $\Upsilon_j\le\gamma\,\Upsilon_{j+1}$. For an infinite track, $\Upsilon$ is bounded above, the transfers being bounded, so iterating this inequality gives $\Upsilon_j\le0$ for every $j$. For a finite track of $n$ periods, the last value is $\Upsilon_n=\gamma\big(\underline b-b_{1t}(t{+}n)+w_{1t}(t{+}n{+}1)\big)\le0$, since working the last period must weakly beat shirking it for the informed agent, who reaches it with probability $\mathcal D_tQ_{t-1}\gamma^{\,n-1}>0$. Either way $\Upsilon_1\le0$, that is, $F_t\le w_{1t}(t{+}1)+b_{1t}(t{+}1)-\underline b$. Adding $\ell_t$ and the first holding constraint gives (N1). Step 3 uses (N1) only in the last period $n_0$ and treats the case of an empty last track separately.

\emph{Step 3: a lower bound on the agent's payoff.} I show that every feasible contract has $U^0_1\ge\min\{R_0,R_\infty\}$, with $R_0$ and $R_\infty$ defined below. Write $P:=P_{n_0-1}$, $Q:=Q_{n_0-1}$, $\varkappa:=Q/P$, and $y:=Y_{n_0}\ge0$. The wages and the severance are nonnegative, enter the payoff with positive weights, and only tighten the holding constraints, (N2), and (N3). Dropping them therefore leaves the payoff bound $U^0_1\ge U(b_0,Y):=\sum_{t\le n_0}\mathcal D_t\big[-c+P_{t-1}b_0(t)+Q_{t-1}Y_t\big]$ and the following consequences of the system: $b_0$ is nonincreasing, $Pb_0(n_0)+Qy\ge c$, and
\begin{equation}\label{eq:oa ps chain}
Y_t\ \ge\ \max\big\{Y_{t+1},\ -c+p_Hb_0(t{+}1)+\gamma\,Y_{t+1}\big\}\qquad(t<n_0).
\end{equation}
The function $U$ is strictly increasing in every $b_0(t)$ and $Y_t$, and the right side of \eqref{eq:oa ps chain} is nondecreasing in $b_0(t{+}1)$ and $Y_{t+1}$.

Suppose first that the $n_0$-informed track is nonempty, so that (N1) at $n_0$ adds $b_0(n_0)\ge\underline b+y$. Write $B(y):=(c-Qy)/P$ and $b_{\min}(y):=\max\{B(y),\underline b+y\}$, so that $b_0(t)\ge b_0(n_0)\ge b_{\min}(y)$ for all $t$. Let $\tilde Y(y)$ solve the chain $\tilde Y_{n_0}:=y$ and $\tilde Y_t:=\max\{\tilde Y_{t+1},\,-c+p_Hb_{\min}(y)+\gamma\tilde Y_{t+1}\}$. Backward induction on \eqref{eq:oa ps chain} gives $Y_t\ge\tilde Y_t(y)$, so $U^0_1\ge V(y):=U(b_{\min}(y),\tilde Y(y))$, and it remains to minimize $V$ over $y\ge0$. Note that $B(y)\ge\underline b+y$ if and only if $y\le y_{\mathrm{cap}}:=(c-P\underline b)/(P+Q)$, which is positive since $P<p_H$. On $[0,y_{\mathrm{cap}}]$, $b_{\min}(y)=B(y)$, and the margin $m_t:=B(y)-\underline b-\tilde Y_t$ determines which branch attains the maximum in the chain. The work branch weakly exceeds the shirk branch at $t$ exactly when $m_{t+1}\ge0$, and along the work branch $m_t=\gamma\,m_{t+1}$ by $c=p_H\underline b$. Since $m_{n_0}=B(y)-\underline b-y\ge0$, backward induction shows that the work branch attains the maximum at every $t$, so $\tilde Y_t=-c+p_HB(y)+\gamma\tilde Y_{t+1}$, with $m_t=\gamma^{\,n_0-t}m_{n_0}\ge0$ at every $t$. Along it each $\tilde Y_t$ is affine in $y$ with coefficient $\gamma^{\,n_0-t}(1+\varkappa)-\varkappa$, and $B(y)$ has the slope $-\varkappa$, so $V$ is affine on $[0,y_{\mathrm{cap}}]$ with slope
\[
\Delta:=\sum_{t\le n_0}\mathcal D_t\Big[Q_{t-1}\,(1+\varkappa)\,\gamma^{\,n_0-t}-\varkappa\,\big(P_{t-1}+Q_{t-1}\big)\Big].
\]
For $y>y_{\mathrm{cap}}$, instead, $b_{\min}(y)=\underline b+y$ and the chain is flat, $\tilde Y_t=y$, since $-c+p_H(\underline b+y)+\gamma y=y$. Then $V(y)=\sum_{t\le n_0}\mathcal D_t\big[-c+P_{t-1}(\underline b+y)+Q_{t-1}y\big]$ has the strictly positive slope $\sum_{t\le n_0}\mathcal D_t(P_{t-1}+Q_{t-1})$, and the two expressions agree at $y_{\mathrm{cap}}$, where $B(y_{\mathrm{cap}})=\underline b+y_{\mathrm{cap}}$. The minimum of $V$ over $y\ge0$ therefore sits at an endpoint of $[0,y_{\mathrm{cap}}]$. It is either $R_0:=V(0)$, where $B(0)=c/P_{n_0-1}$, or $R_\infty:=V(y_{\mathrm{cap}})$, where
\[
B(y_{\mathrm{cap}})=\underline b+y_{\mathrm{cap}}=b_\infty:=\frac{c+Q_{n_0-1}\underline b}{P_{n_0-1}+Q_{n_0-1}},
\]
the bonus of the constant-bonus contract, at which the uninformed agent is just willing to work in period $n_0$.

Suppose instead that the $n_0$-informed track is empty. Then (N1) at $n_0$ is unavailable, and a sharper constraint replaces it. The strategy that shirks in period $n_0$ and reports ``$F$'' in period $n_0$ differs from the obedient strategy only at the on-path pair $(n_0,n_0{-}1)$. There it collects the report lump sum and the track severance, $\ell_{n_0}+w_{1n_0}(n_0{+}1)=y$, in place of the obedient agent's continuation payoff from period $n_0$. Feasibility therefore implies $-c+Pb_0(n_0)+Qy+\nu_{n_0-1}w_0(n_0{+}1)\ge y$. With $y\ge w_0(n_0{+}1)$, the terminal part of (N2), and $1-Q-\nu_{n_0-1}=P$, this rearranges to $b_0(n_0)\ge c/P+y$. Then $b_0(t)\ge c/P$ at every $t$, and backward induction on \eqref{eq:oa ps chain}, from $Y_{n_0}=y\ge0$, gives $Y_t\ge\tilde Y_t(0)$, the chain above at $y=0$, where $b_{\min}(0)=c/P$. Hence $U^0_1\ge V(0)=R_0$. The bound $\min\{R_0,R_\infty\}$ therefore holds for all informed durations $n_1$.

\emph{Step 4: report-free attainment.} For every $y\in[0,y_{\mathrm{cap}}]$, consider the contract that posts
\[
\big(B(y),\;\dots,\;B(y),\;\underline b+y,\;\underline b+y,\;\dots\big),
\]
with the bonus $B(y)$ in each research-stage period $t\le n_0$ and the bonus $\underline b+y$ in every development-stage period $t>n_0$. It sets $n_{1k}=\infty$ for every $k$, pays no wage and no lump sum, and every informed track posts the same bonus as the uninformed track in every period in which both are active. At $y=0$ it is the flattened PBS contract, and at $y=y_{\mathrm{cap}}$ it is the constant-bonus contract, the single bonus $b_\infty$ posted forever. I show that it is feasible with the agent's payoff $V(y)=R_0+\Delta y$. The posted bonus sequence is nonincreasing, since $B(y)\ge\underline b+y$ on $[0,y_{\mathrm{cap}}]$, and no wage is paid, so a held success can only be disclosed later at a weakly lower bonus. Disclosing at once is therefore optimal at every private history, and every success pays the current posted bonus. The agent's problem is thus that of the main model, with the values $\mathcal W_t$, $X_t$ and $V^0$ of Step 1 of Theorem~\proofref{thm:main result}'s proof. The values $F_t$ of the informed tracks satisfy $F_t=-c+p_HB(y)+\gamma F_{t+1}$ for $t<n_0$ and $F_{n_0}=\sum_{j\ge1}\gamma^{\,j-1}\big(p_H(\underline b+y)-c\big)=y$, so they reproduce the chain of Step 3, $F_t=\tilde Y_t(y)$, and the obedient agent's payoff is $V(y)$. Working in every period is optimal for an informed agent. Indeed, when the informed agent's continuation payoff is $W$, working a period with bonus $b$ weakly beats shirking it if and only if $b-\underline b\ge W$. In the development stage, every effort rule of an informed agent collects $p_H(\underline b+y)-c=p_Hy$ in each worked period and reaches its $i$-th worked period with probability $\gamma^{\,i-1}$. So each such effort rule is worth at most $\sum_{i\ge1}\gamma^{\,i-1}p_Hy=y$, which working in every period attains. The informed agent's continuation payoff in every development-stage period is therefore $y$, and the condition holds with equality there. In a research-stage period $2\le t\le n_0$, the continuation payoff is $F_t$, so the condition reads $m_t\ge0$ for the margin of Step 3. Backward induction over the finitely many research-stage periods then gives $\mathcal W_t=F_t$ and $X_t=Y_t=F_t$ for every $t$, so reporting at once is a best response. Two arguments complete the proof of feasibility.

\emph{Coupling.} A strategy that reports ``$F$'' while uninformed in some period $s<n_0$ is weakly dominated by the strategy that copies its efforts period by period, sends no report before period $n_0$, and reports in period $n_0$ if the copied strategy has reported by then. The contract is report-free, so on every path the two strategies collect the same bonus in every period. The fake options can therefore be dropped from every pair before period $n_0$.

\emph{Induction.} Let $J^{\mathrm{dev}}(d)$ be the optimal value of an uninformed agent who enters the development stage on a track after $d$ silent worked periods. The stage is stationary and $\delta=1$, so a shirked period changes neither his belief nor any transfer. Only the sequence of his worked periods matters, and therefore
\[
J^{\mathrm{dev}}(d)=\max\big\{0,\ \xi^{\mathrm{dev}}(d)+\nu_dJ^{\mathrm{dev}}(d{+}1)\big\},
\qquad
\xi^{\mathrm{dev}}(d):=-c+P_d(\underline b+y)+Q_d\,y,
\]
where $\mu_d$, $P_d$, $Q_d$ and $\nu_d$ are the belief and the probabilities of Appendix~\ref{app:preliminaries} after $d$ silent worked periods, and a finding hands him the continuation payoff $y$ of an informed agent. The flow $\xi^{\mathrm{dev}}(d)$ is decreasing in $d$, since $P_d$ and $Q_d$ are decreasing (Fact 1), and satisfies $\xi^{\mathrm{dev}}(n_0{-}1)=-c+P\underline b+(P+Q)y\le0$ on $[0,y_{\mathrm{cap}}]$. Every effort rule after $d\ge n_0-1$ silent worked periods thus collects a nonnegatively weighted sum of nonpositive flows, so $J^{\mathrm{dev}}(d)=0$ for every $d\ge n_0-1$. Define the value of working in every period through $n_0$, $\tilde J(t,d):=\xi_t(\mu_d)+\nu_d\tilde J(t{+}1,d{+}1)$ on the pairs $(t,d)$ with $d\le t-1$, where $\xi_t(\mu):=-c+P(\mu)B(y)+Q(\mu)F_t$. The terminal value $\tilde J(n_0{+}1,d):=J^{\mathrm{dev}}(d)$ is the value of reporting ``$F$'' in period $n_0$ after $d$ silent worked periods. I claim that $\tilde J(t,d)\ge\tilde J(t{+}1,d)$ at every pair, so working in every period dominates the shirk branch everywhere. At $t=n_0$, the shirk branch is worth $J^{\mathrm{dev}}(d)$, obtained by faking in period $n_0$. If $J^{\mathrm{dev}}(d)>0$, then $\tilde J(n_0,d)-J^{\mathrm{dev}}(d)=\xi_{n_0}(\mu_d)-\xi^{\mathrm{dev}}(d)=P_d\big(B(y)-\underline b-y\big)\ge0$, by $F_{n_0}=y$. If $J^{\mathrm{dev}}(d)=0$, then $\tilde J(n_0,d)\ge\xi_{n_0}(\mu_d)\ge\xi_{n_0}(\mu_{n_0-1})=0$, since $\xi_{n_0}$ is nondecreasing in the belief and $\xi_{n_0}(\mu_{n_0-1})=0$ by the definition of $B(y)$. For $t<n_0$,
\[
\tilde J(t,d)-\tilde J(t{+}1,d)=Q(\mu_d)\,p_H\big(B(y)-\underline b-F_{t+1}\big)+\nu_d\big[\tilde J(t{+}1,d{+}1)-\tilde J(t{+}2,d{+}1)\big]\ \ge\ 0,
\]
the first term by $m_{t+1}\ge0$ and the second by induction. With the fake options before period $n_0$ dropped by the coupling, backward induction on the best-response recursion then gives $V^0(t,d)=\tilde J(t,d)$ at every pair, so $V^0(1,0)=\tilde J(1,0)=U^0_1+\mathcal D_{n_0}\nu_{n_0-1}J^{\mathrm{dev}}(n_0)=U^0_1$. The contract is feasible.

\emph{Step 5: the sign of $\Delta$.} By the reach identities, $\mathcal D_tQ_{t-1}=\mu_0q_H\gamma\,\gamma'^{\,t-1}$ and $\mathcal D_t\big(P_{t-1}+Q_{t-1}\big)=\mathcal D_t-\mathcal D_{t+1}$. Summing the geometric series $\sum_{t\le n_0}\gamma^{\,n_0-t}\gamma'^{\,t-1}=(\gamma^{\,n_0}-\gamma'^{\,n_0})/(\gamma-\gamma')$ and using $\gamma-\gamma'=\gamma q_H$ gives
\[
\Delta=(1+\varkappa)\,\mathcal K-\varkappa\,\mathcal N,
\qquad
\mathcal K:=\mu_0\big(\gamma^{\,n_0}-\gamma'^{\,n_0}\big),
\qquad
\mathcal N:=1-\mathcal D_{n_0+1}.
\]
Here $\mathcal K$ is the probability that a finding, but no success, has arrived by period $n_0$, and $\mathcal N$ is the probability that a success or a finding arrives by period $n_0$. Multiply $\Delta$ by $\mathcal D_{n_0}P_{n_0-1}/\mu_0>0$. Using $(1+\varkappa)P_{n_0-1}=1-\nu_{n_0-1}$ and $\mathcal D_{n_0}Q_{n_0-1}=\mu_0q_H\gamma\,\gamma'^{\,n_0-1}$, the product is
\[
\big(\gamma^{\,n_0}-\gamma'^{\,n_0}\big)\big(\mathcal D_{n_0}-\mathcal D_{n_0+1}\big)-q_H\gamma\,\gamma'^{\,n_0-1}\big(1-\mathcal D_{n_0+1}\big),
\]
which is affine in $\mu_0$ by $\mathcal D_t=\mu_0\gamma'^{\,t-1}+(1-\mu_0)(1-p_L)^{t-1}$. Call its values $h_0$ at $\mu_0=0$ and $h_1$ at $\mu_0=1$. At $\mu_0=1$, factoring $\gamma'^{\,n_0-1}$,
\[
h_1=\gamma'^{\,n_0-1}\Big[\big(\gamma^{\,n_0}-\gamma'^{\,n_0}\big)(1-\gamma')-(\gamma-\gamma')\big(1-\gamma'^{\,n_0}\big)\Big]\ <\ 0,
\]
since $(\gamma^{\,n_0}-\gamma'^{\,n_0})/(\gamma-\gamma')=\sum_{j<n_0}\gamma^{\,n_0-1-j}\gamma'^{\,j}<\sum_{j<n_0}\gamma'^{\,j}=(1-\gamma'^{\,n_0})/(1-\gamma')$, termwise by $\gamma<1$ and strictly at $j=0$ because $n_0>1$. At $\mu_0=0$, substituting $\gamma'=\gamma(1-q_H)$ and factoring $\gamma^{\,n_0}$,
\[
h_0=\gamma^{\,n_0}\big(1-(1-q_H)^{n_0}\big)\big(1-(1-p_L)^{n_0}\big)\Big[f(1-p_L)-f(1-q_H)\Big],
\]
where $f(x):=(1-x)x^{\,n_0-1}/(1-x^{\,n_0})=1/\sum_{i=0}^{n_0-1}x^{-i}$ is strictly increasing on $(0,1)$, since each summand $x^{-i}$ with $i\ge1$ is strictly decreasing. The two middle factors are positive since $q_H,p_L\in(0,1)$, so $\mathrm{sign}(h_0)=\mathrm{sign}\big((1-p_L)-(1-q_H)\big)=\mathrm{sign}(q_H-p_L)$. Hence the following holds for every prior $\mu_0\in(0,1)$. If $q_H\le p_L$, then $h_0\le0$ and $h_1<0$, so $\Delta<0$ and $R_\infty<R_0$. If $q_H>p_L$, the affine expression has the single zero $\bar\mu:=h_0/(h_0-h_1)\in(0,1)$, with $\Delta>0$ and $R_0<R_\infty$ at $\mu_0<\bar\mu$, and $\Delta<0$ and $R_\infty<R_0$ at $\mu_0>\bar\mu$.

\emph{Step 6: the informed durations and the assembly.} The obedient agent discloses every success at once, so every schedule's surplus is that of the main model. By Steps 1 to 3, every feasible contract at the schedule $(n_0,n_1)$ pays the agent at least $\min\{R_0,R_\infty\}$, for all informed durations $n_1$. The surplus rises with every informed duration, a worked informed period yielding $p_H\pi-c>0$, so $S(n_0,n_1)\le S(n_0,\infty\mathbf 1)$. Among contracts with uninformed duration $n_0$, the principal's payoff is therefore at most $S(n_0,\infty\mathbf 1)-\min\{R_0,R_\infty\}$. By Step 4, the flattened PBS contract earns the principal $S(n_0,\infty\mathbf 1)-R_0$ and the constant-bonus contract earns $S(n_0,\infty\mathbf 1)-R_\infty$, both at the schedule $(n_0,\infty\mathbf 1)$. By Step 5, if $q_H\le p_L$, the constant-bonus contract attains the principal's optimal payoff at every prior. If $q_H>p_L$, the flattened PBS contract attains it at priors below the threshold $\bar\mu$, and the constant-bonus contract at priors above it, both at $\mu_0=\bar\mu$. This is the proposition. $\hfill\square$

\subsection{Proof of Propositions \ref*{prop:bad news} and \ref*{prop:bad news efficient}}\label{app:proof bad news}

Throughout this section, the silence factor $\rho$ of Section~\ref{sec:bad news} satisfies $\rho\ge1$, and $\bar b=c/p_L$. The notation below is local to this section. With $q_H=0$ and $q_L\in(0,1)$, a worked period at belief $\mu$ yields a success, a finding but no success, or silence with probabilities
\[
P(\mu):=p_L+(p_H-p_L)\mu,\qquad Q(\mu):=(1-\mu)\,q_L(1-p_L),\qquad \nu(\mu):=\mu\nu_H+(1-\mu)\nu_L,
\]
where $\nu_H:=1-p_H$ and $\nu_L:=(1-p_L)(1-q_L)$ are the probabilities of silence conditional on the state, and $\rho=\nu_H/\nu_L$. Silence multiplies the likelihood ratio of $\omega=H$ to $\omega=L$ by $\rho\ge1$, so the on-path belief $\mu_t$ after $t$ silent worked periods is nondecreasing in $t$ and interior. Write $P_t$ and $\nu_t$ for the on-path probabilities and $\mathcal D_t:=\delta^{t-1}\prod_{s\le t-2}\nu_s$ for the discounted probability of reaching period $t$ on the uninformed track, as in (Fact 4).

\emph{Proposition~\ref{prop:bad news}.} Let $p_L\pi<c$, and fix any $\delta\in(0,1]$. Schedules with $n_0\le\infty$ are admitted under the model's maintained bounds, and every value is taken to be a supremum over strategies. First, at any schedule $(n_0,\mathbf 0)$, the myopic contract is feasible with $U^0_1=0$. As a direct mechanism at this schedule, it posts $b_0(t)=c/P_{t-1}$ in every period $t\le n_0$ and sets every other transfer to zero. Its informed tracks are empty, so a report ends all transfers, while an informed agent who does not report faces the posted bonus. The obedient agent's period payoff is $P_{t-1}b_0(t)-c=0$ for every $t$, so $U^0_1=0$. To verify (IC), fix any strategy and condition on a history at which the agent works in period $t$. If he is uninformed after $d\le t-1$ silent worked periods, he holds the belief $\mu_d$ and perceives the success probability $P_d$, so his expected period payoff is $c[P_d/P_{t-1}-1]\le0$, since $\rho\ge1$ implies $\mu_d\le\mu_{t-1}$ and $P$ increases in the belief. If he is informed, it is $p_Lb_0(t)-c<0$, since $b_0(t)=c/P_{t-1}<c/p_L$ at an interior belief. Shirked periods and reports pay nothing, so every strategy's value is a sum of nonpositive conditional period payoffs. The exchange of conditioning and summation is legitimate because the positive part of any path is a single bonus, at most $c/P_0$. Hence $V^0(1,0)\le0=U^0_1$, and (IC) holds with equality. Second, no contract at any schedule earns the principal more. Under (LL) the agent's payoff is nonnegative, since shirking in every period collects nonnegative transfers. With the total surplus under the obedient strategy grouped by the arrival period, every entered informed track contributes nonpositively, because a worked informed period yields $p_L\pi-c<0$. The principal's payoff is therefore at most $S_0(n_0):=\sum_{t\le n_0}\mathcal D_t(P_{t-1}\pi-c)$, hence at most the supremum of $S_0$ over all uninformed durations up to $\infty$. This supremum is finite, since $\mathcal D_t$ decays geometrically, by $p_L>0$. It is attained at the duration $0$ or $\infty$, since the summands change sign at most once, from negative to positive, as $P_{t-1}$ is nondecreasing. It equals the first-best surplus, since informed work is wasteful, and the myopic contract at an attaining duration collects it and leaves the agent zero payoff. The same argument applies when the finding is contractible. Indeed, the agent's payoff is again nonnegative under (LL), and the surplus is again at most the first-best surplus. The myopic contract, which recommends no work after a finding, remains feasible and leaves the agent zero payoff, because the agent then has only effort deviations, a subset of those ruled out above. This proves Proposition~\ref{prop:bad news}.

\emph{Proposition~\ref{prop:bad news efficient}.} Henceforth $\delta=1$ and $c<p_L\pi$. Since even the informed track is worth running, infinite schedules are admitted, $n_0\le\infty$ and $n_{1k}\le\infty$. The transfer bounds that the model maintains on infinite tracks are not imposed here. Dropping them only enlarges the class of contracts, so the upper bounds on the principal's payoff derived below remain valid in the model's class, and the two contracts that attain them post bounded bonuses. Payoffs, which are undiscounted expected totals, remain well defined in $[-\infty,\infty]$. Indeed, a worked period ends the relationship with probability at least $p_L>0$, whatever the agent knows, so under every strategy the expected number of worked periods is at most $1/p_L$. The negative part of every payoff sum, which comes from effort costs, is thus summable. Every such sum therefore has a limit in $(-\infty,\infty]$, which is also the value that the main text assigns to an infinite sum. Every surplus is finite by Step 1 below, and a contract with an infinite payoff to the agent earns the principal $-\infty$ and is ignored. A contract implements a schedule if it satisfies (LL) and (IC) at that schedule, as in the first step \eqref{eq:first step}. Define, for $n\le\infty$,
\[
\mathcal H_n:=\bar b\,(p_H-p_L)\sum_{j<n}\nu_H^{\,j}=\mathcal H_\infty\big(1-\nu_H^{\,n}\big),
\qquad
\mathcal H_\infty:=\frac{\bar b\,(p_H-p_L)}{p_H},
\]
and the threshold
\[
\pi_c:=\bar b+\frac{\mu_0\,\mathcal H_\infty\,(1-\nu_L)}{(1-\mu_0)\,(1-p_L)\,q_L}.
\]

\emph{Step 1: reach identities and the surplus split.} For a belief $\mu$ with likelihood ratio $\theta$, write $\mu^{(j)}$ for the belief with likelihood ratio $\rho^{\,j}\theta$. The Bayes identities $\nu(\mu)\,\mu^{(1)}=\mu\,\nu_H$ and $\nu(\mu)\,(1-\mu^{(1)})=(1-\mu)\,\nu_L$, multiplied over the beliefs $\mu,\mu^{(1)},\dots,\mu^{(d-1)}$, give
\begin{equation}\label{eq:app xsurvival}
\prod_{j<d}\nu\big(\mu^{(j)}\big)\cdot\mu^{(d)}=\mu\,\nu_H^{\,d},
\qquad
\prod_{j<d}\nu\big(\mu^{(j)}\big)\cdot\big(1-\mu^{(d)}\big)=(1-\mu)\,\nu_L^{\,d},
\end{equation}
which become $\mathcal D_{d+1}\mu_d=\mu_0\nu_H^{\,d}$ and $\mathcal D_{d+1}(1-\mu_d)=(1-\mu_0)\nu_L^{\,d}$ at $\mu=\mu_0$, since $\mu^{(d)}$ is then the on-path belief $\mu_d$. Now group the total surplus under the obedient strategy by the arrival period, as in Appendix~\ref{app:preliminaries}. By \eqref{eq:app xsurvival}, the probability mass with $\omega=H$ reaching the uninformed period $t$ is $\mu_0\nu_H^{\,t-1}$ at the flow $p_H\pi-c$, and the mass with $\omega=L$ is $(1-\mu_0)\nu_L^{\,t-1}$ at the flow $p_L\pi-c$. The period-$k$ informed track carries the mass $(1-\mu_0)\nu_L^{\,k-1}q_L(1-p_L)$, all with $\omega=L$, and works its $n_{1k}$ periods at the flow $p_L\pi-c$ and the survival probability $1-p_L$. Summing the geometric series, $S(n_0,n_1)=S_0(n_0)+S_1(n_0,n_1)$ with
\[
\begin{gathered}
S_0(n_0)=\mu_0(p_H\pi-c)\frac{1-\nu_H^{\,n_0}}{p_H}+(1-\mu_0)(p_L\pi-c)\frac{1-\nu_L^{\,n_0}}{1-\nu_L},\\
S_1(n_0,n_1)=(1-\mu_0)(p_L\pi-c)\frac{(1-p_L)q_L}{p_L}\sum_{k=1}^{n_0}\nu_L^{\,k-1}\big(1-(1-p_L)^{n_{1k}}\big),
\end{gathered}
\]
every term finite and, by $c<p_L\pi$, every increment of $S_0$ strictly positive. At the fully infinite schedule, $S_1(\infty,\infty\mathbf 1)=(1-\mu_0)(p_L\pi-c)(1-p_L)q_L/[p_L(1-\nu_L)]$ is affine and strictly increasing in $\pi$. It equals $\mu_0\mathcal H_\infty$ exactly at $\pi=\pi_c$, by the definition of $\pi_c$.

\emph{Step 2: the shirk-then-fake bound.} Every contract implementing a schedule $(n_0,n_1)$ satisfies
\begin{equation}\label{eq:app fake bound}
U^0_1\ \ge\ \mu_0\,\sup_{k\le n_0}\mathcal H_{n_{1k}},
\end{equation}
with the right side taken to be $0$ when every track is empty, in which case $U^0_1\ge0$ under (LL), since shirking in every period collects nonnegative transfers. Fix a track $k$ with $n:=n_{1k}\ge1$ periods, indexed by their position $j$, and with transfers $b_j:=b_{1k}(k{+}j)$ and $w_j:=w_{1k}(k{+}j)$, the severance counted as the last wage entry. Write $s_j:=w_j+p_Lb_j-c$ for the informed flow at position $j$, and $A^L_j$ for the value of working in every period from position $j$ on, conditional on $\omega=L$. Since $q_H=0$, $A^L_j$ is also the informed agent's value. The bound is trivial if $U^0_1=\infty$, so let $U^0_1<\infty$. The obedient agent enters the track with positive probability and collects $\ell_k+A^L_1$ there, and every payoff sum has a summable negative part, so $A^L_1$ is finite, and so is every $A^L_j$. The informed agent reaches position $j$ with probability $(1-\mu_0)\nu_L^{\,k-1}q_L(1-p_L)^{\,j}>0$, so (IC) against shirking once there and working afterward, and against shirking forever from there, implies
\[
b_j\ \ge\ \bar b+A^L_{j+1}
\qquad\text{and}\qquad
A^L_j\ \ge\ \sum_{i\ge j}w_i\ \ge\ 0 ,
\]
where the first follows from $s_j+(1-p_L)A^L_{j+1}\ge w_j+A^L_{j+1}$. Next, let $A^H_j$ be the value of working in every period from position $j$ on, conditional on $\omega=H$, given by $A^H_j=w_j+p_Hb_j-c+\nu_HA^H_{j+1}$ and well defined in $[0,\infty]$, since each flow dominates $s_j\ge p_LA^L_{j+1}\ge0$. Working in every period of the track is one effort rule of a faker. Under this effort rule, a real finding does not change his payoff, because it leaves him the continuation payoff $A^L_{j+1}$ that he was collecting anyway. His value at belief $\mu$ is therefore at least $\mu A^H_1+(1-\mu)A^L_1$. The difference $\Delta^H_j:=A^H_j-A^L_j$ satisfies, by $\nu_H-(1-p_L)=-(p_H-p_L)$ and the first inequality above,
\[
\Delta^H_j=(p_H-p_L)\big(b_j-A^L_{j+1}\big)+\nu_H\Delta^H_{j+1}\ \ge\ \mathcal H_1+\nu_H\Delta^H_{j+1},
\]
which, applied repeatedly from $j=1$, gives $\Delta^H_1\ge\mathcal H_n$ if $n<\infty$. If $n=\infty$, the tails $\nu_H^{\,i}A^L_{i+1}\le(1-p_L)^{\,i}A^L_{i+1}$ vanish because the series $A^L_1$ converges, so $\Delta^H_1\ge\mathcal H_\infty$. Finally, consider the deviation that shirks in every period $t\le k$, reports ``$F$'' in period $k$, and plays the track as a faker. Shirking preserves the prior at zero cost, and without discounting the delay is free. This deviation earns at least $\ell_k+A^L_1+\mu_0\mathcal H_n\ge\mu_0\mathcal H_n$, so \eqref{eq:app fake bound} follows from (IC).

\emph{Step 3: feasibility of the myopic contract and of posting $\bar b$.} On a schedule with $n_1\equiv\mathbf 0$, the myopic contract is feasible with $U^0_1=0$, by the first half of Proposition~\proofref{prop:bad news}'s proof above, applied with $\delta=1$.

On the fully infinite schedule, the contract that posts $\bar b$ in every period of every track is feasible with $U^0_1=\mu_0\mathcal H_\infty$. An informed agent earns zero from every action. A worked period yields $p_L\bar b-c=0$, a shirked period yields $0$, and a report pays nothing and leads to a track of zero value. Every strategy of an informed agent is thus worth zero, so obedient effort is optimal for him and reporting at once is a best response. While the agent is unreported and uninformed, his history matters only through the period and the number of silent worked periods, and a real finding is worth zero and is reported at once without loss. His strategies therefore reduce to those that work a set of uninformed periods and fake at most once. Working a period at belief $\mu'$ yields $P(\mu')\bar b-c=\mu'\mathcal H_1$. By \eqref{eq:app xsurvival} with any starting belief $\mu$, the $i$-th worked period contributes $\mu\nu_H^{\,i-1}\mathcal H_1$ wherever it sits, since shirked periods cost neither reach probability nor discounting at $\delta=1$. A faker's value at belief $\mu$ is therefore $\mu\mathcal H_i$ from working $i$ periods, at most $\mu\mathcal H_\infty$. A strategy working $i$ uninformed periods and then faking is worth $\mu_0\mathcal H_i+\mu_0\nu_H^{\,i}\mathcal H_\infty=\mu_0\mathcal H_\infty$, and one that never fakes is worth $\mu_0\mathcal H_i\le\mu_0\mathcal H_\infty$. The obedient agent works in every period, and every finding branch is worth zero, so he earns $\sum_{t\ge1}\mu_0\nu_H^{\,t-1}\mathcal H_1=\mu_0\mathcal H_\infty$. Hence (IC) holds.

\emph{Step 4: the bounds below and above $\pi_c$ and the assembly.} Write $\bar{\mathcal H}:=\sup_{k\le n_0}\mathcal H_{n_{1k}}\in[0,\mathcal H_\infty]$, so that \eqref{eq:app fake bound} reads $U^0_1\ge\mu_0\bar{\mathcal H}$. Note that $1-(1-p_L)^{n_{1k}}\le1-\nu_H^{\,n_{1k}}=\mathcal H_{n_{1k}}/\mathcal H_\infty\le\bar{\mathcal H}/\mathcal H_\infty$ for every $k$, since $1-p_L>\nu_H$. If $\pi\le\pi_c$, Step 1 gives
\[
S_1(n_0,n_1)\ \le\ S_1(\infty,\infty\mathbf 1)\,\frac{\bar{\mathcal H}}{\mathcal H_\infty}\ \le\ \mu_0\bar{\mathcal H}\ \le\ U^0_1,
\]
where the first inequality follows from the display of $S_1$ and $\sum_k\nu_L^{\,k-1}\le1/(1-\nu_L)$, and the second uses $S_1(\infty,\infty\mathbf 1)\le\mu_0\mathcal H_\infty$ at $\pi\le\pi_c$. The principal's payoff is therefore always at most $S_0(n_0)\le S(\infty,\mathbf 0)$. If $\pi\ge\pi_c$, comparing the split term by term gives
\[
S(\infty,\infty\mathbf 1)-S(n_0,n_1)\ \ge\ S_1(\infty,\infty\mathbf 1)\Big(1-\frac{\bar{\mathcal H}}{\mathcal H_\infty}\Big)\ \ge\ \mu_0\big(\mathcal H_\infty-\bar{\mathcal H}\big),
\]
where the first inequality holds since $S_0(n_0)\le S_0(\infty)$ and the term of $S_1$ for each track $k$ satisfies the inequality noted at the start of this step, subtracted from one. The second inequality uses $S_1(\infty,\infty\mathbf 1)\ge\mu_0\mathcal H_\infty$ at $\pi\ge\pi_c$. The principal's payoff is therefore always at most $S(n_0,n_1)-\mu_0\bar{\mathcal H}\le S(\infty,\infty\mathbf 1)-\mu_0\mathcal H_\infty$. By Step 3, below $\pi_c$, the myopic contract at the schedule $(\infty,\mathbf 0)$ attains the bound $S(\infty,\mathbf 0)$, which proves part (i). Above $\pi_c$, the contract that posts $\bar b$ in every period attains the bound $S(\infty,\infty\mathbf 1)-\mu_0\mathcal H_\infty$, which proves part (ii). $\hfill\square$

\end{appendices}

\end{document}